\documentclass[onecolumn,aps,pre,groupedaddress,showpacs,floatfix]{revtex4-1}
\usepackage{times}
\usepackage[autostyle]{csquotes}
\usepackage{subcaption}
\usepackage{graphicx,color,colortbl}
\usepackage{amsmath}
\usepackage{booktabs}
\usepackage{slashbox}
\usepackage{hyperref}
\usepackage{xcolor}
\usepackage{framed}
\usepackage{blindtext}
\usepackage{pagecolor,lipsum}%
\usepackage[most]{tcolorbox}
\usepackage[autostyle]{csquotes}
\usepackage{etoolbox,lineno}
\usepackage{afterpage}
\usepackage{graphicx}
\usepackage{epstopdf}
\usepackage{subcaption}
\usepackage{hyperref}
\usepackage{bm,upgreek}

\usepackage{amsfonts} 

\begin{document}

\title{Infection Kinetics of Covid-19: Is Lockdown a Potent Containment Tool?}

\author{Amit K Chattopadhyay}
\affiliation{                    
Mathematics and Aston Institute of Materials Research (AMRI), Aston University, Aston Triangle, Birmingham, B4 7ET, United Kingdom}
\email{a.k.chattopadhyay@aston.ac.uk} 
\author{Debajyoti Choudhury}
\affiliation{Department of Physics \& Astrophysics, University of Delhi, Delhi 110007, India}
\email{debajyoti.choudhury@gmail.com}
\author{Goutam Ghosh}
\affiliation{                    
Gandhi Institute of Engineering and Technology University, Gunupur 765022, Odisha, India}
\email{vc@giet.edu}
\author{Bidisha Kundu}
\affiliation{Formerly, Department of Aerospace Engineering, Indian Institute of Science, Bangalore 560012, India}
\email{bidishakundu.iisc@gmail.com}
\author{Sujit Kumar Nath}
\affiliation{School of Computing, University of Leeds, Leeds LS2 9JT, UK \\ Faculty of Biological Sciences, University of Leeds, Leeds LS2 9JT, UK}
\email{s.k.nath@leeds.ac.uk}

\begin{abstract}
\noindent
Covid-19 is raging a devastating trail with the highest mortality-to-infected ratio ever for a pandemic. Lack of vaccine and therapeutic has rendered social exclusion through lockdown as the singular mode of containment. Harnessing the predictive powers of Machine Learning within a 6 dimensional infection kinetic model, depicting interactive evolution of 6 infection stages - healthy susceptible ($H$), predisposed comorbid susceptible ($P$), infected ($I$), recovered ($R$), herd immunized ($V$) and mortality ($D$) - the model, PHIRVD, provides the first accurate mortality prediction of 18 countries at varying stages of strategic lockdown, up to 30 days beyond last data training. PHIRVD establishes mortality-to-infection ratio as the correct pandemic descriptor, substituting reproduction number, and highlights the importance of early and prolonged but strategic lockdown to contain secondary relapse. \\

\noindent
{\bf Significance Statement}: \\
\noindent
1. accurate prediction of the day-by-day mortality profiles of 18 countries, 30 days beyond the last data of data training, \\
\noindent
2. precise quantification of the impact of early-vs-later lockdown impositions, \\
\noindent
3. accurate prediction of secondary relapse timelines, \\
\noindent
4. establishment of mortality-to-infected ratio as the correct pandemic descriptor substituting the popular choice of reproduction number, a proven failure in predicting future infection kinetics and secondary surge. \\

\noindent
The outcomes have potential to redefine healthy policy landscape, particularly in light of secondary relapse and possible future SARS-COV/Ebola group incursion. 
\end{abstract}
\date{\today}


\maketitle

\newpage

\section*{Introduction}

\noindent
Deadlier than all pandemics in the last 100 years, barring 
HIV, Covid-19 rages on despite imposition of movement restrictions as
well as clinical testing and community health
measures\cite{non-pharma1,non-pharma2}. As of 4 August 2020, SARS-COV-2 has infected ca 18.5 million worldwide with ca 700,000 dead. Covid-19 containment has been a major strategic issue for governments worldwide, with particular emphasis on the correct lockdown timing and span. Alarming belated infection spurt have been registered in over-populated countries like India, Brazil and Iran with early and extensive lockdowns. While the low mortality rates exhibited by low-resourced yet densely populated Asian
countries have been attributed to the relative youth of the
populations~\cite{young}, rich and sparsely populated Sweden depicts an
alarming dead-to-infected ratio in contrast to its European neighbours
\cite{sweden}.

Quarantine has been advised as the best infection control
measure~\cite{control1, control2}. This has led to key questions 
as to the ideal start point and the absolute span of the ensuing
lockdown. Major cases in support of lockdown are Vietnam
and Cuba, that have claimed almost no death~\cite{vietnam, cuba}, although
such claims have been questioned~\cite{doubt}. In countries like
Italy, the UK, the US, Sweden and Brazil, with strategic reluctance
for early lockdown, comparatively softer prohibition lockdown
protocols have admittedly transpired to gruesome statistics. On the other hand,
European countries like Germany, the Netherlands, Belgium and France
as also non-European countries like Australia, New Zealand and Korea
who enforced early lockdowns
initially registered remarkably low infection and mortality rates
\cite{corona-propagation}, with $1.0<R_0<2.0$ during lockdown, that spiked later (\url{www.worldometers.info}). Many suffered post-lockdown relapse \cite{covid-relapse1,
  covid-relapse2} with a sudden spurt in infection
\cite{postpandemic}. Regions like India, Iran and New York State, with
variable quarantine measures, have all seen late infection
surges. While India resorted to an early clampdown with an early
withdrawal, New York State resorted to a late lockdown, but both with
similar numerical implications, a feature attributed to inevitable
movement of migrant workers~\cite{Lancet-kucharski}.

Analyzes of the SARS epidemic of 2003 showed that case isolation and
contact tracing \cite{non-pharma1, contact-tracing}, while highly effective if implemented at early
stages, become ineffectual if the basic infection spread occurs before
symptomatic detection \cite{sars1, sars2}. This finding was
revisited in Covid-19 transmission kinetics \cite{Nature-He}
pointing to the importance of appropriate early (pre-symptomatic)
stage strategizing. Other studies stress the importance of combining
isolation \cite{Lancet-Hellewell}, social distancing 
with widespread testing \cite{Nature-Giordano} and contact tracing
\cite{non-pharma2}. Initial predictive models 
\cite{Lancet-kucharski} used data from Wuhan and Italy \cite{Nature-Giordano}. Both efforts suffer
from a lack of robustness due to inaccurate future prediction that is reliant on sparse data, devoid of any inherent ML training protocol.The first predictive study used a Bayesian inference structure on a simplistic SIRV model \cite{Science-Dehning, Jo}, using infection statistics from Germany. While a move in the right direction, it suffered from two key deficiencies: lack of a time evolving death rate as an independent dynamical variable and over-reliance on infection statistics in predicting mortality rate. \cite{Lancet-Hellewell} addressed this, but it lacked the probabilistic kernel of \cite{Science-Dehning}. Another issue that has often been overlooked is the role of a containment strategy in counterpoising the contagion of the disease by identifying and blocking the key nodal links in the complex signaling network defining the chemical pathways \cite{Matamalas2018}.

\section*{Results}

\subsection*{Infection Kinetics of Healthy and Comorbid Susceptible}
COVID-19 infection propagation epidemiology clearly points to the need for analyzing the vastly different infection and mortality profiles of the healthy versus the comorbid susceptible groups. Our key target is to study this interactive infection propagation ad then predict future mortality and infection profiles, emphasizing mortality as the key policy indicator. The present article is the first to marry a robust Susceptible(S)-Infection(I)-Recovered(R)-Vaccinated(V) (SIRV) structure
\cite{Aliou} with a Machine Learning (ML) prediction kernel, using a
multi-layered error filtration structure, to generate a predictive model called PHIRVD (see Materials and Methods).  PHIRVD delivers three major successes at an unprecedented level of accuracy: prediction of the number of infected and dead
over the next 30 days (validated using sparse data) for each of the 18 countries
considered, a comparative analysis of the impact of lockdown using
multiple withdrawal dates for 6 worst-hit countries with high ongoing
infection rates, and a detailed temporal profile of future
reproduction numbers that can be (and have been) verified against real
data. PHIRVD also establishes mortality-to-infection ratio as the key dynamic pandemic descriptor instead of reproduction number.

\noindent
\subsection*{\bf Mathematical Model - PHIRVD} 
\noindent 
Our compartmentalised Covid-19 pandemic kinetics uses a 
6-dimensional dynamical system combining
SIR and SEIR kernels (30, 31), called PHIRVD:

\begin{eqnarray}
\label{dynamics0}
\frac{dH}{dt}&=& -\beta_1 H I {+\ q_{1H}R + q_{2H}V } - {h_{2v} H} -
\gamma H, \nonumber \\
\label{dynamics1} 
\frac{dP}{dt}&=&-\beta_2 P I -(\gamma+\delta) P {+\ q_{1P}R + q_{2P}V
}-p_{2v} P, \nonumber \\
\label{dynamics2} 
\frac{dI}{dt}&=&(\beta_1 H + \beta_2 P + \beta_3 R) I -(\gamma+\zeta)
I - w I, \nonumber \\
\label{dynamics3}
\frac{dR}{dt}&=& w I - \beta_3 R I - \gamma R - q_{1H} R - q_{1P}R, \nonumber \\
\label{dynamics4} 
\frac{dV}{dt}&=& {-(q_{2H}+q_{2P})V } -\gamma V + h_{2v} H + p_{2v} P,
\nonumber \\
\label{dynamics5} 
\frac{dD}{dt}&=&\gamma(H+R+V) + (\gamma+\delta)P + (\gamma+\zeta) I.
\label{dynamics6}
\label{eq1}
\end{eqnarray} 
%

\begin{figure} 
\centering \includegraphics[width=4in,height=3.0in]{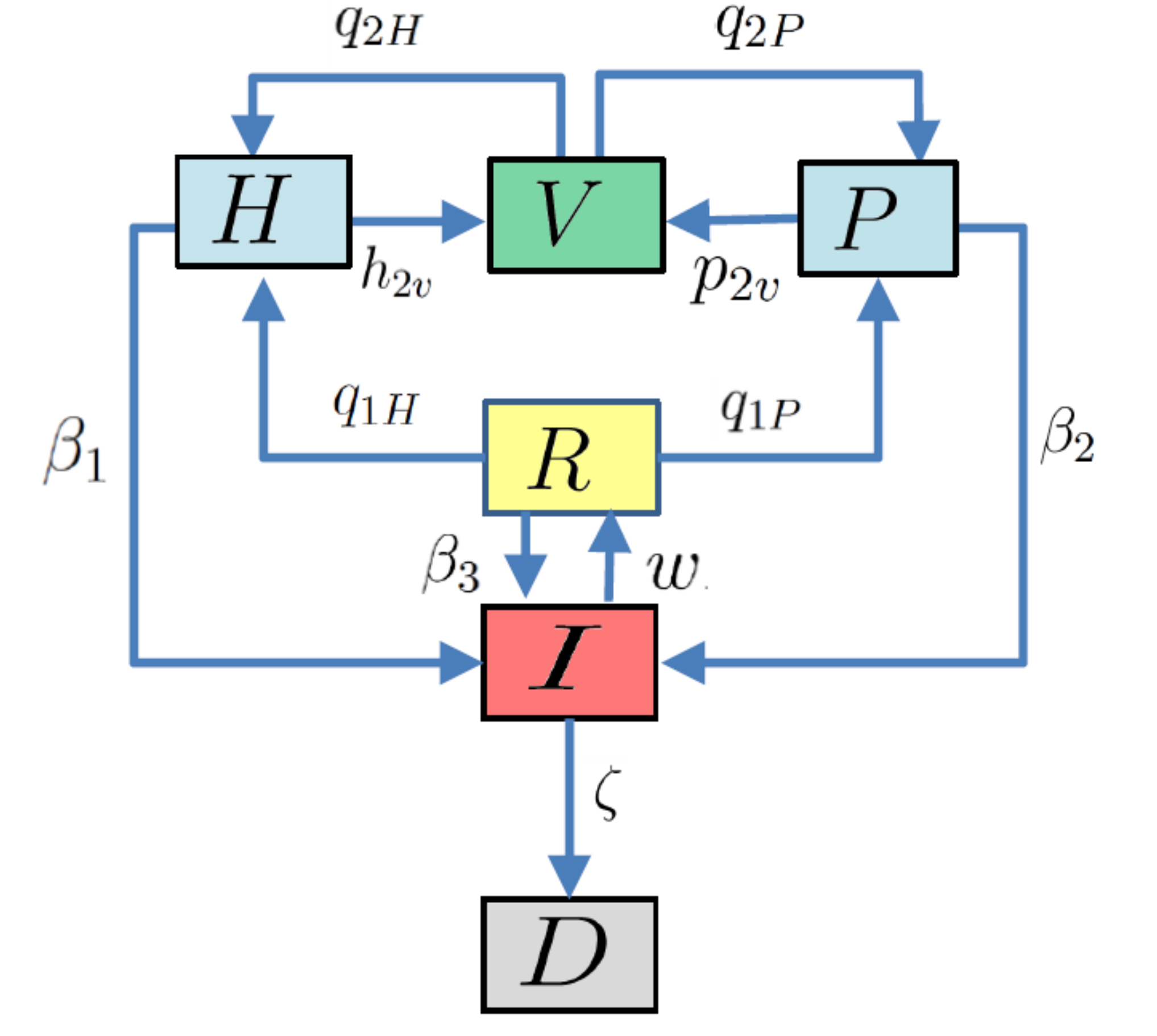}
\caption{Schematic diagram outlining the infection kinetic profile of our model PHIRVD:
Healthy Susceptible ($H$); Predisposed Comorbid Susceptible ($P$); Infected ($I$); Recovered ($R$); Herd immunized ($V$) and Dead ($D$). 
}
\label{flowchart}
\end{figure}
The parameters in the model characterize the infection rate of healthy
agents ($\beta_1$), infection rate of agents with pre-existing health
conditions ($\beta_2$), relapse rate ($\beta_3$), conversion rates of
recovered to healthy susceptible ($q_{1H}$) and previously
\enquote{immuned} to healthy susceptible ($q_{2H}$), conversion rates
of recovered to pre-existing susceptible ($q_{1P}$) and previously
\enquote{immuned} to pre-existing susceptible ($q_{2P}$), death rate
due to non-Covid interference ($\gamma$), additional death rate due to
agents with pre-existing conditions ($\delta$) and that due to
infected ($\zeta$), recovery rate ($w$), rate at which healthy
($h_{2v}$) and pre-existing susceptible ($p_{2v}$) groups are
quarantined. Our focus being Covid-19 infection and mortality
statistics, we neglect death ($\gamma=0$) and infection rate
($\delta=0$) due to all non-Covid causes. Furthermore, data strongly suggests that $\beta_1 \ll \beta_2$. Hence, the
infection rate of $H$-group is considered to be a small fraction
($\lambda$) of the $P$-group, i.e. $\beta_1=\beta_2 \lambda$. 
The death variable $D$ thus acts like a \enquote{sink} of
the dynamical system ensuring a population conservation inbuilt within
the model ($H+P+I+R+V+D$ = constant).

In training our model, we find it useful to define an extra
variable $I_c(t)$, which represents the cumulative number of
those infected upto a given date. In other words, it includes not
only those who are currently infected, but also those who have since
recovered or died, {\it i. e.} $\frac{dI_c}{dt}=(\beta_1 H  + \beta_2 P + \beta_3 R) I.$
Since we have considered relapse in our model, it is to be noted
that $I_c(t) \neq I(t) + R(t) + D(t)$.

\subsection*{Data repositories}
Identifying the infection kinetics of Covid-19 as an interactive evolution process involving six time evolving population density variables: healthy susceptible ($H$), susceptible with pre-existing conditions or comorbidity ($P$), infected ($I$), recovered ($R$),
naturally immuned (i.e. a clone for vaccinated $V$) and dead ($D$), the 
PHIRVD model uses statistics from the Johns Hopkins Covid-19 database
\cite{John-Hopkins-repository} to accurately predict mortality and infection statistics of 18 Asian, European and American countries. Data threshold was set beyond the first 19 days of low (or no) infection, followed by data training between 10 February 2020 to 29 June 2020. Results were later cross-verified from other
databases e.g. US: \url{https://usafacts.org}; EU:
\url{https://data.europa.eu/}; UK:
\url{https://coronavirus.data.gov.uk/}; India:
\url{https://www.covid19india.org/}. 
The Bayesian Markov Chain Monte Carlo (MCMC) \cite{MCMC} infrastructure in PHIRVD
trains the repository data to probabilistically predict the 17 parameters of the infection kinetic model (see Materials and Methods). Unlike previous predictive Machine Learning models
\cite{Lancet-kucharski, Lancet-Hellewell, Nature-Giordano, Science-Dehning, Jo, Aliou}, this structure allows more dynamic
adaptive control of the infection kinetic estimation resulting in a
highly accurate predictive module. 

\subsection*{Mortality and Infection: Prediction against Reality}
The 18 countries or regions under study were divided into 4 infection classes, the first three based on decreasing mortality-to-infection ratio for countries past their infection peak: UK, Netherlands, Sweden, New
York State ({\bf Class A}); Germany, Korea, Australia, Russia, Vietnam ({\bf Class B}); and
Italy, Spain, Hubei ({\bf Class C}). Class {\bf Class D} comprises India, Poland, Iran, France, Portugal and Brazil, with ongoing infection regimes. We deliberately chose New York State instead of the entire United
States due to its high population density and tourist/ worker traffic
that is quite different from the national average.

\begin{figure}[ht]
    \begin{subfigure}[t]{0.48\textwidth}
      \includegraphics[width=\textwidth,height=0.3\textheight]{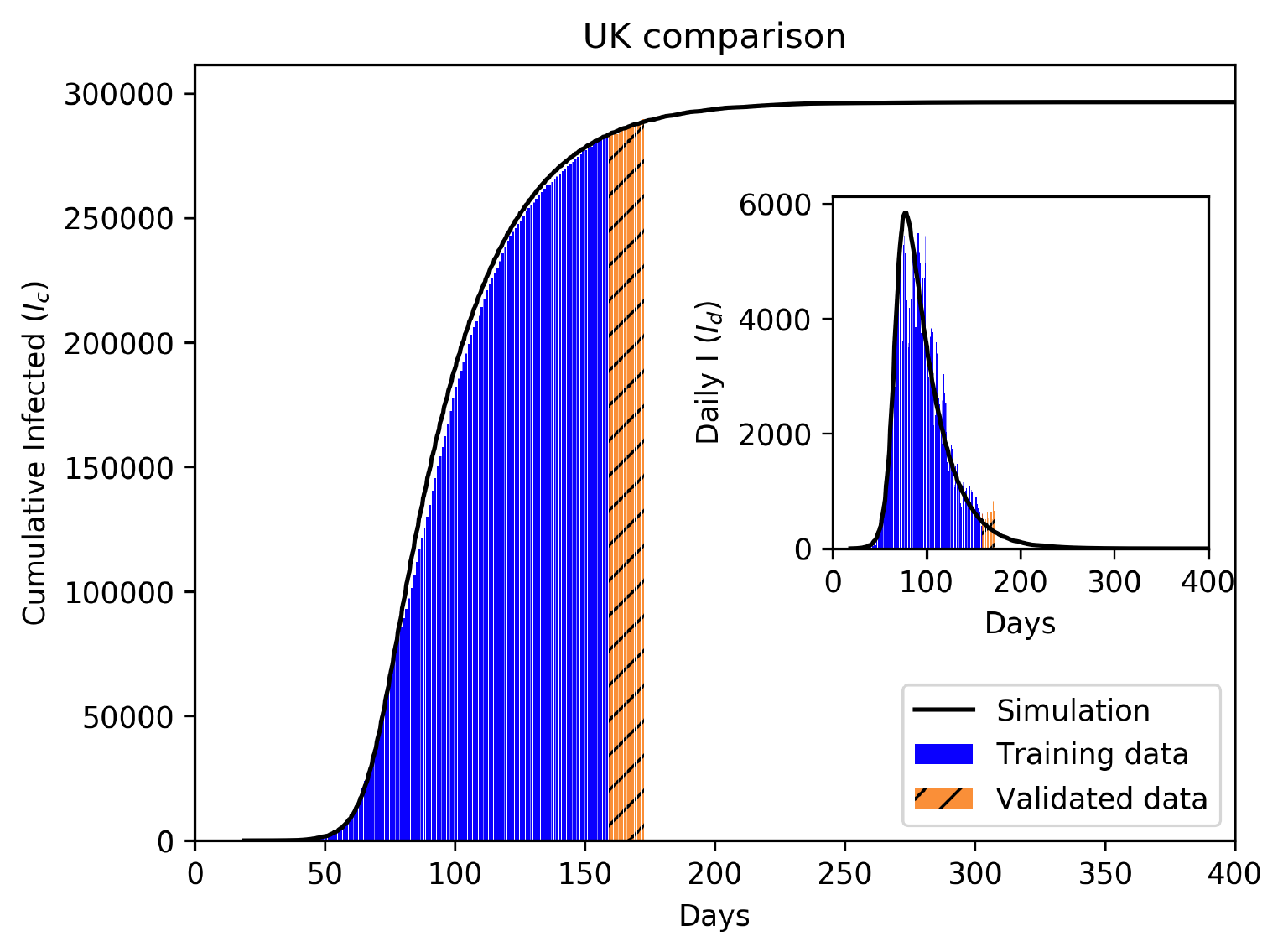}
      \caption{UK infection profiles.}
    \end{subfigure}
    \hfill
    \begin{subfigure}[t]{0.48\textwidth}
      \includegraphics[width=\textwidth,height=0.3\textheight]{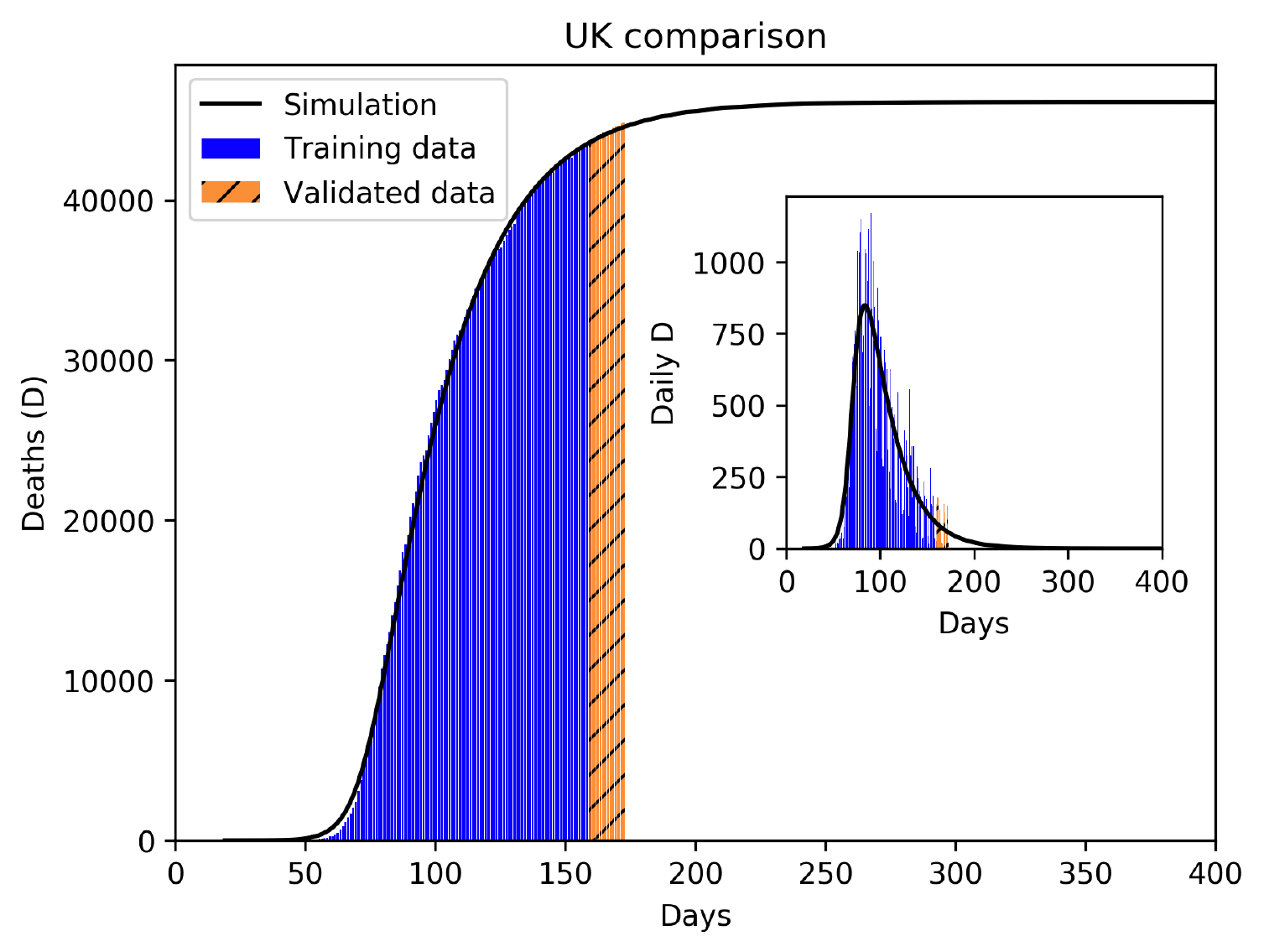}
      \caption{UK mortality profiles.}
    \end{subfigure}
\caption{Infection (Fig 2a) and mortality (Fig 2b) epidemiology for
  the UK (Class A). Outsets represent cumulative
  statistics while the insets are for daily updates in the number of
  infected and death respectively. Here \enquote{0} marks 22 January 2020; data training between 10 February to 29 June 2020.}
\label{fig_classA_uk}
  \end{figure}

\begin{figure}[h!]
    \begin{subfigure}[t]{0.48\textwidth}
      \includegraphics[width=\textwidth,height=0.3\textheight]{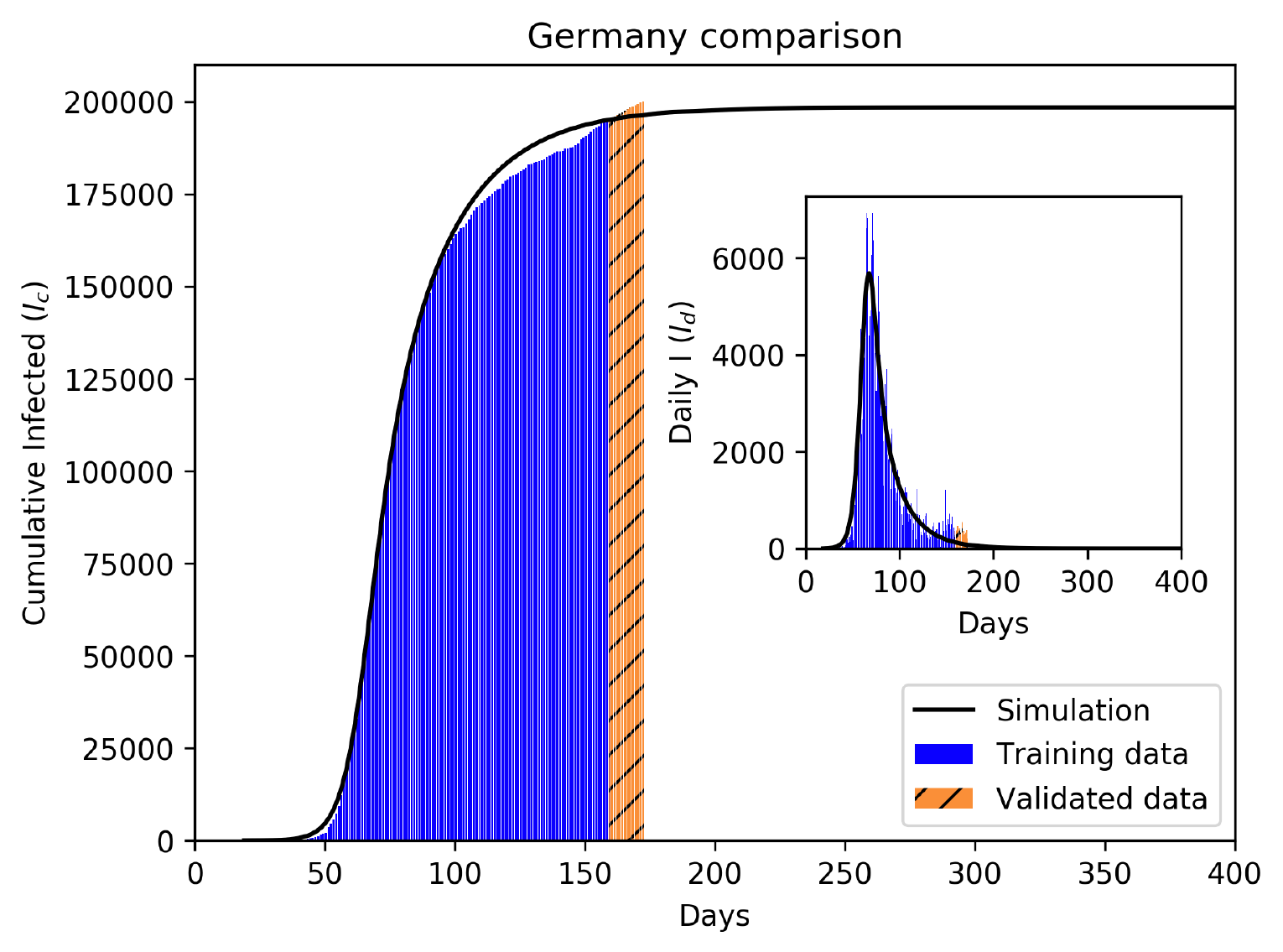}
      \caption{Germany infection profiles.}
    \end{subfigure}
    \hfill
    \begin{subfigure}[t]{0.48\textwidth}
      \includegraphics[width=\textwidth,height=0.3\textheight]{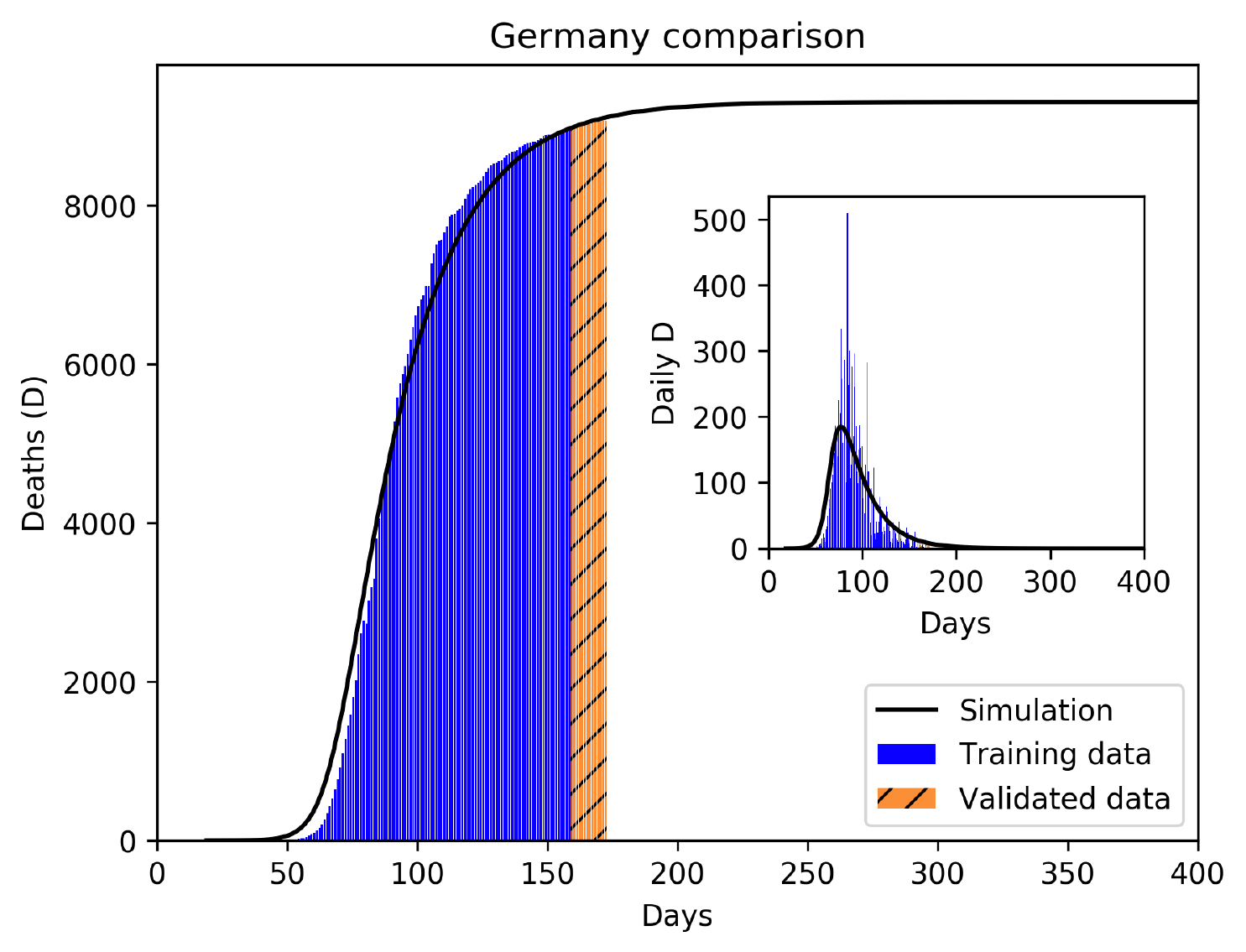}
      \caption{Germany mortality profiles.}
    \end{subfigure}
\caption{Infection (Fig 3a) and mortality (Fig 3b) epidemiology for
  Germany (Class B). Outsets represent cumulative
  statistics while the insets are for daily updates in the number of
  infected and death respectively. Here \enquote{0} marks 22 January 2020; data training between 10 February to 29 June 2020.}
\label{fig_classB_germany}
  \end{figure}
%
%
\begin{figure}[h!]
    \begin{subfigure}[t]{0.48\textwidth}
      \includegraphics[width=\textwidth,height=0.3\textheight]{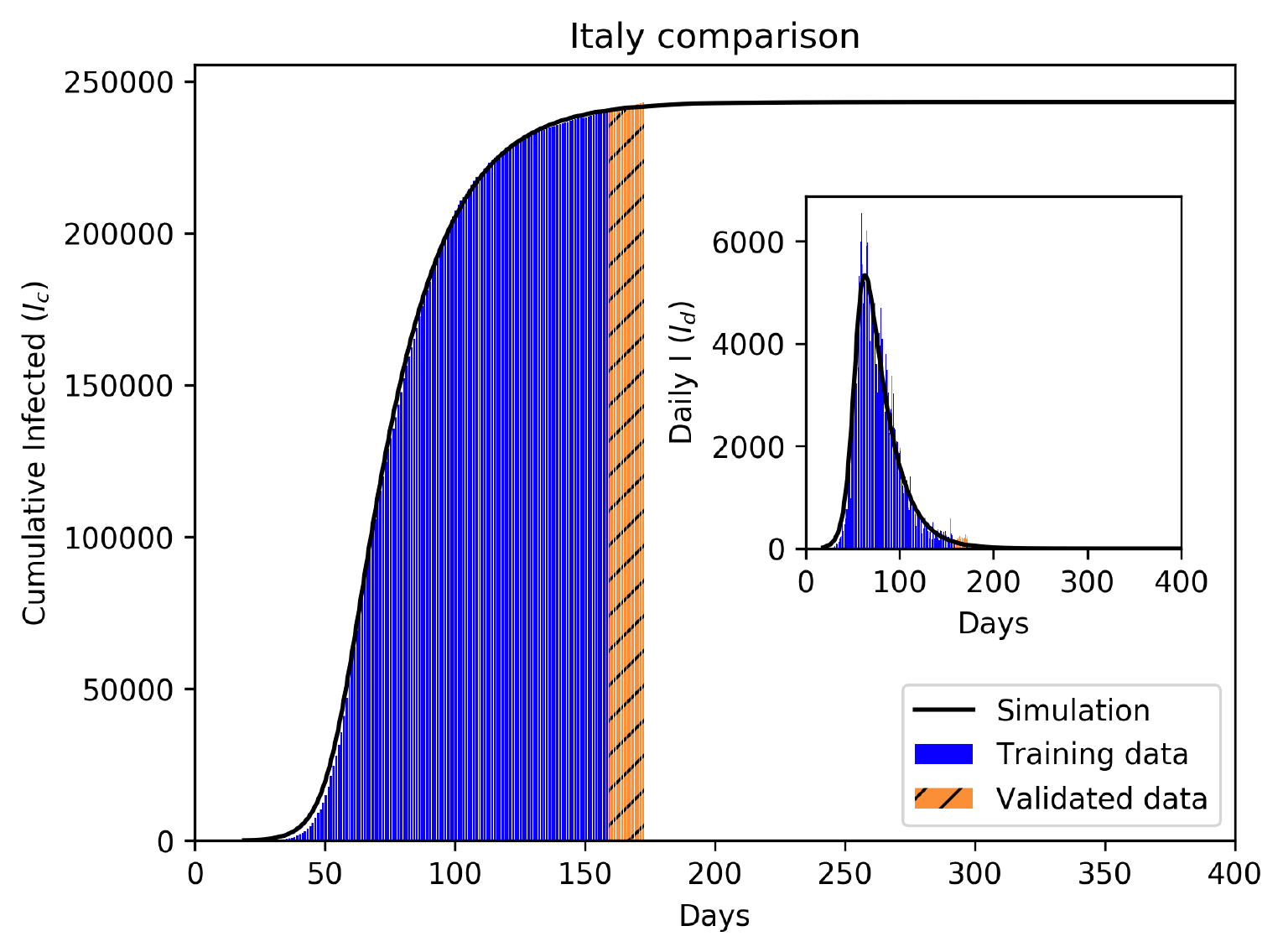}
      \caption{Italy infection profiles.}
    \end{subfigure}
    \hfill
    \begin{subfigure}[t]{0.48\textwidth}
      \includegraphics[width=\textwidth,height=0.3\textheight]{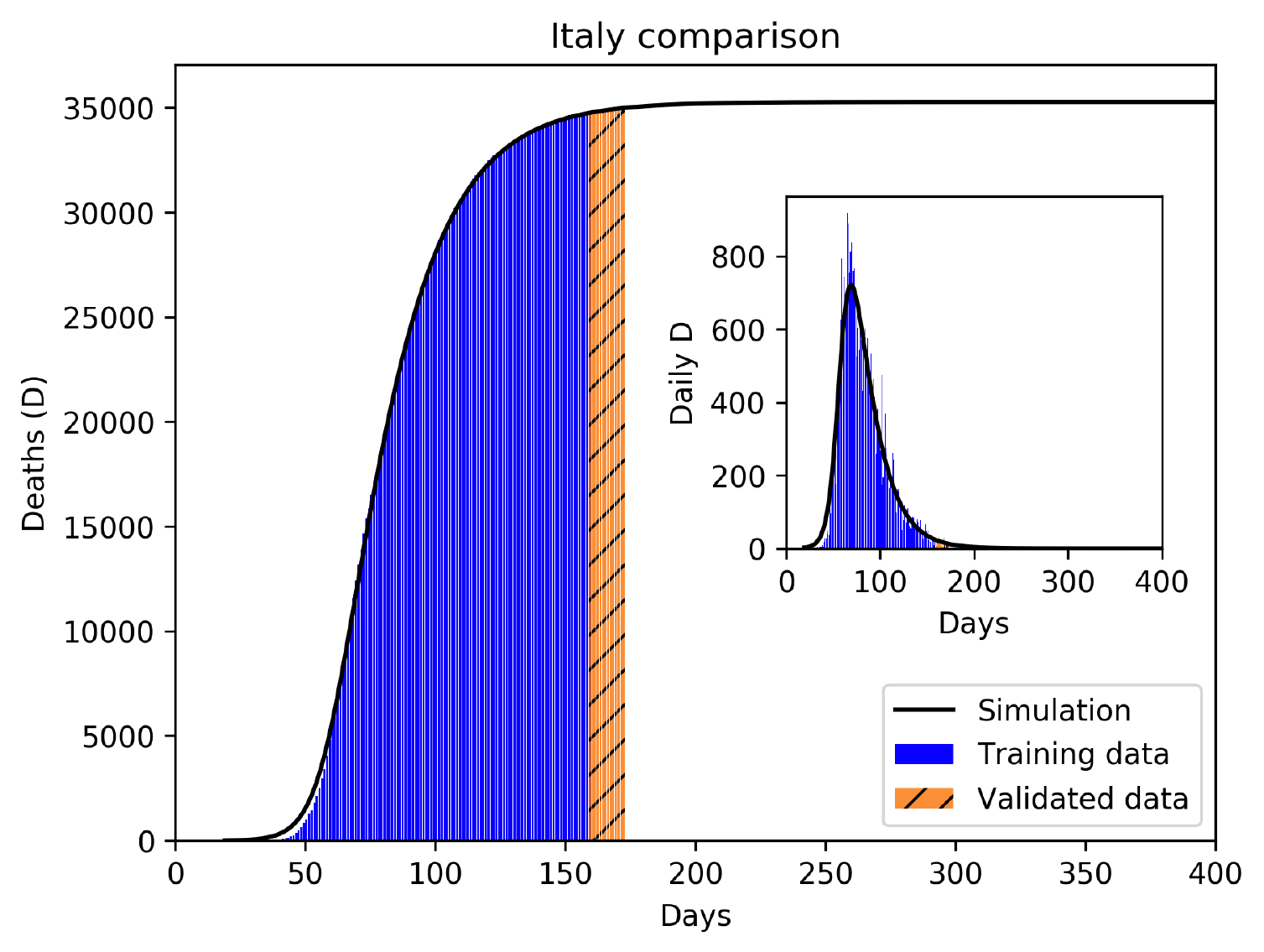}
      \caption{Italy mortality profiles.}
    \end{subfigure}
\caption{Infection (Fig 4a) and mortality (Fig 4b) epidemiology for
  Italy (Class C). Outsets represent cumulative
  statistics while the insets are for daily updates in the number of
  infected and death respectively. Here \enquote{0} marks 22 January 2020; data training between 10 February to 29 June 2020.}
\label{fig_classC_italy}
  \end{figure}
%
%
\begin{figure}[h!]
    \begin{subfigure}[t]{0.48\textwidth}
      \includegraphics[width=\textwidth,height=0.3\textheight]{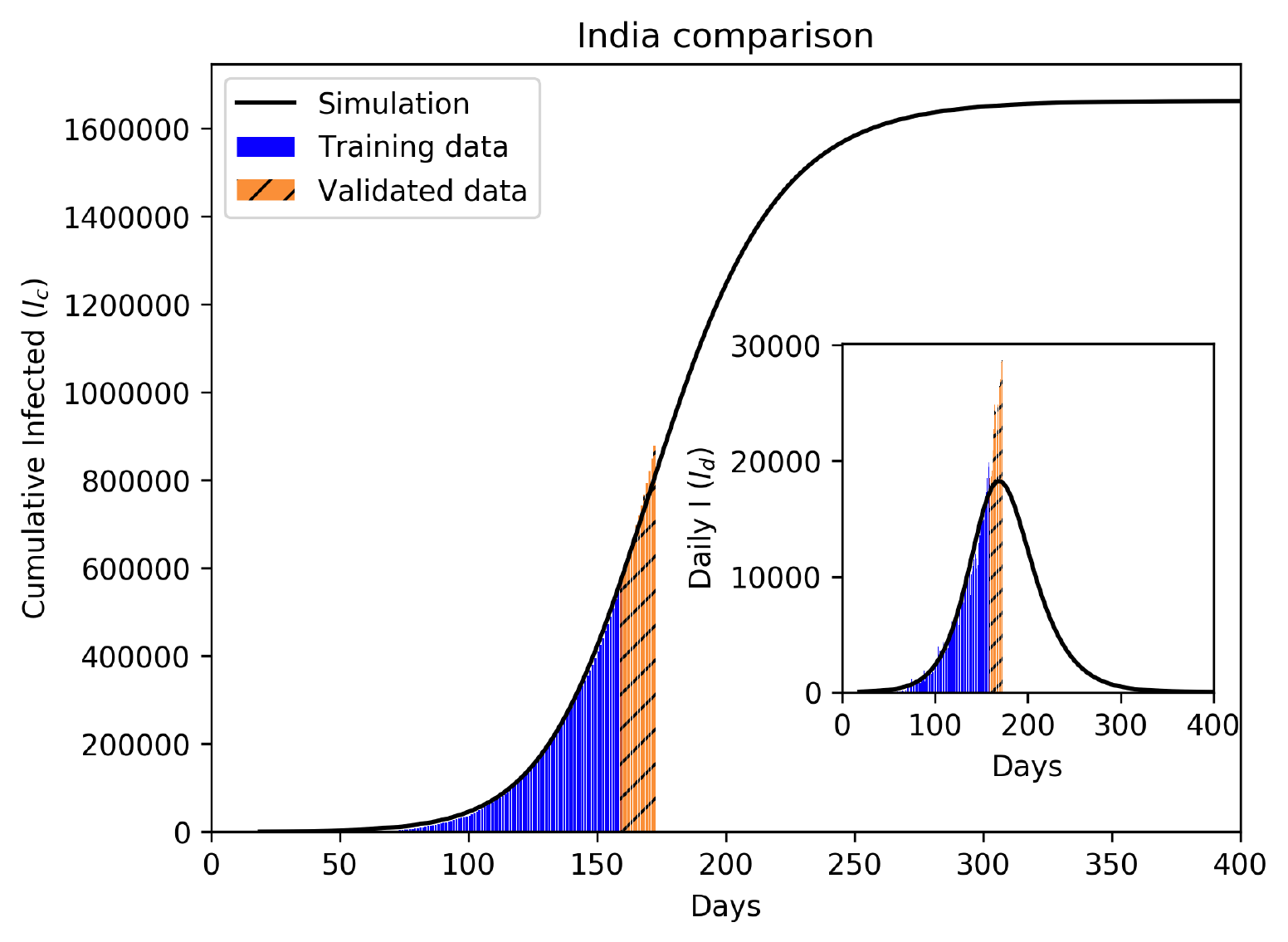}
      \caption{India infection profiles.}
    \end{subfigure}
    \hfill
    \begin{subfigure}[t]{0.48\textwidth}
      \includegraphics[width=\textwidth,height=0.3\textheight]{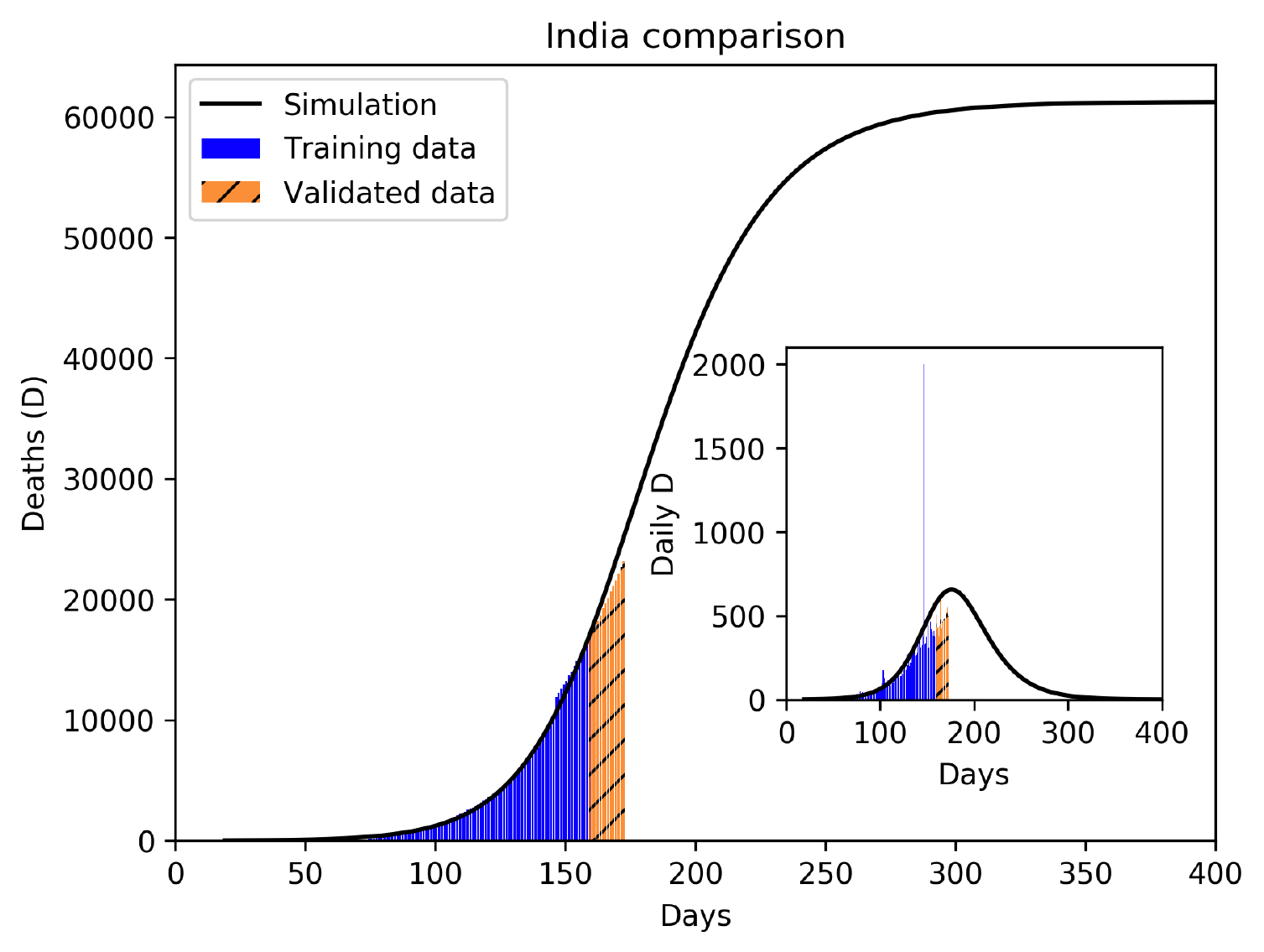}
      \caption{India mortality profiles.}
    \end{subfigure}
\caption{Infection (Fig 5a) and mortality (Fig 5b) epidemiology for
  India (Class D).  Outsets represent cumulative
  statistics while the insets are for daily updates in the number of
  infected and death respectively. Here \enquote{0} marks 22 January 2020; data training between 10 February to 29 June 2020.}
\label{fig_classD_india}
  \end{figure}
%
%
With the number of reported cases being
highly dependent on the number of daily testings, not necessarily in agreement
with the actual disease propagation dynamics, we observe
some deviations between the simulated $I(t)$ and the actual
number of reported cases.  On the other hand, $D(t)$ is less affected
by the testing rate. Since we are using mortality statistics with the same
weightage as the infected data, we prioritize mortality prediction. We
note that daily training of any epidemiological model will invariably
achieve better data match, as many studies have shown, but {\it they all
lack the key predictive ability, that our ML embedded propagation
kinetic model thrives on}.

Comparative statistics for our Class A representative, the UK, is shown in Figure
\ref{fig_classA_uk}. The blue region marks the training zone that
fixes the parameters. Based on the highest
mortality to infection ratio in each group, the representative
countries for the other 3 classes are Germany (Class B), Italy (Class
C), India (Class D). Figures \ref{fig_classB_germany}, \ref{fig_classC_italy} and \ref{fig_classD_india} represent infection statistics for Class B (Germany), C (Italy) and D (India) respectively (other plots in Appendix II). Chi-square tested (see Materials and Methods for Chi-squared statistic used) accuracy chart in Table \ref{table_pvalue_infected_4countries} clearly points to the veracity of the accuracy claim made. On the other hand, Vietnam presents an interesting case. With a
reported zero mortality rate notwithstanding high population density,
it has been repeatedly cited as an example of early quarantine
success. The model tracks even such an exceptional case
    to a moderate level of
    accuracy (in Appendix II). 
    The outsets and insets respectively outline the
cumulative versus the daily infection traffic. Details for other countries, for 4 infection
classes, are provided in Appendix II.

Table \ref{table_firstvalidation} presents a comparative chart of the PHIRVD model predictions versus real
data, separately for the numbers of infected and dead, for countries
representing the 4 classes with data trained between 10 February to 29 June: Class A (UK), Class B (Germany), Class C
(Italy) and Class D (India). Futuristic prediction is shown until 12 July. For other countries in each individual class, with data training between 10 February to 10 May, 30 days' prediction until 9 June establishes the predictive strength of this model (see Tables S2-S5, Appendix III), error validated as shown in Table S1 (see Appendix I).
\begin{table}[ht]
\centering
\begin{tabular}{c | c|c | c|c  cc}
\toprule \hline
Country & \multicolumn{2}{c|}{Daily New Infected} &
\multicolumn{2}{c}{Daily New Death} \\ \cline{2-5}
 &  $\epsilon $ & p-value &   $\epsilon $ & p-value  \\
\midrule \hline
UK & 0.26&0.23 & 0.36&0.14\\
Germany&0.42  & 0.18 & 0.45 & 0.25\\
Italy&  0.32& 0.22& 0.3 &  0.28 \\
India& 0.52 &  0.25& 0.52 & 0.38  \\
 \hline \bottomrule
\end{tabular}
\caption{p-Values for daily new infected and dead for Class A-D representative countries between 11 Feb to 16 June 2020.}
\label{table_pvalue_infected_4countries}
\end{table}
 \begin{table}[ht]
\centering
\resizebox{\linewidth}{!}{%
\begin{tabular}{|c|c|c|c|c|c|c|c|c|c|c|c|c|c|c|c|c|c|c|c|c|c|c|c|c|c|c|c|c|c|c|c|c}
\toprule \hline
 \backslashbox{\tabular{@{}l@{}}Days\endtabular}{ Country}  & \multicolumn{4}{c|}{UK}& \multicolumn{4}{c|}{Germany}& \multicolumn{4}{c|}{Italy}& \multicolumn{4}{c|}{India}\\
 \cline{2-17}
 & \multicolumn{2}{c|}{Infected} &\multicolumn{2}{c|}{Death}
 & \multicolumn{2}{c|}{Infected} &\multicolumn{2}{c|}{Death}
 & \multicolumn{2}{c|}{Infected} &\multicolumn{2}{c|}{Death}
 & \multicolumn{2}{c|}{Infected} &\multicolumn{2}{c|}{Death}
 \\ \cline{2-17}
 & Data & Simulation & Data & Simulation  & Data & Simulation & Data & Simulation & Data & Simulation & Data & Simulation& Data & Simulation& Data & Simulation\\
\midrule
\hline
30/06/20&403&472&155&91&376&126&14&12&142&121&23&23&18641&17298&507&570\\
\hline
01/07/20&60&455&176&88&475&121&5&11&182&116&21&22&19160&17471&434&579\\
\hline
02/07/20&4&439&89&84&477&117&11&11&201&110&30&21&20903&17628&379&588\\
\hline
03/07/20&502&424&136&81&410&112&4&11&223&106&15&20&22771&17767&442&596\\
\hline
04/07/20&624&408&67&79&418&108&10&10&235&101&21&19&24850&17889&613&604\\
\hline
05/07/20&516&394&22&76&325&104&3&10&192&96&7&18&24248&17992&425&612\\
\hline
06/07/20&352&380&16&73&541&100&0&9&208&92&8&17&22251&18078&466&618\\
\hline
07/07/20&581&366&155&70&279&96&10&9&137&88&30&17&22753&18145&483&625\\
\hline
08/07/20&630&353&126&68&356&92&14&9&193&84&15&16&24879&18194&487&631\\
\hline
09/07/20&642&341&85&66&302&89&11&8&214&81&12&15&26506&18224&475&636\\
\hline
10/07/20&512&329&48&63&331&85&6&8&276&77&12&14&27114&18236&519&640\\
\hline
11/07/20&820&317&148&61&377&82&7&8&188&74&7&14&28606&18230&550&645\\
\hline
12/07/20&650&306&21&59&210&79&1&7&234&70&9&13&28732&18206&501&648\\
 \hline \bottomrule
\end{tabular}
}
\caption{Validation of Daily new Infected and Death: UK (Class A), Germany (Class B), Italy (Class C), India (Class D)}
\label{table_firstvalidation}
\end{table}

 \begin{figure}
 \begin{center}
      \includegraphics[width=0.6\textwidth,height=0.3\textheight]{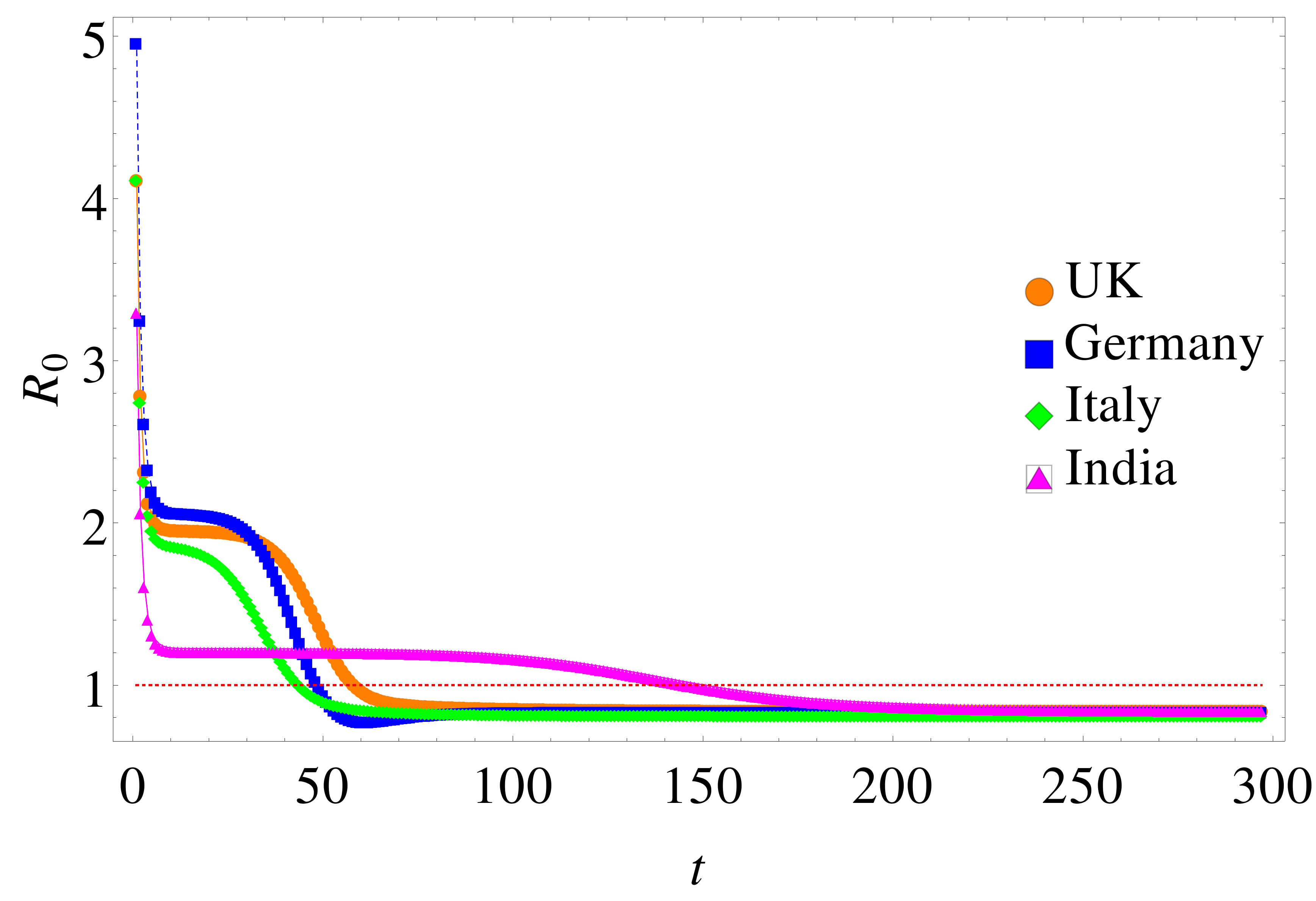}
  \end{center}
\caption{Daily temporal evolution of the basic reproduction rate for countries from Class A (UK),
  Class B (Germany), Class C (Italy) and Class D (India). The dotted
  line sets the pandemic threshold; count \enquote{0} starts at 14 February 2020, excluding data for the first 19 days (statistics recorded 22 January 2020 onwards) due to low infection, and additional 4 days of gestation. MCMC training between 10 February 2020 to 29 June 2020.}
\label{fig_R0ABCD}
  \end{figure}

The expected number of secondary cases produced from each infected individual is traditionally defined as the basic reproduction number. A linear analytical estimation of reproduction number $R_{\text{e}}$ at fixed point leaves us with only 8 independent parameters characterizing the secondary infection kinetics (note that a linear analytical formulation does not provide an exact quantitative estimate but a proportional one only). The detailed calculation of $R_{\text{e}}$ is provided in the Materials and Methods section.
Figure \ref{fig_R0ABCD} depicts the time evolution of
secondary infection for the 4 representative countries from infection
classes A-D, represented by the basic reproduction number $R_0$
\cite{R0-Seth, R0-Nishiura,R0-Ferguson} (see Materials and Methods). $R_0$ kinetics of all
other countries are provided in Appendix I. Class A countries
consistently show the sharpest drop in $R_0$ and the flattest
stability period, followed by progressive decrease in $R_0$ decay rate
and gestation span for classes B, C and D respectively. The point of
note here is that while Germany and Italy show higher levels of
infection than the UK, the gestation period for the UK is a lot larger
than both. India shows a similar trend although the absolute numbers
for India are a lot lower than the other three, indicating a
complicated relationship between Full Width at Half Maximum (FWHM) and
gestation period.
 %
 
\section*{Discussion}
\noindent
Combining conventional infection kinetic modeling with a predictive Bayesian MCMC, PHIRVD quantifies the impact of lockdown as a containment tool. It precisely estimates mortality statistics for 18 countries, accurate upto the next 30 days, beyond the last date of data training. Ideal lockdown imposition and withdrawal times have been predicted and validated, including for ongoing regimen e.g. India. PHIRVD also predicts secondary relapse timings and establishes mortality-to-infection ratio as the key pandemic predictive descriptor instead of reproduction number. PHIRVD is also capable of analyzing the impact of migration, an ongoing project. Our findings clearly suggest that phased lockdown is a potent containment tool but needs to be strategically imposed, where the correct implementation and withdrawal times are paramount. Secondary infection and mortality prediction will be key to future strategic quarantine imposition and analyzing impact of future therapeutics. 

PHIRVD leads to three key outcomes. First,
we present highly accurate probabilistic predictions for the numbers
of infected and dead for each country for a total of 18 countries,
typically 3 weeks beyond the last date of (Machine Learned) data
training. We can safely claim that this is the first, inherently
probabilistic COVID-19 model that can claim such high levels of
accuracy over such an extended time period (upto 30 days) probing in to the future
and that too for all countries considered. 

Second, we can precisely predict ideal lockdown withdrawal
dates for the countries. The full simulations plots (in Appendix II) 
clearly outlines how an increasing infection profile initially matches
with decreasing numbers of pre-existing susceptible and increasing
statistics for the recovered, that then slows down as the infection
peak arrives, eventually to tail off in to a no-infection
landscape. While the qualitative trends are similar for all classes
(A, B, C, D) of countries, the impact of lockdown on the first peak,
and then a second (relapse) peak, hint at the internal health versus
econometrics of the countries concerned. To prove this point, we compare infection (and mortality) propagation
kinetics of 2 chosen countries for two different dates, one on the recess (UK: Figure 7), the other with
uprising infection level (India: Figure 8). As opposed to
the recent furore about school children being exposed to the Covid-19
menace as a result of early lockdown withdrawal, our result clearly
shows that there is practically no difference in mortality between a
withdrawal on June 1, 2020 as against a later withdrawal e.g. July 1,
2020 (although a withdrawal on May 1 would have been
  disastrous). The 1 June (almost equally safe)
withdrawal would, of course, be favoured on economic and social grounds.
  
\begin{figure}[ht]
       \begin{subfigure}[t]{0.48\textwidth}
      \includegraphics[width=\textwidth,height=0.3\textheight]{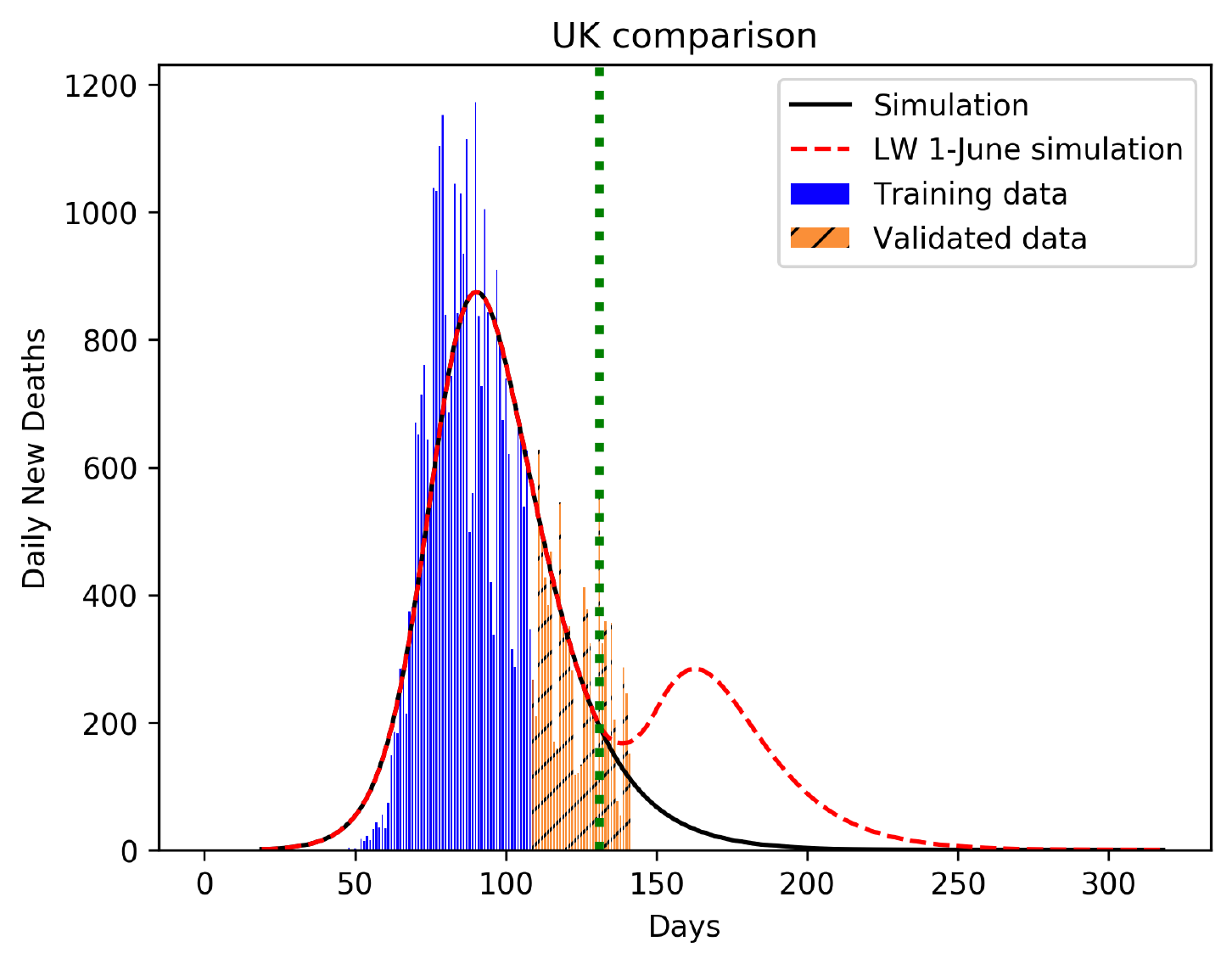}
      \caption{UK: 1 June 2020}
    \end{subfigure}
           \begin{subfigure}[t]{0.48\textwidth}
      \includegraphics[width=\textwidth,height=0.3\textheight]{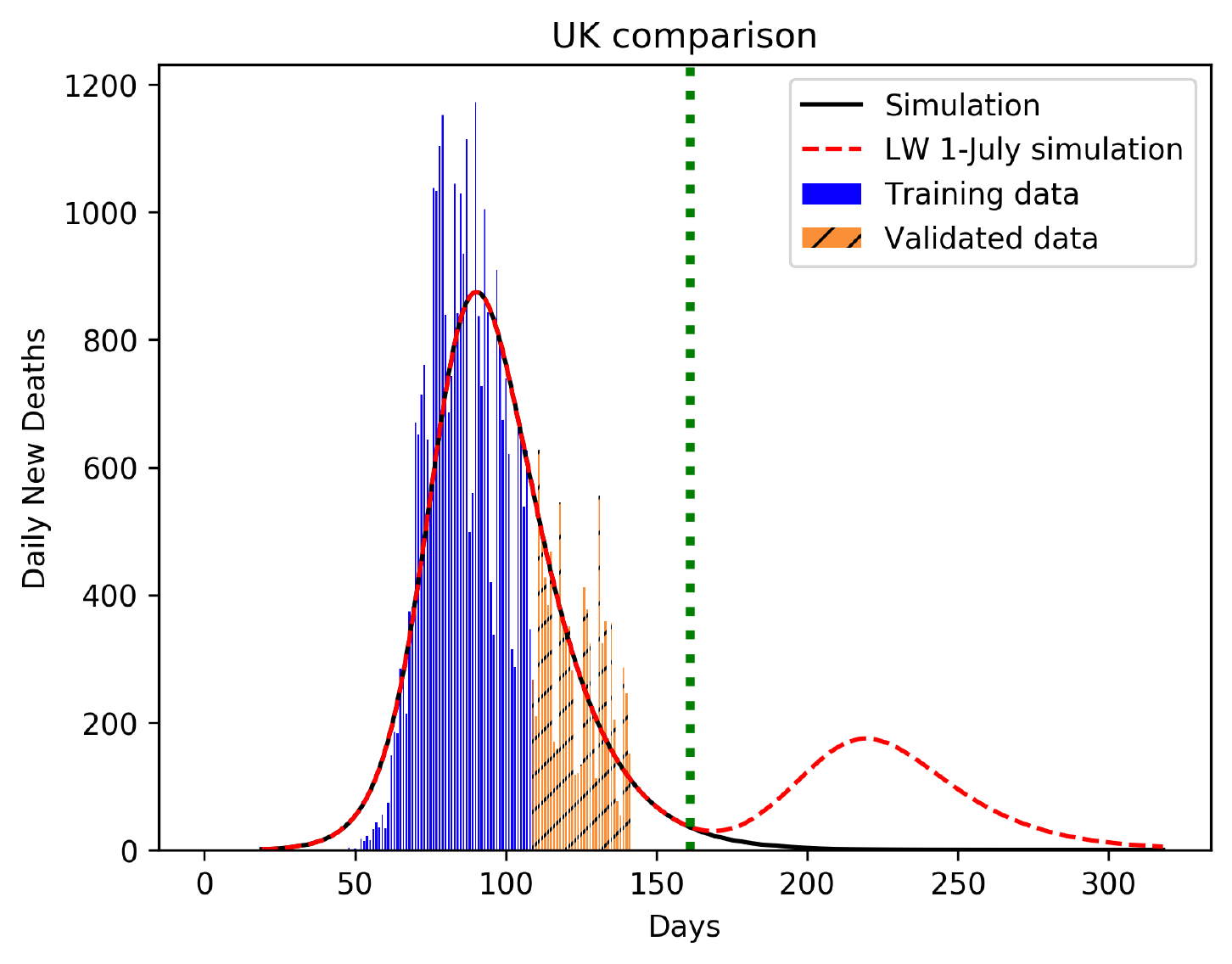}
      \caption{UK: 1 July 2020}
    \end{subfigure}
\caption{Lockdown withdrawal dates compared for the UK (partial decclared lockdown withdrawal on 23 March 2020). Analysis is based on daily mortality statistics. The perpendicular dotted line represents lockdown withdrawal date. Here \enquote{0} marks 22 January 2020.}
\label{fig_uklock}
  \end{figure}

\begin{figure}[ht]
       \begin{subfigure}[t]{0.48\textwidth}
      \includegraphics[width=\textwidth,height=0.3\textheight]{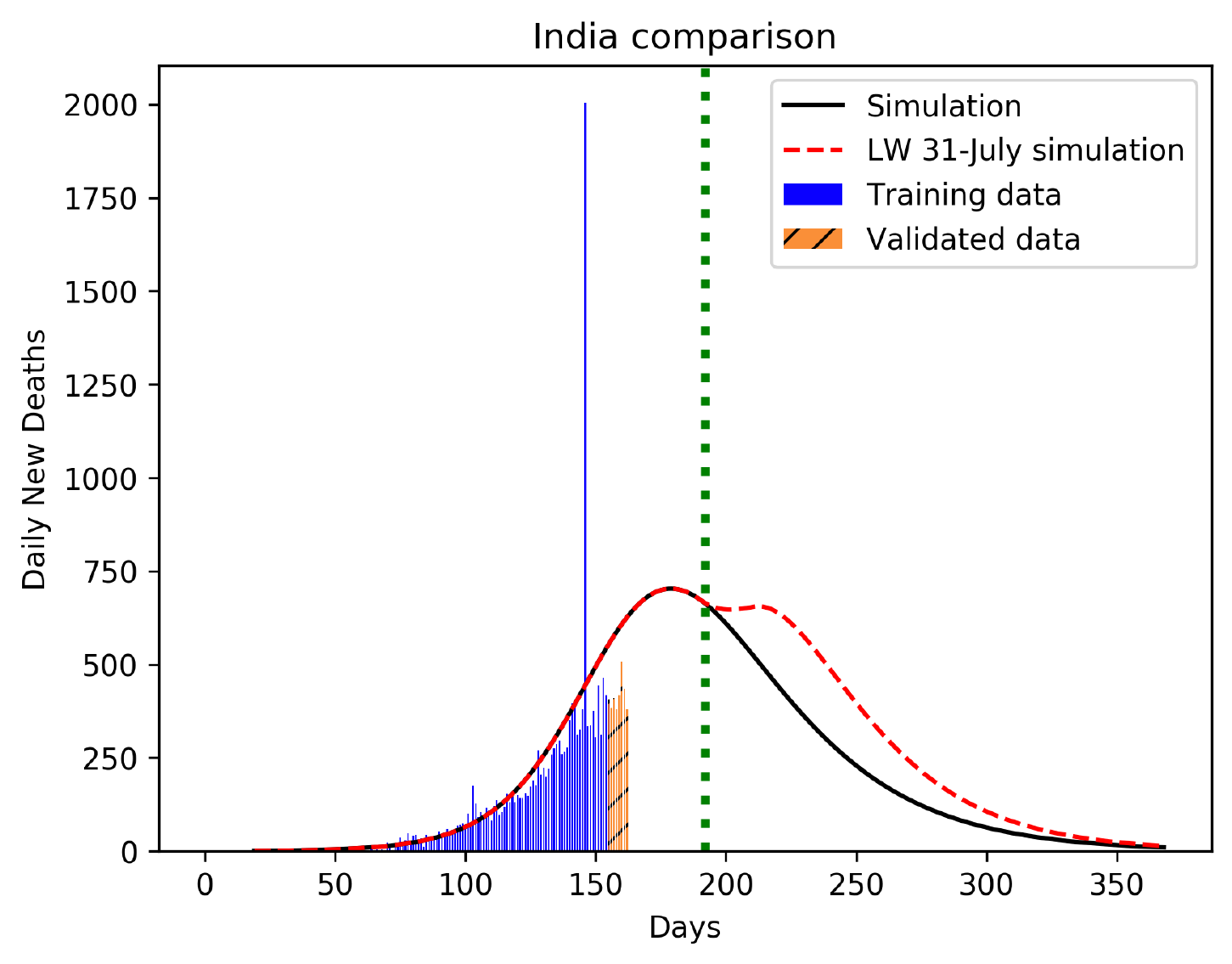}
      \caption{India: 31 July 2020}
    \end{subfigure}
           \begin{subfigure}[t]{0.48\textwidth}
      \includegraphics[width=\textwidth,height=0.3\textheight]{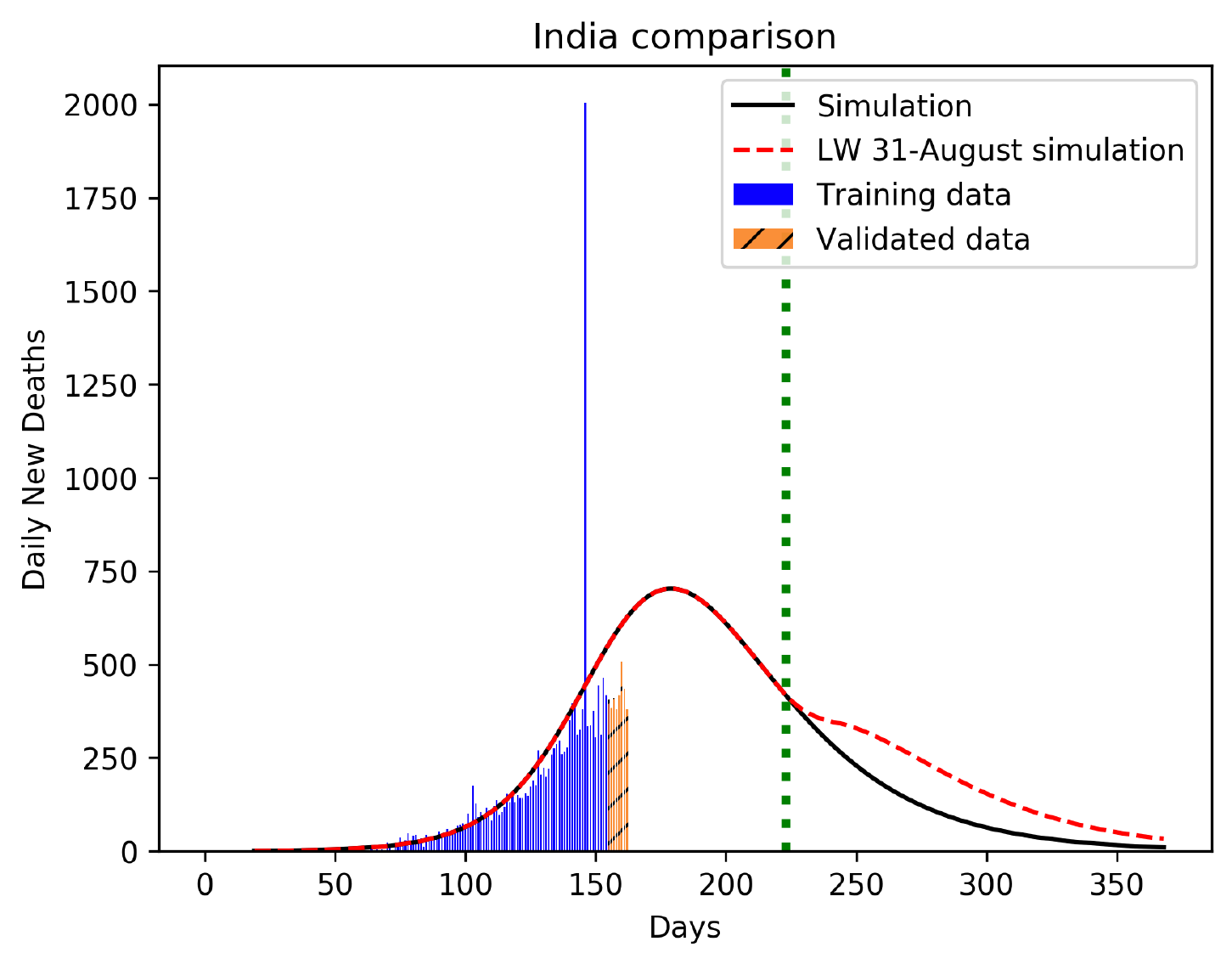}
      \caption{India: 31 August 2020}
    \end{subfigure}
\caption{Lockdown withdrawal dates compared for India (partial lockdown withdrawal on 4 July 2020). Analysis is based on daily mortality statistics. The perpendicular dotted line represents lockdown withdrawal date. Here \enquote{0} marks 22 January 2020.}
\label{fig_indialock}
  \end{figure}

   The third key outcome of our analysis is the establishment of mortality:infection ratio as the key descriptor of pandemic over and above reproduction number, that has conventionally been used for the purpose. The proof of this is in the accurate prediction of the secondary infection relapse time that the reproductive number fails to predict. As can be seen from Figures
   \ref{fig_uklock}a and \ref{fig_uklock}b, this
   relapse time period could be deferred with a late lockdown
   withdrawal on July 1 (as compared to June 1) although the peak
   mortality rates are not hugely different (ca 200 at 1 July compared
   to ca 400 at 1 June). Using 1 July 2020 as the UK lockdown withdrawal date, there is a clear signature of secondary relapse in the first week of September (identified as the second peak in Figure \ref{fig_uklock}. The Indian situation is clearly more
   challenging, though, as shown in Figure \ref{fig_indialock}. While perhaps economically unsustainable,
   India could benefit with a lockdown even beyond 31 July, 2020. For other nations
   like Iran, Portugal, France and Poland, our predictions of
   non-trivial secondary relapses (all in late June) match almost perfectly with data,
   both infected and dead. 


A real point of contention amongst politicians, health professionals
and medical scientists has, for long, been
the correct lockdown implementation and withdrawal times. In
statistical parlance, this effectively amounts to an estimation of the
FWHM as has been estimated for Wuhan at
2.6 weeks from initial infection \cite{Tomie}. To analyze these counterclaims, we incorporate the effects of withdrawal of lockdown as a country specific, dynamically evolving quantity. 

In palpable absence of any functional vaccine or therapeutics, these results can provide a remarkable gateway to pandemic strategizing, going beyond its immediate relevance.  What quarantine strategy to choose and when to implement or withdraw it is of crucial importance, for which our model can serve as a future benchmark. 

\section*{Materials and Methods}

\subsection*{Motivation of the PHIRVD model}
\noindent
PHIRVD uniquely combines a dynamically evolving infection propagation model that tracks the phenomenology of infection kinetics with a probabilistic predictive algorithm, the latter chosen as a Bayesian Markov Chain Monte Carlo (MCMC) kernel. The Bayesian MCMC is used to train past data to predict time independent generic parameters that can predict the future statistics. The choice is guided by the strength of Bayesian MCMC in a range of dynamical modeling studies in complementary fields (32,33).

\subsection*{Reproduction number $R_{\text{e}}$ at fixed point }
For $\gamma=0, \delta=0$, from Eq. (\ref{eq1}) the disease free equilibrium (DFE) or fixed point is given by 
$P^* =H^*\frac{h_{2v} q_{2P}}{p_{2v} q_{2H}}$, $I^*=0, R^*=0$, $V^*=H^* \frac{h_{2v}}{ q_{2H}}$. To evaluate the reproduction number $R_{\text{e}}$, we have to break the equation of $ \frac{dI}{dt}$ into two parts $\mathcal{F}, \mathcal{V} $, i.e.,
 
\begin{eqnarray}
  \frac{dI}{dt}=\mathcal{F}-\mathcal{V}
\end{eqnarray}
where $\mathcal{F}=(\beta_1 H  + \beta_2 P + \beta_3 R) I$ and $\mathcal{V}=(\zeta+w) I $.
Now, $F=\frac{\partial \mathcal{F}}{\partial I}|_{DFE}$ and $\Sigma=\frac{\partial \mathcal{V}}{\partial I}|_{DFE}$. Then $
R_{\text{e}}=\frac{F}{\Sigma}=\frac{ H^* \left(\frac{\beta_{2} h_{2v}
   q_{2P}}{p_{2v} q_{2H}}+\beta_{1}\right)}{\zeta +\omega
   }
$.
\subsection*{ Lockdown Dynamics} 
During the time period, over which we trained our model, most of the 
countries (except Sweden), of our interest, were under lockdown. 
Therefore, we studied the effects of withdrawal/relaxation of 
lockdown for some countries by introducing a
time varying parameter $L(t)$ in the model in Eq. (\ref{eq1}) substituting $\beta_{1,2,3}$
with $\beta_{1,2,3}\:L(t)$ respectively, where
$L(t) = 1,\:\:\text{for} \:\:t \leq
t_0\:\:\text{and}\:\:\alpha\:\:\text{for}\:\:t \geq t_0+k$. For $t_0<t<t_0+k$, $L(t) = \frac{1}{k} [
  \alpha(t-t_0)+(t_0+k-t) ]$. Here $t_0$ marks the lockdown
withdrawal time point, $k$ is the approximate time duration during
which the susceptible and infected population mixes well (e.g. within
one week or one month etc.) and $\alpha$ is the parameter quantifying
the homogeneity of mixing.  The largest $\alpha$ value (sum of $H$, $P$,
$I$, $R$, $D$ and $V$) implies that almost all susceptible have been in contact
with an infected person. The function $L(t)$ is such that before
lockdown withdrawal, it does not alter the contact probability while
after withdrawal, it linearly increases from the value $1$
to $\alpha$ over a time interval of $k$ days, ensuring that the
contact probability between susceptible and infected increases from a low to a
high value within this period. 

\subsection*{ Parameter Estimation}
The Bayesian MCMC data training leading to
supervised learning is itself conducted in two steps using a
double-filtration process. First, infection data alone are used to arrive at
a preliminary set of values, characterizing each country. The said
values are then filtered through combined infected and mortality
statistics for a second training to sequentially converge to a preset
upper limit. The training schedule is repeated multiply to ensure
accurate predictions of the training dataset. Estimation of the equilibrium reproduction number is strategically used to reduce the effective parameter space from 13 to 8 parameters, perfectly conforming with the Bayesian MCMC prediction which shows that value fluctuations with other parameters do not contribute much to the infection kinetics. The model clearly separates the $H$ and $P$ infection classes to reflect their differential levels of
infection and mortality.  Another constituent is the death rate
kinetics embedded in the central structure. 
The infection propagation model outlined in Eq. (\ref{eq1}) is a
multi-parameter model whose parameters are evaluated using predictive
data modeling within the Bayesian MCMC construct. Similar
structures have been selectively used in 
(19,23)
albeit for single-country specific models without any explicit
mortality dynamics. Over-reliance on infection statistics has often
led to incorrect estimation for mortality statistics, whose accurate
prediction is our first key target, an aim that is remarkably well
served by our ML-embedded compartmentalised model. We present both the
cumulative and daily (inset plots) statistics of infected population
over $400$ days, data trained between 10 February 2020 to 29 June 2020 (140 days) and
then predicted up to the next 8 weeks (shown up to 12 July 2020 in Table 1).

\subsection*{The Bayesian Markov Chain Monte Carlo (MCMC) algorithm: }
To understand how the algorithm uses the data to determine the
  parameters, it is useful to recall some elements of Bayesian
  statistics (32, 33).
  Let $\bm{D}=(D_1, D_2, ..., D_n)$ represent the
  full data vector that is being used to train the algorithm. For our
  case, the subscripts run over both the time intervals (daily) as
  well as the data types, such as $I_c(t_i)$ and $D(t_i)$. Similarly,
  let $\bm{\Theta}=(\theta_1, \theta_2, ..., \theta_\alpha)$
  represent the vector of parameters. A key ingredient is the prior
  probabibility distribution ({\em Bayesian priors}) for each
  $\theta_i$. While the absence of any knowledge of the system would
  call for a prior that is flat in the physically allowed region, the
  incorporation of such knowledge (which, in the present context,
  could be divined from the analysis of, say even part of the data for
  a single country in a given class) quickly gives the prior a somewhat
  peaked structure. In other words, one could as well start with a
  normal-distributed prior, {\em viz.}, $\bm{\Theta} 
  \sim N(\bm{\Theta_0,\sigma})$, where the vector
  $\bm{\Theta_0}$ represents the mean of the parameters
  and $\bm{\sigma}=(\sigma_1, \sigma_2, ..., \sigma_\alpha)$ the
  standard deviation. As it turns out, the dependence of the final result
  on the prior
  is quite insignificant.Given a 
  $\bm{\Theta}$, it is straightforward to calculate the
  conditional probability $\mathbb{P}(\bm{D|\Theta})$ of obtaining
  a realization $\bm{D}$ for the data. Using Bayes' theorem, the
  posterior probability for $\bm{\Theta}$ given the data is expressed as
\begin{eqnarray}
\mathbb{P}(\bm{\Theta}|D)=\frac{\mathbb{P}(\bm{D|\Theta})\mathbb{P}(\bm{\Theta})}
{\mathbb{P}(\bm{D})},
\end{eqnarray}
where $\mathbb{P}(\bm{D)=\int_\Omega \mathbb{P}(\bm{D}|\bm{\Theta}})
\mathbb{P}(\bm{\Theta})d\bm{\Theta}$,
with $\Omega$ denoting the whole parameter space. This, immediately
  leads us to the likelihood ratio of two parameter vectors
$\bm{\Theta_1}$ and $\bm{\Theta_2}$, namely
\begin{equation}
  \frac{\mathbb{P}(\bm{\Theta_2|D})}{\mathbb{P}(\bm{\Theta_1|D})}
=\frac{\mathbb{P}(\bm{D|\Theta_2})\mathbb{P}(\bm{\Theta_2})}
{\mathbb{P}(\bm{D|\Theta_1})\mathbb{P}(\bm{\Theta_1})} \ .
\end{equation} 
We now resort to a 3-step algorithm: 
\begin{enumerate}
\item Choose parameters (including initial conditions) through
a random walk in the parameter space. The nature of the
  random walk is determined by the prior probability distributions for
  the parameters, including initial conditions. 
  \item Calculate the likelihood ratio function for the parameters, given the data. 
\item Decide whether to accept the suggested parameter set or not. 
\end{enumerate}
\noindent
 {\bf Step 1:}\\
\noindent
 Let $\bm{S_i}=(S_{i1}, S_{i2}, ..., S_{in})$ be
 the simulated vector at the
 $i^{th}$ step for parameter values
$\bm{\Theta_i}=(\theta_{i1}, \theta_{i2}, ..., \theta_{i\alpha})$.
Compared to the total population, the data $I_c(t), D(t)$ {\em etc.} are quasi-continuous and can be assumed to be drawn
from a Normal distribution with respective standard deviations $\bm{\Gamma}=(\gamma_1, \gamma_2, ...,
\gamma_n)$ and means $\bm{S_i}=(S_{i1}, S_{i2}, ...,
S_{in})$. Therefore, the posterior probability (or likelihood, in case
of continuous probability density) of the parameter vector
$\bm{\Theta_i}$ is,
\begin{eqnarray}
  \displaystyle
  \mathbb{P}(\bm{\Theta_i|D})=\frac{\mathbb{P}(\bm{D|\Theta_i})
\mathbb{P}(\bm{\Theta_i})}{\mathbb{P}(\bm{D})}
  = (2\pi)^{-(n+\alpha)/2}
  \left[\prod_{j=1}^n\gamma_j\prod_{\beta=1}^\alpha\sigma_\beta\mathbb{P}(\bm{D})\right]^{-1}
  \, \exp\left(\frac{-1}{2}\sum_{j=1}^n\left(\frac{S_{ij}-D_j}{\gamma_j}\right)^2\right) \ . 
\label{likelihood1}
\end{eqnarray}
Next, we execute
a random walk in $\bm{\Theta}$-space with distribution 
$N(\bm{\Theta_i,\sigma})$ to find $\bm{\Theta_{i+1}}$, and calculate 
again the posterior likelihood function, with the simulated data vector 
$\bm{S_{i+1}}$, corresponding to the parameter vector $\bm{\Theta_{i+1}}$
as 
\begin{footnotesize}
\begin{equation}
\begin{array}{rcl}
  \displaystyle
  \mathbb{P}(\bm{\Theta_{i+1}|D})
  &=& \displaystyle \frac{\mathbb{P}(\bm{D|\Theta_{i+1}})
\mathbb{P}(\bm{\Theta_{i+1}})}{\mathbb{P}(\bm{D})} \\[2ex]
  & = & \displaystyle (2\pi)^{-(n+\alpha)/2}
  \left[\prod_{j=1}^n\gamma_j\prod_{\beta=1}^\alpha \sigma_\beta\mathbb{P}(\bm{D})\right]^{-1}
  \, 
   \exp\left( - \, 
  \frac{1}{2}\sum_{j=1}^n\left(\frac{S_{(i+1)j}-D_j}{\gamma_j}\right)^2
  - \,\frac{1}{2}\sum_{\beta=1}^\alpha\left(\frac{\theta_{(i+1)\beta}-\theta_{i\beta}}{\sigma_\beta}\right)^2
  \right) \ . 
\end{array}
\label{likelihood2}
\end{equation}
\end{footnotesize}

%
\noindent
{\bf Step 2:}\\
The likelihood ratio is now calculated to be
$  \mathbb{P}(\bm{\Theta_{i+1}|D}) / \mathbb{P}(\bm{\Theta_i|D})$. 

%
\noindent
{\bf Step 3:}\\ Next, we generate a uniform random number $r \sim U[0,1]$.  If $r <
\mathbb{P}(\bm{\Theta_{i+1}|D})/\mathbb{P}(\bm{\Theta_i|D})$,
we accept $\bm{\Theta_{i+1}}$, otherwise we go back to Step 1 and repeat the procedure.  \\ 
We have used cumulative infected and dead data as the vector
$\bm{D}$ and we normalize (as described above) the data vector
$\bm{D}$, as well as the simulated vector $\bm{S_i}$ at every
step, before calculating the likelihood ratio in Step 2 above.
We have used $\sigma = (\bm{\sigma_P},
\bm{\sigma_{IC}})$, where $\bm{\sigma_P} = (0.01, 0.01, 0.01,
0.01, 0.01, 0.01, 0.01, 0.01, 0.01, 0.01, 0.01)$ only for parameters
part, $\bm{\sigma_{IC}} = (0.1, 0.1, 0.001, 0.0, 0.0, 0.0)$ for
initial data part, and $\bm{\Gamma}=(\gamma_1, \gamma_2, ...,
\gamma_n)$, where $\gamma_j = (0.1-0.05)(j-1)/(n-1)+0.05$.  The
initial days (where the numbers are low) in the data are given
relatively smaller weightage than the later days for fitting, as the
noise level is higher initially, than the signal.

\subsection*{Estimation of the reproduction number kinetics}
\noindent
Understandably, the basic reproduction number $R_0$
  is no longer a constant. Defining $R_0(t)$ as the average number of secondary
  infections from a primary case at a given epoch $t$, and similarly
  $I_d(t)$ as the  number of daily new cases, we have
\begin{eqnarray} 
I_d(t) &=& \int_0^{\infty} R_0(t) ~ I_d(t-\tau) ~ g(\tau) ~ d\tau, 
\label{rnot1}
\end{eqnarray} 
where $g(\tau)$ is the probability density function of the generation
time $\tau$, defined as the time required for a new secondary
infection to be generated from a primary infection. In other words,
$\tau$ is the time interval between the onset of a primary case to the
onset of a secondary case, generated from this primary case.  As is
reported 
(26), the mean generation time is approximately
$6.5$ days, we assume $g(\tau)$ has a Gamma distribution 
with $g(\tau)
= \mathrm{Gamma}(6.5, 0.62)$. We represent $R_0(t)$ as a function of
time as
 
\begin{eqnarray} 
R_0(t) &=& \frac{I_d(t)}{\int_0^{\infty}  I_d(t-\tau) ~ g(\tau) ~ d\tau}. 
\label{rnot2}
\end{eqnarray} 
We approximate the denominator of equation (\ref{rnot2}) directly from our 
simulated data, by a discrete sum, and evaluate $R_0$ at $n^{th}$ day as  
\begin{eqnarray} 
R_0(n) &=& \frac{I_d(t)}{\int_0^{\infty}  I_d(t-\tau) ~ g(\tau) ~ d\tau}   
\approx \frac{I_d(n)}{\displaystyle \sum_{\tau=0}^{n-1}I_d(n-\tau) ~ g(\tau)}. 
\label{rnotdiscrete}
\end{eqnarray} 

\subsection*{Statistical error estimation and $p$-values}
Using the Chi-square statistic as $\chi^{2} \equiv \sum\limits_{i=1}^n
\bigg(\frac{D_{i}-S_{i}}{\epsilon S_{i}+1}\bigg)^{2}$ ($0<\epsilon<1$), where $D_i$ are observed data and $S_i$ the simulated data for the $\text{i}^{\text{th}}$ day, we quantify the accuracy of our model fitting 
with the real data. Understandably, the data for daily new infections and 
daily new deaths are contaminated by noise, more severely than the corresponding cumulative data. Hence, a Chi-square test applied on cumulative data will always give a high $p$-value. However, to test the 
power of our predictive machine learning algorithm, we calculated the $p$-values on daily new data of deaths and infected. Assuming the real data are drawn from a normal distribution with mean value same as the simulated data, and with a standard deviation equal to some fraction of the simulated data, we derive our Chi-square statistic. Although, the real data of infected and dead are always positive, as the infection increases, this assumption is very well valid, except for a very small time interval at the starting of infection in a population. 

\section*{Data and materials availability}
\noindent
Data from the Johns Hopkins repository (\url{https://github.com/CSSEGISandData/Covid-19}) were used, together with country specific repositories, e.g. US: \url{https://usafacts.org}; EU:
\url{https://data.europa.eu/}; UK:
\url{https://coronavirus.data.gov.uk/}; India:
\url{https://www.covid19india.org/}. All the epidemiological
information we used is documented in the Extended Data and Supplementary
Tables. 
 The codes and relevant files are made available through the Aston Data Repository.
\section*{Author Contributions}
AKC and DC designed the core model, sequentially modified by SKN. SKN led the MCMC computation
and model simulation, while AKC and BK led the analytical sections. DC and GG, together with SKN and BK,
were in charge of comparative statistical error estimation. All authors wrote and approved the manuscript. All authors have identical contribution towards the final output. 
\section*{Acknowledgments} 
AKC acknowledges Darren Flower for his comments and advice on the manuscript.


\newpage
\section*{\LARGE Supporting Information Appendix (SI)}

\noindent
{\bf Data training}: In order to establish the predictive strength of this machine Learning enforced model, in the appendices, we trained the data between 10 February to 10 May and predicted for the next 30 days (until 9 June 2020). This is uniformly done for all countries.

\renewcommand{\thefigure}{S\arabic{figure}}
\setcounter{figure}{0}

\subsection*{Appendix I: Reproduction Number Dynamics for Class A, B, C, D Countries} 
\begin{figure}[h!]
    \begin{subfigure}[t]{0.48\textwidth}
      \includegraphics[width=\textwidth,height=0.28\textheight]{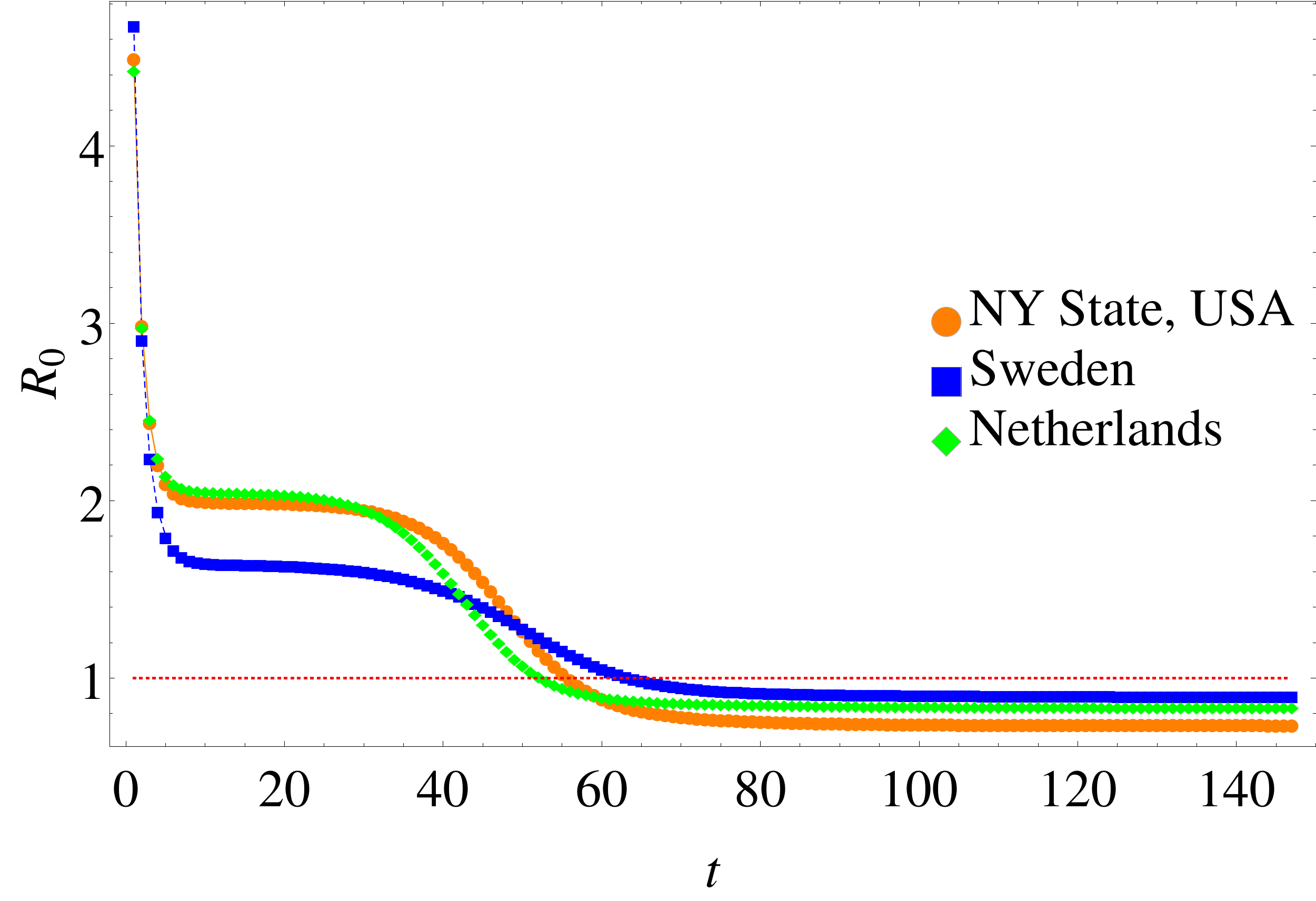}
      \caption{Class A countries}
    \end{subfigure}
    \hfill
     \begin{subfigure}[t]{0.48\textwidth}
      \includegraphics[width=\textwidth,height=0.28\textheight]{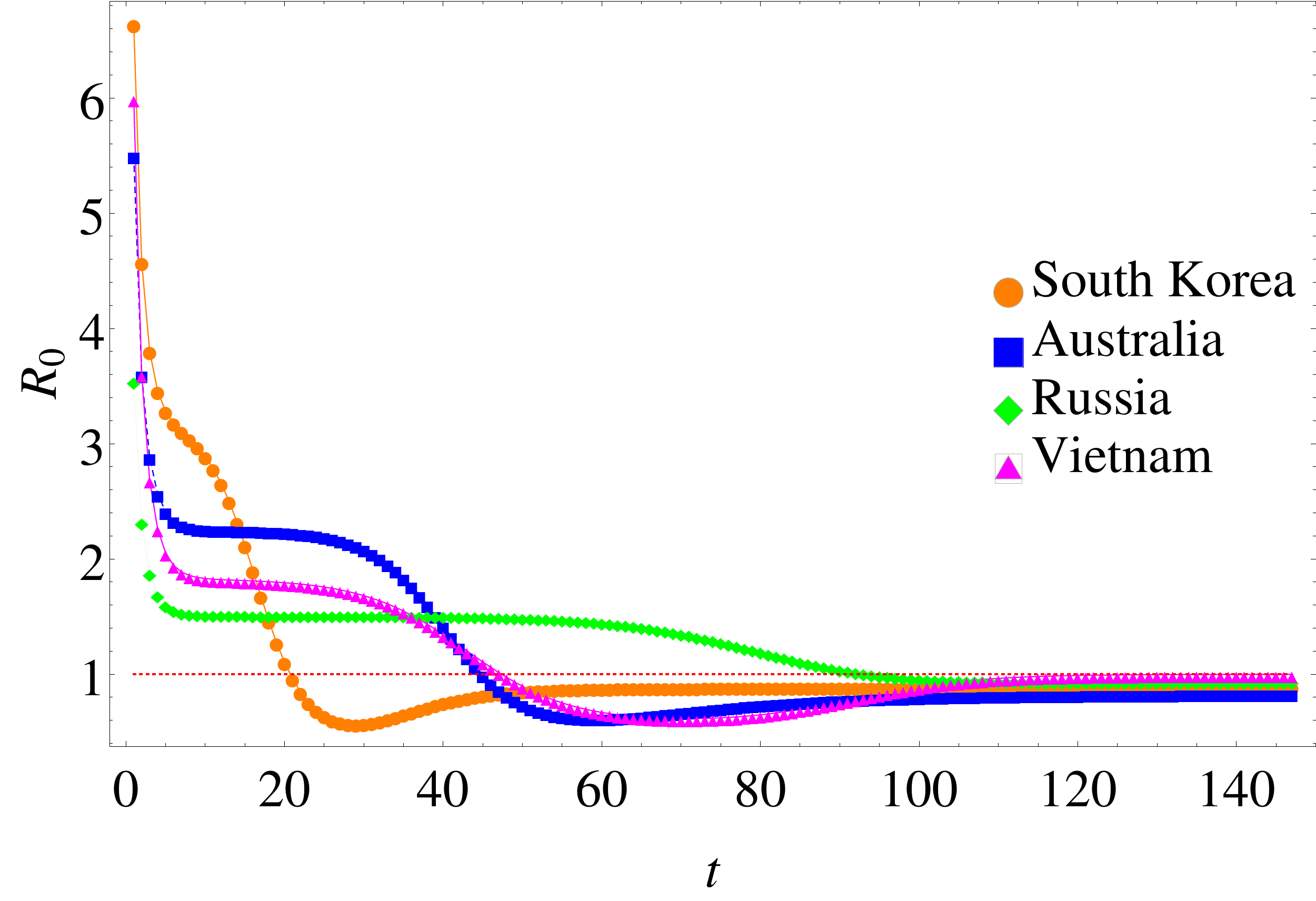}
      \caption{Class B countries and Vietnam (for comparison)}
    \end{subfigure}
       \begin{subfigure}[t]{0.48\textwidth}
      \includegraphics[width=\textwidth,height=0.28\textheight]{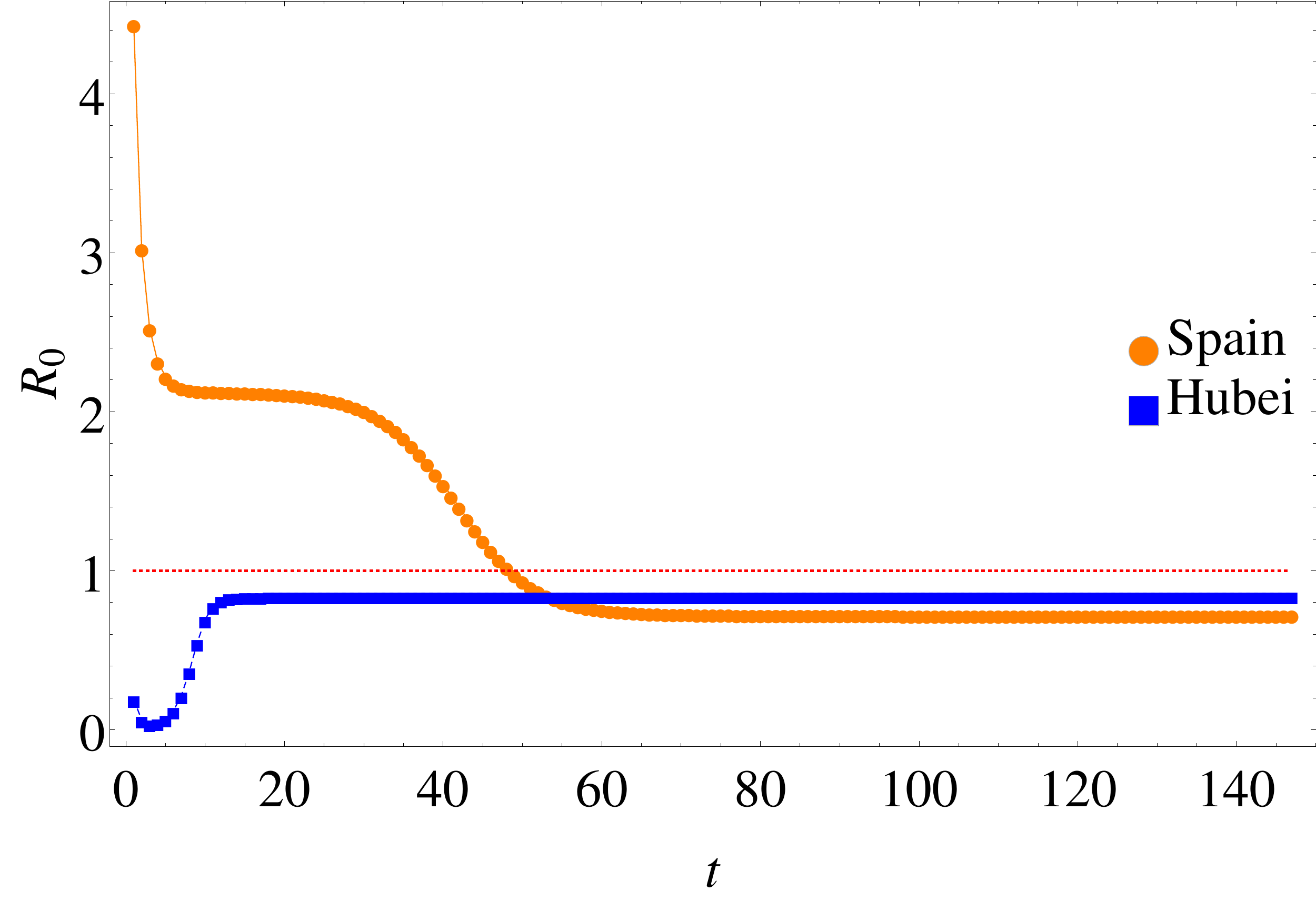}
      \caption{Class C countries}
    \end{subfigure}
    \hfill
    \begin{subfigure}[t]{0.48\textwidth}
      \includegraphics[width=\textwidth,height=0.28\textheight]{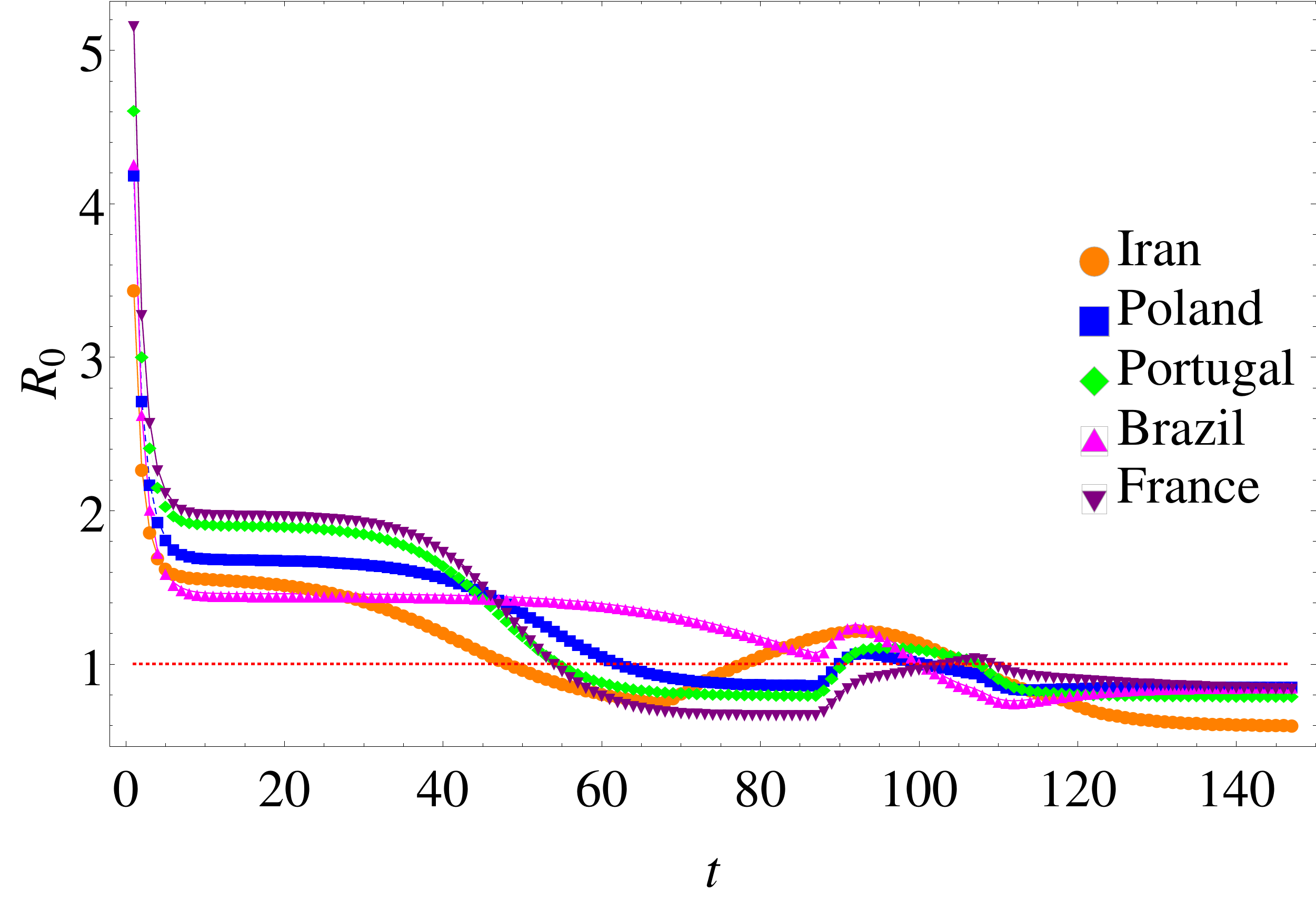}
      \caption{Class D countries}
    \end{subfigure}
\caption{Temporal evolution of the basic reproduction rate for all 4 infection classes, on a day by day basis. The dotted line at $R_0=1$ points to the optimum above which the epidemic to pandemic regime starts.}
\label{fig8}
\end{figure}

\makeatletter
\@addtoreset{table}{section}
\makeatother
\renewcommand\thetable{S1}

\begin{table}[h!]
\centering
\begin{tabular}{c | c|c | c|c  cc}
\toprule \hline
Country & \multicolumn{2}{c|}{Daily New Infected} &
\multicolumn{2}{c}{Daily New Death} \\ \cline{2-5}
 &  $\epsilon$ & p-value  & $\epsilon$  & p-value\\
\midrule \hline
%
Australia &0.48 & 0.31& 0.12& 0.18\\
Korea& 0.6 &0.48 & 0.3 &0.19    \\
NY state, USA& 0.55& 0.24& 0.45&0.17 \\
Poland &0.45 &0.67&0.25 & 0.28 \\
Russia &0.55 & 0.14 & 0.25&0.19\\
Belgium & 0.55 &  0.46& 0.35 & 0.26 \\
Brazil&  0.6& 0.37& 0.45 &  0.46 \\
Hubei& 1 & 0 & 1 &0   \\
Portugal&0.5 & 0.17&0.19 &0.69\\
Spain&0.75 & 0.19&0.5 &0.87\\
Sweden &0.6 &0.28 &0.5 &0.2\\
Vietnam &0.9 &0 &0.1 &1\\
Netherlands & 0.4 &  0.63& 0.31 & 0.41 \\
Iran & 0.45 &0.21  &0.35  &0.37   \\
 \hline \bottomrule
\end{tabular}
\caption{p-Values for daily new infected and dead for other Class A-D countries between 10 Feb to 10 May 2020. The
  statistic $\chi_D^{2} \equiv \sum\limits_{i=1}^n
\bigg(\frac{D_{i}-S_{i}}{\epsilon S_{i}+1}\bigg)^{2}$ ($0<\epsilon<1$) represents the chi-square value, where $D_i$ are observed data and $S_i$ the simulation data for the $\text{i}^{\text{th}}$ day.}
\label{table_pvalue_infected}
\end{table}
%
\clearpage

\subsection*{Appendix II: Infection and mortality plots for countries in Classes A, B, C and D}
\begin{figure}[h!]
    \begin{subfigure}[t]{0.48\textwidth}
      \includegraphics[width=\textwidth,height=0.3\textheight]{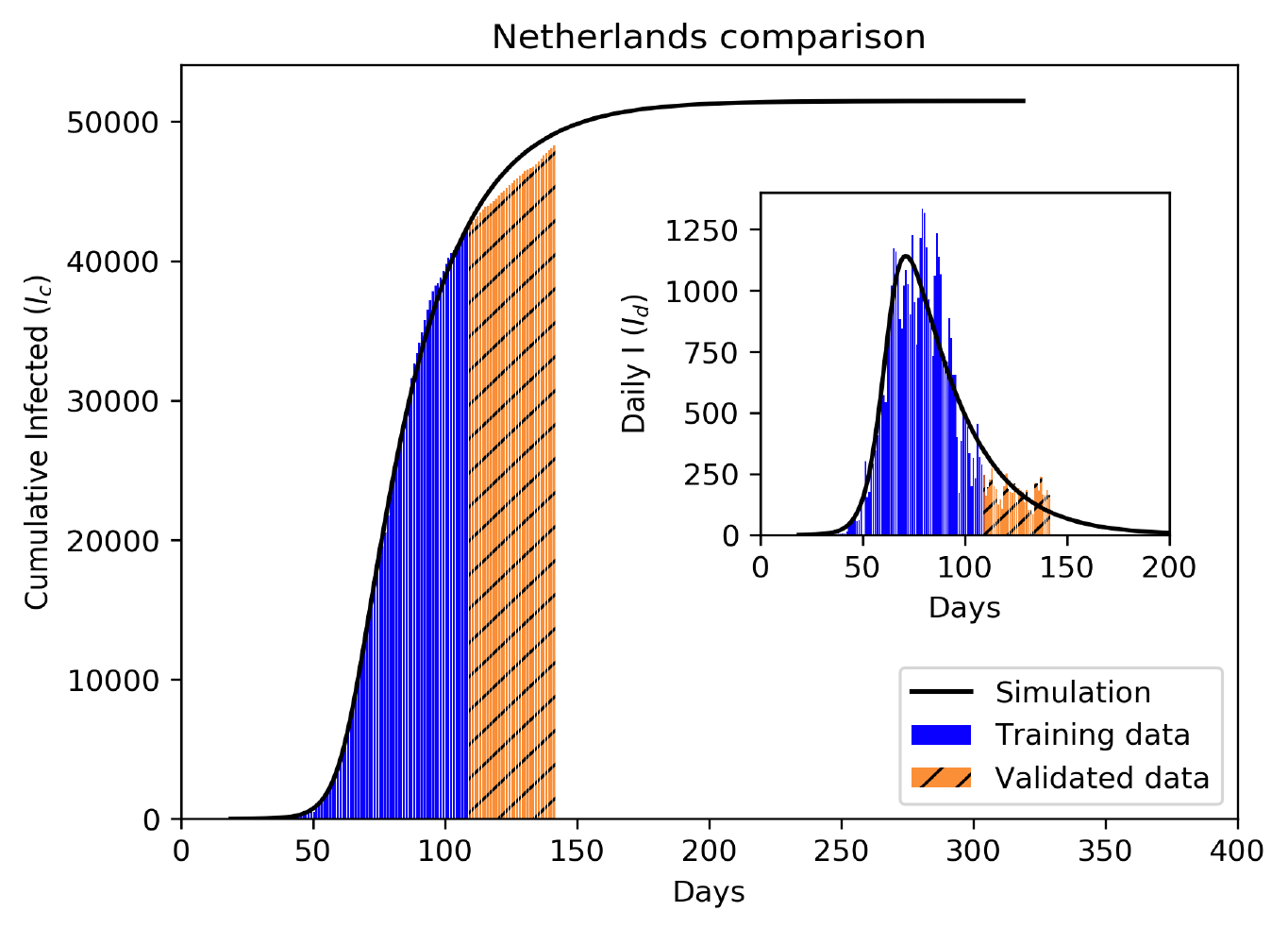}
      \caption{Netherlands infection profiles.}
    \end{subfigure}
    \hfill
    \begin{subfigure}[t]{0.48\textwidth}
      \includegraphics[width=\textwidth,height=0.3\textheight]{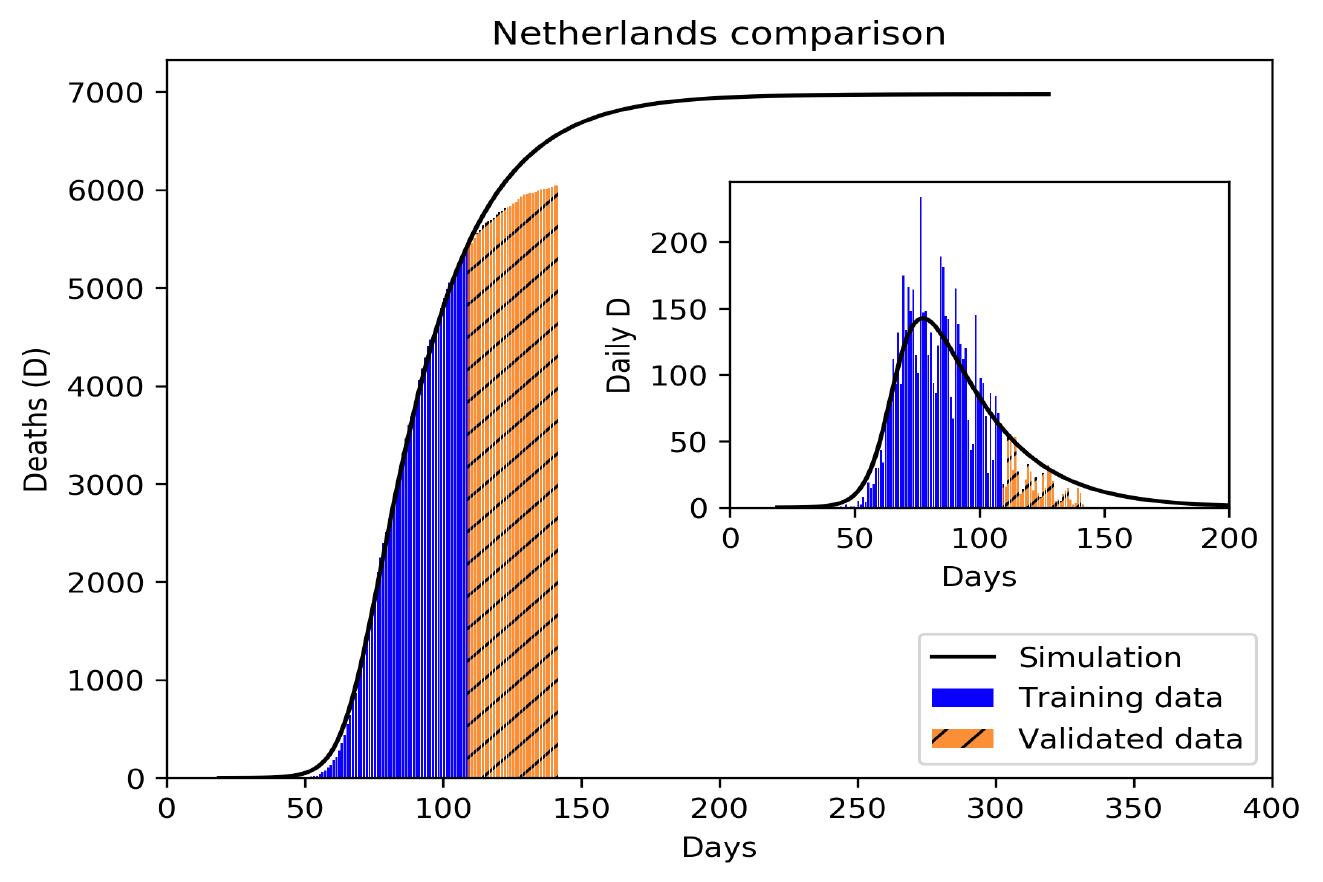}
      \caption{Netherlands mortality profiles.}
    \end{subfigure}
\caption{Infection (S2a) and mortality (S2b) epidemiology for Netherlands (Class A). The outsets all represent the cumulative statistics while the insets are for daily updates in the number of infected and death respectively.}
\label{fig_classA_netherlands}
  \end{figure}
  
  \begin{figure}[h!]
    \begin{subfigure}[t]{0.48\textwidth}
      \includegraphics[width=\textwidth,height=0.3\textheight]{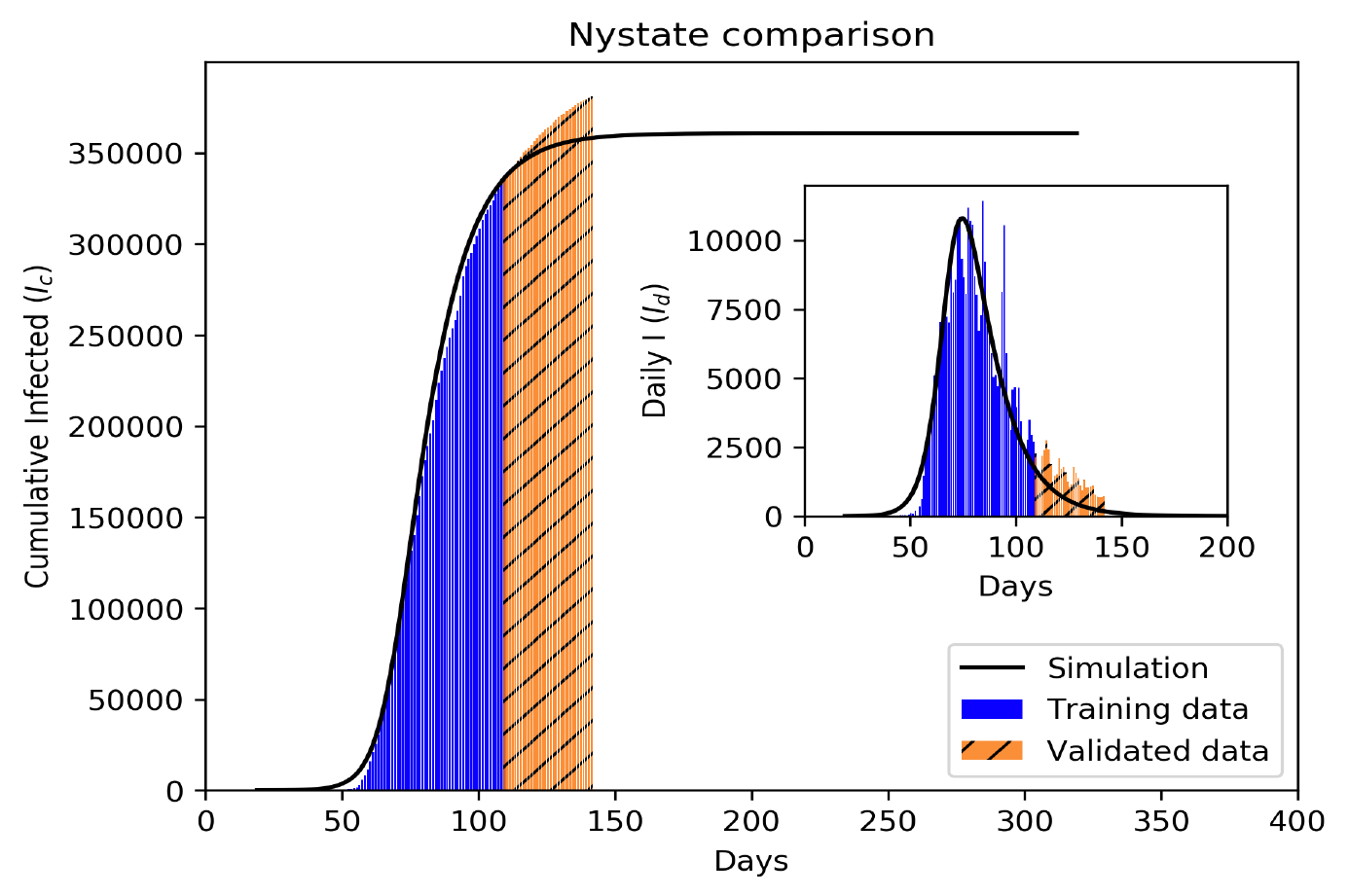}
      \caption{New York State infection profiles.}
    \end{subfigure}
    \hfill
    \begin{subfigure}[t]{0.48\textwidth}
      \includegraphics[width=\textwidth,height=0.3\textheight]{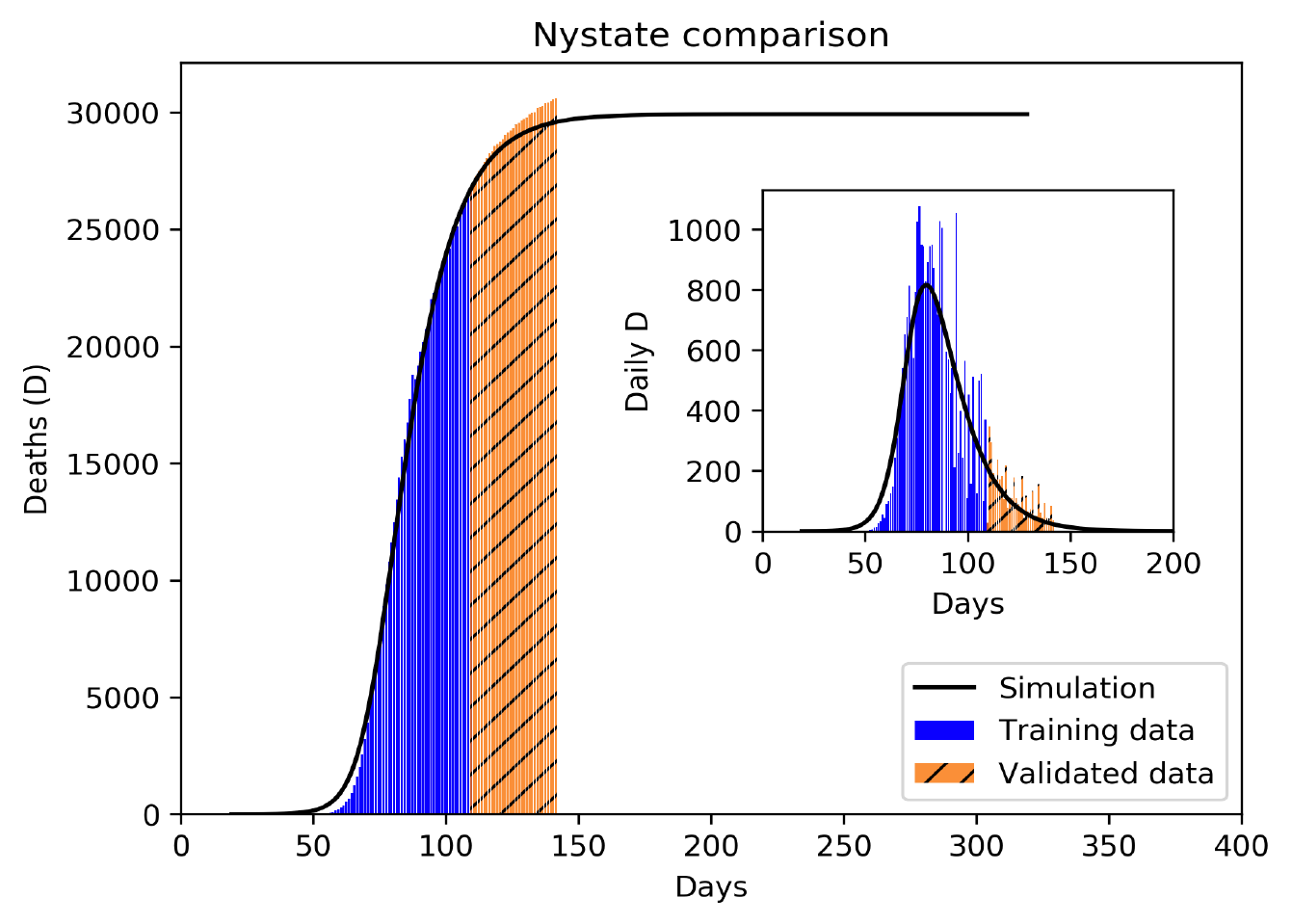}
      \caption{New York State mortality profiles.}
    \end{subfigure}
\caption{Infection (S3a) and mortality (S3b) epidemiology for New York State (Class A). The outsets all represent the cumulative statistics while the insets are for daily updates in the number of infected and death respectively.}
\label{fig_classA_nystate}
  \end{figure}
  
  \begin{figure}[h!]
    \begin{subfigure}[t]{0.48\textwidth}
      \includegraphics[width=\textwidth,height=0.3\textheight]{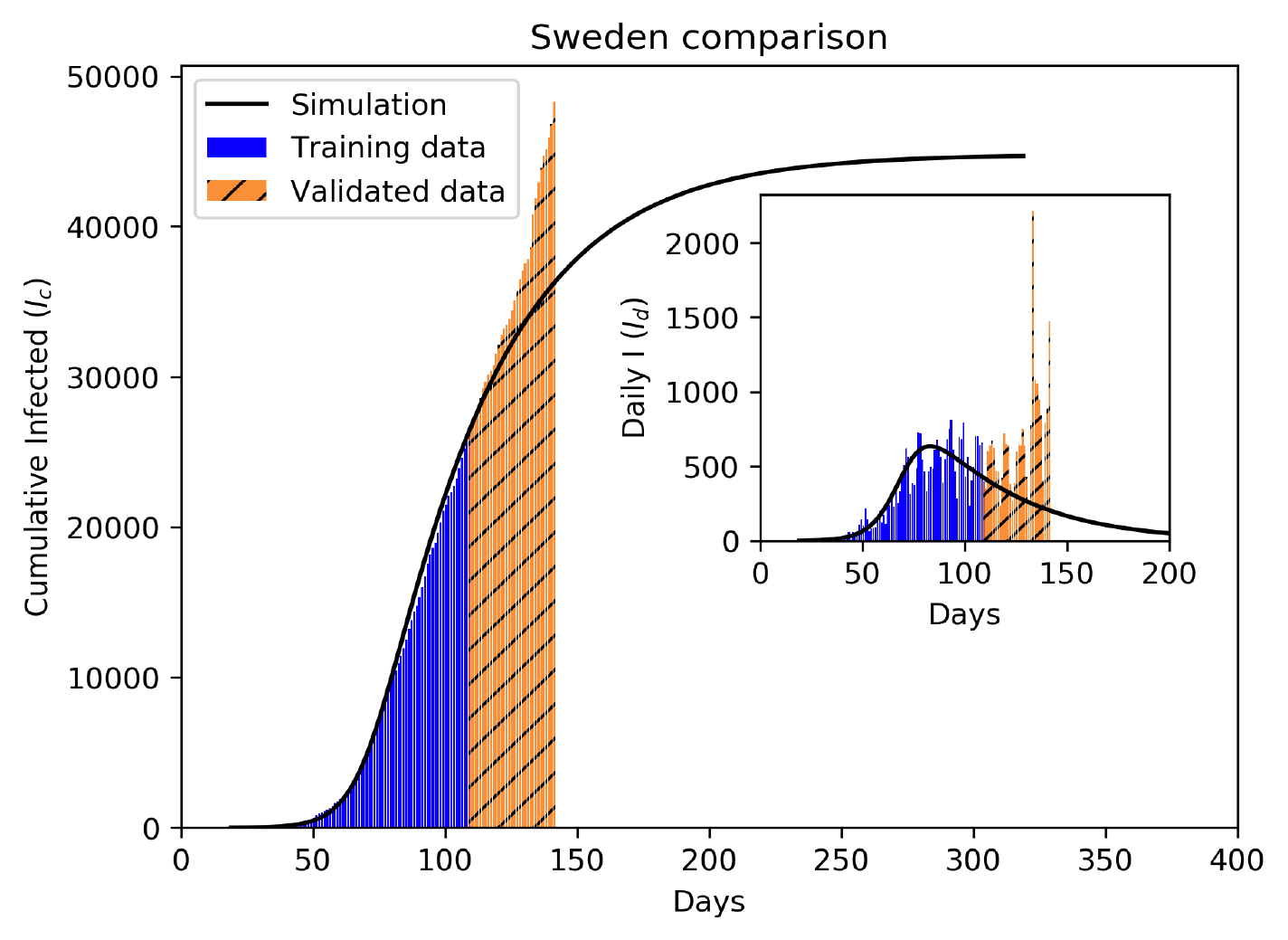}
      \caption{Sweden infection profiles.}
    \end{subfigure}
    \hfill
    \begin{subfigure}[t]{0.48\textwidth}
      \includegraphics[width=\textwidth,height=0.3\textheight]{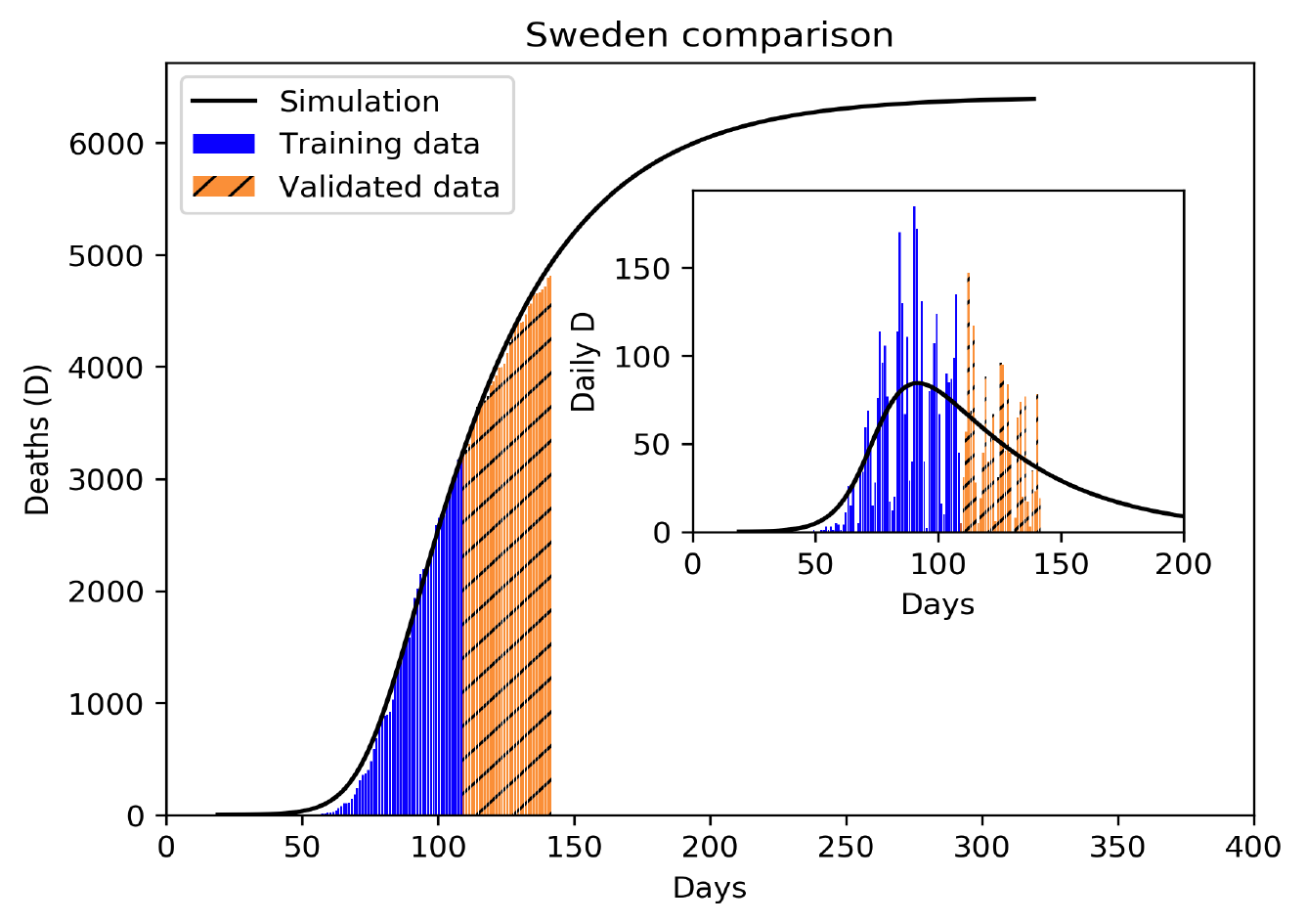}
      \caption{Sweden mortality profiles.}
    \end{subfigure}
\caption{Infection (S4a) and mortality (S4b) epidemiology for Sweden (Class A). The outsets all represent the cumulative statistics while the insets are for daily updates in the number of infected and death respectively.}
\label{fig_classA_sweden}
  \end{figure}
%
%
%
%
%

%
    \begin{figure}[h!]
    \begin{subfigure}[t]{0.48\textwidth}
      \includegraphics[width=\textwidth,height=0.3\textheight]{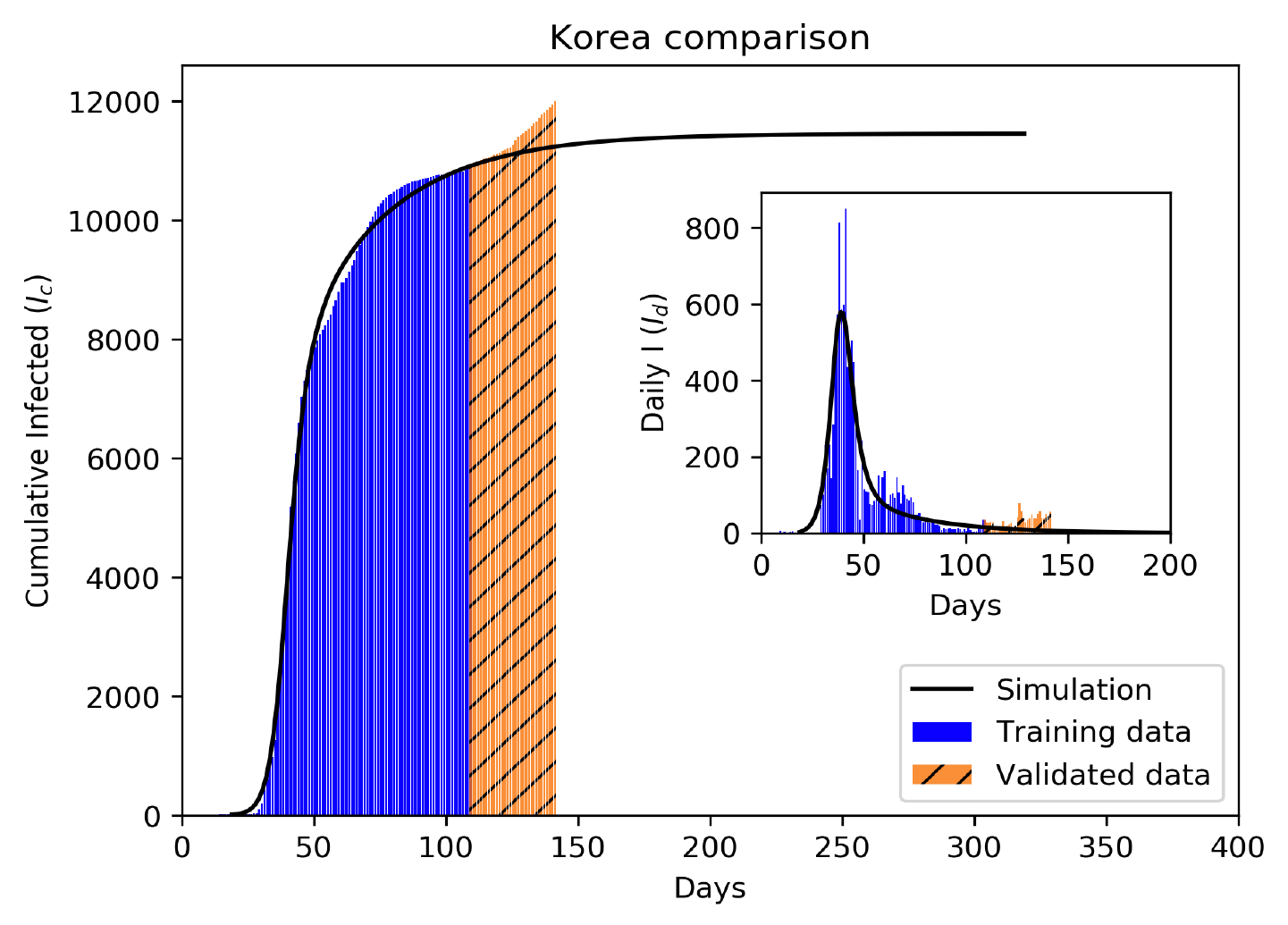}
      \caption{Korea infection profiles.}
    \end{subfigure}
    \hfill
    \begin{subfigure}[t]{0.48\textwidth}
      \includegraphics[width=\textwidth,height=0.3\textheight]{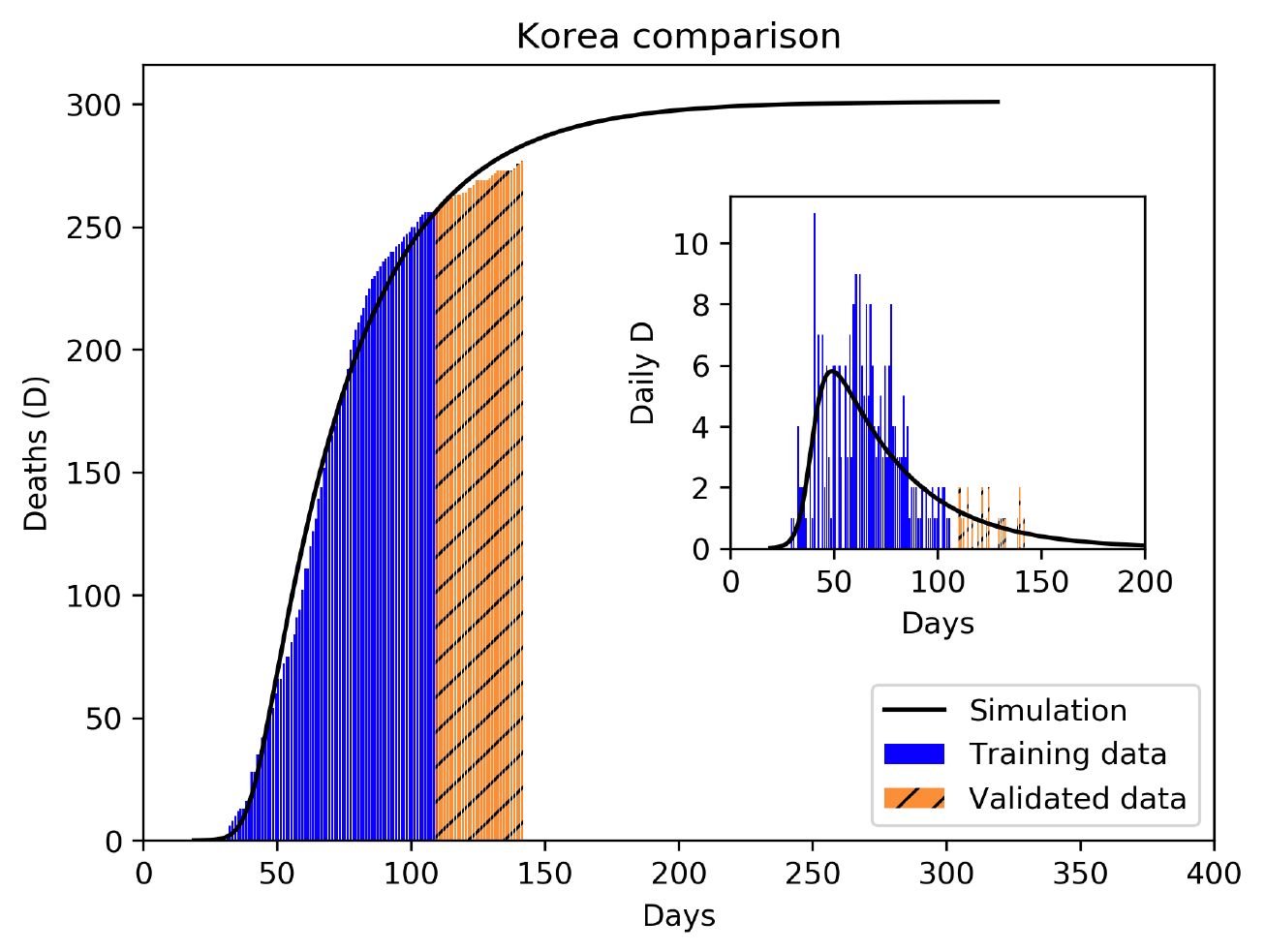}
      \caption{Korea mortality profiles.}
    \end{subfigure}
\caption{Infection (S5a) and mortality (S5b) epidemiology for Korea (Class B). The outsets all represent the cumulative statistics while the insets are for daily updates in the number of infected and death respectively.}
\label{fig_classB_korea}
  \end{figure}
  
   \begin{figure}[h!]
    \begin{subfigure}[t]{0.48\textwidth}
      \includegraphics[width=\textwidth,height=0.3\textheight]{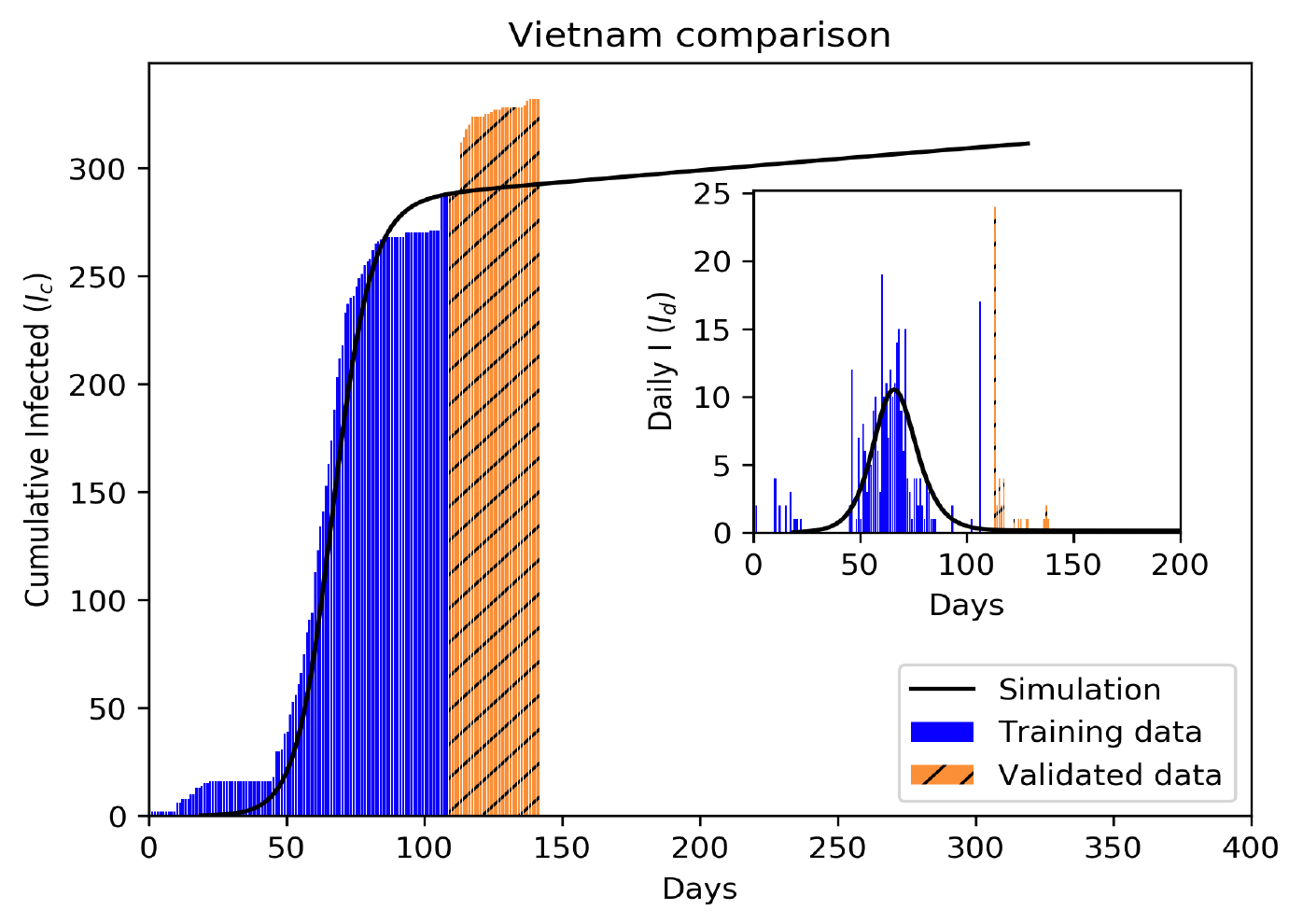}
      \caption{Vietnam infection profiles.}
    \end{subfigure}
    \hfill
    \begin{subfigure}[t]{0.48\textwidth}
      \includegraphics[width=\textwidth,height=0.3\textheight]{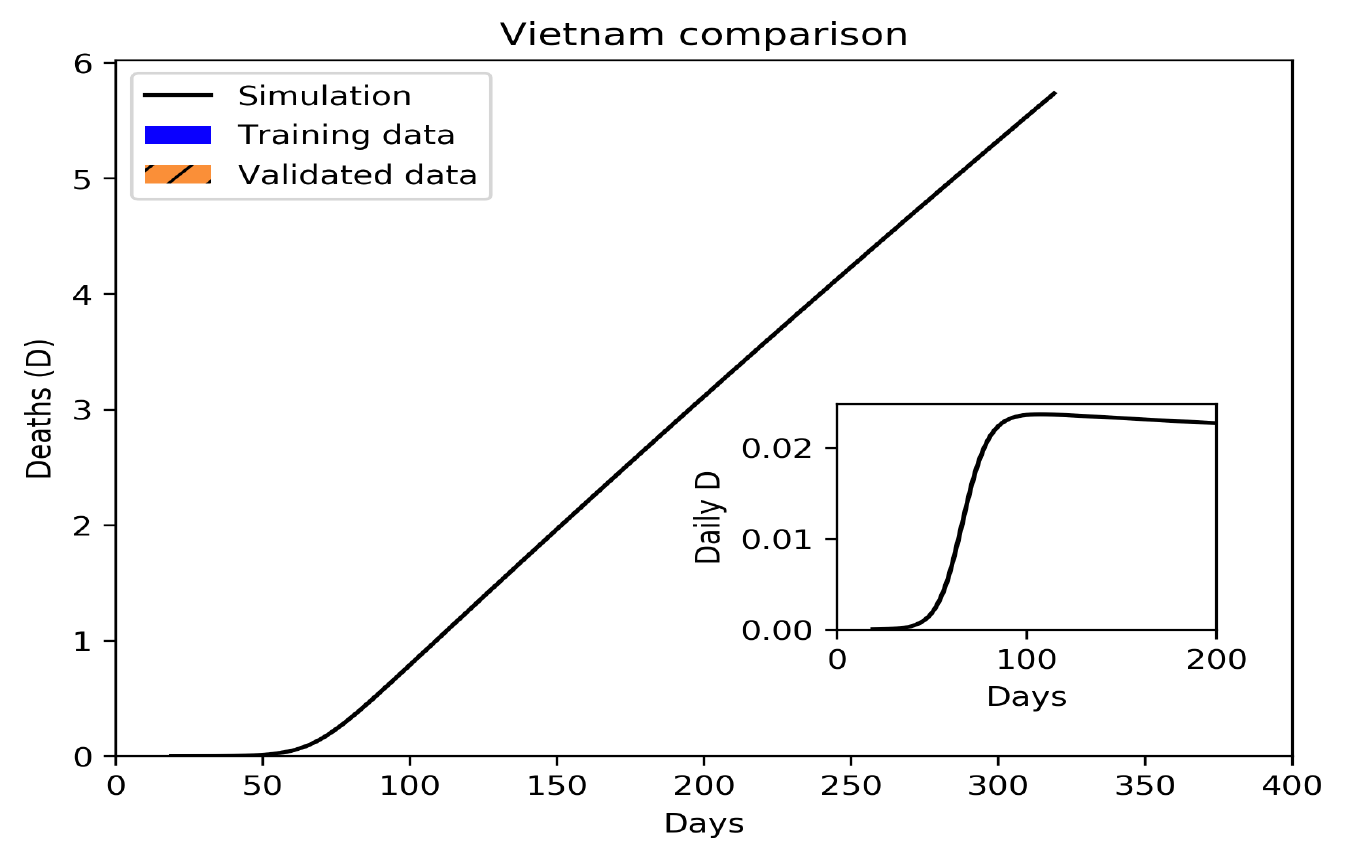}
      \caption{Vietnam mortality profiles.}
    \end{subfigure}
\caption{Infection (S6a) and mortality (S6b) epidemiology for Vietnam. The outsets all represent the cumulative statistics while the insets are for daily updates in the number of infected and death respectively.}
\label{fig_Vietnam}
  \end{figure}
  
   \begin{figure}[h!]
    \begin{subfigure}[t]{0.48\textwidth}
      \includegraphics[width=\textwidth,height=0.3\textheight]{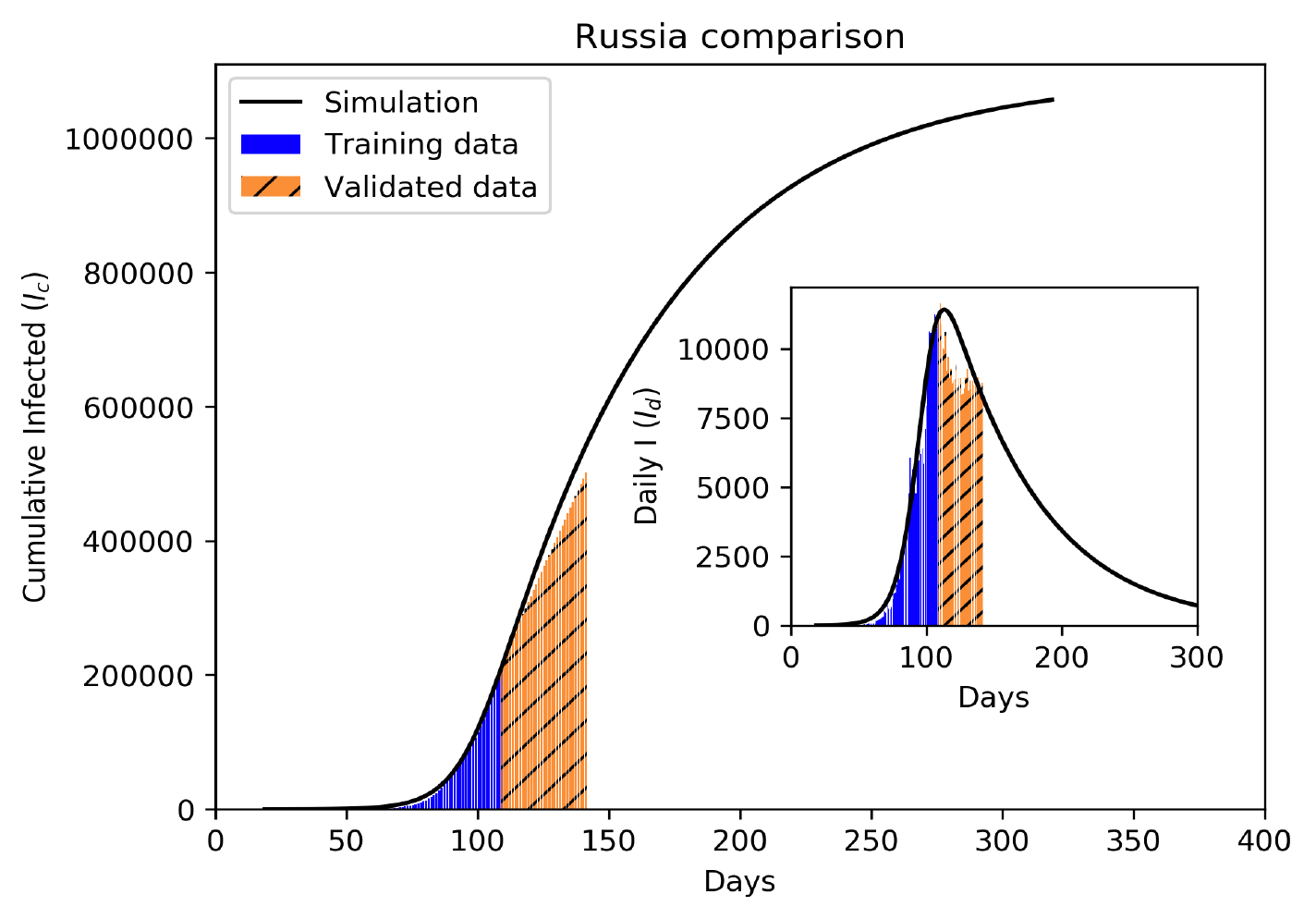}
      \caption{Russia infection profiles.}
    \end{subfigure}
    \hfill
    \begin{subfigure}[t]{0.48\textwidth}
      \includegraphics[width=\textwidth,height=0.3\textheight]{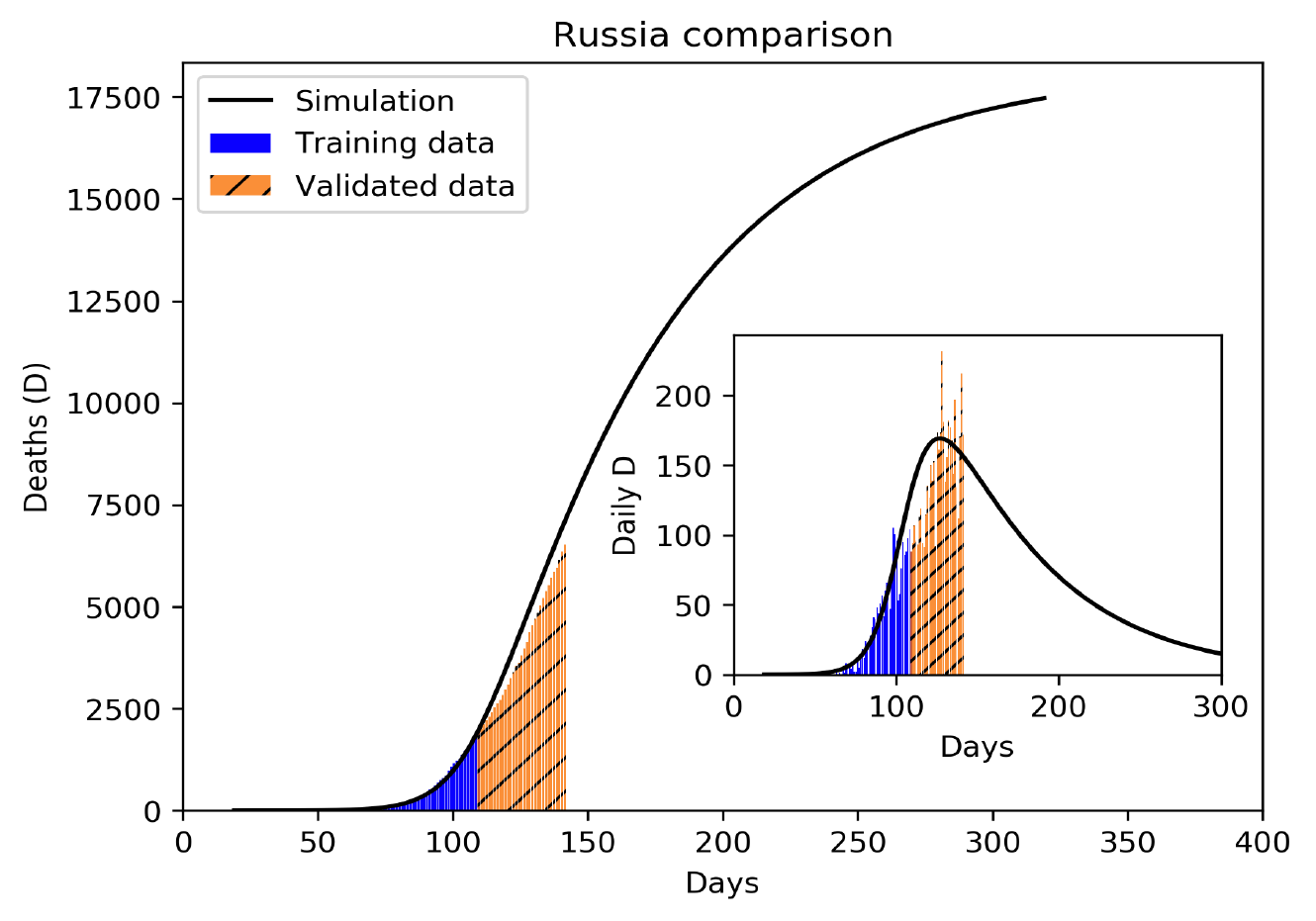}
      \caption{Russia mortality profiles.}
    \end{subfigure}
\caption{Infection (S7a) and mortality (S7b) epidemiology for Russia (Class B). The outsets all represent the cumulative statistics while the insets are for daily updates in the number of infected and death respectively.}
\label{fig_classB_russia}
  \end{figure}
  
     \begin{figure}[h!]
    \begin{subfigure}[t]{0.48\textwidth}
      \includegraphics[width=\textwidth,height=0.3\textheight]{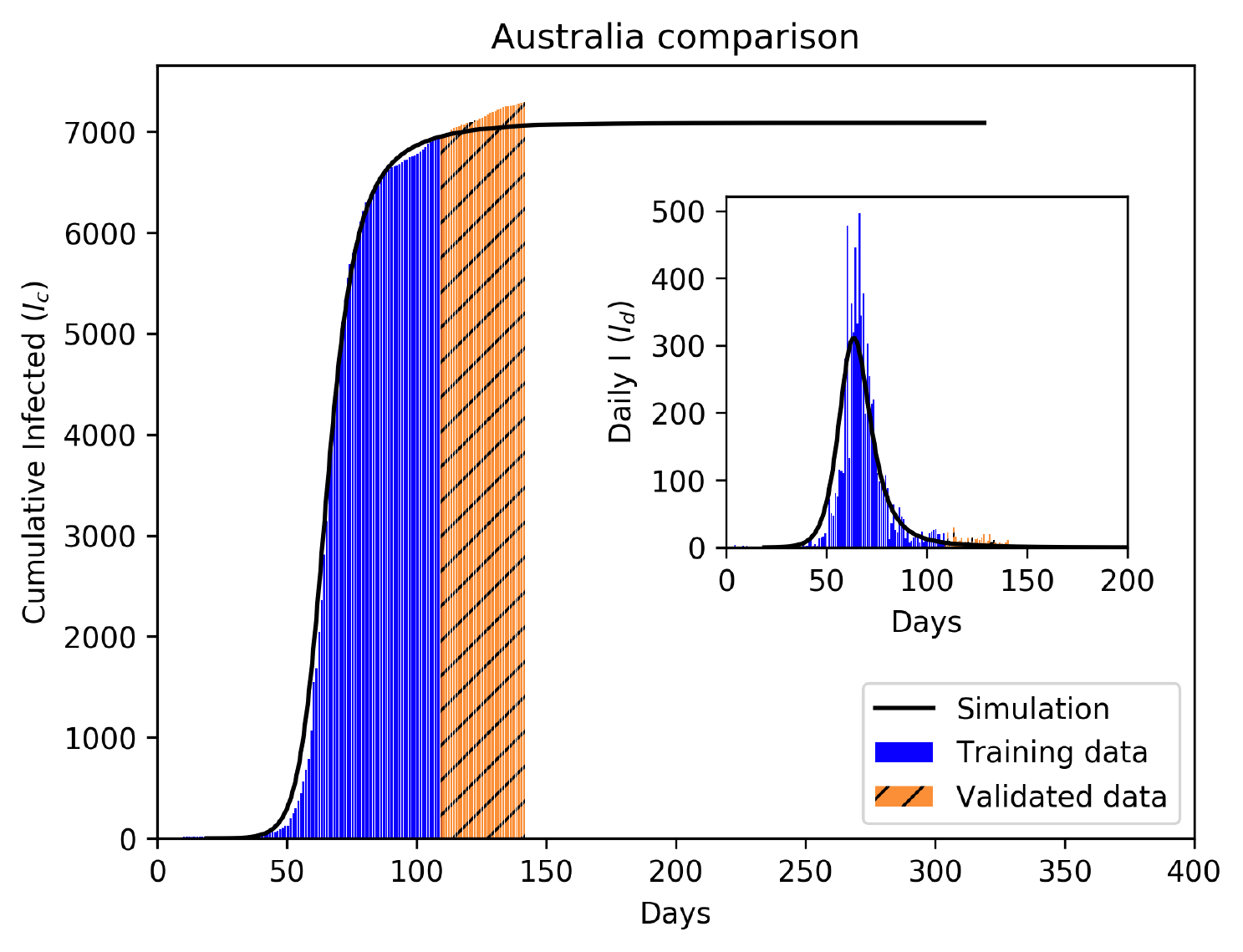}
      \caption{Australia infection profiles.}
    \end{subfigure}
    \hfill
    \begin{subfigure}[t]{0.48\textwidth}
      \includegraphics[width=\textwidth,height=0.3\textheight]{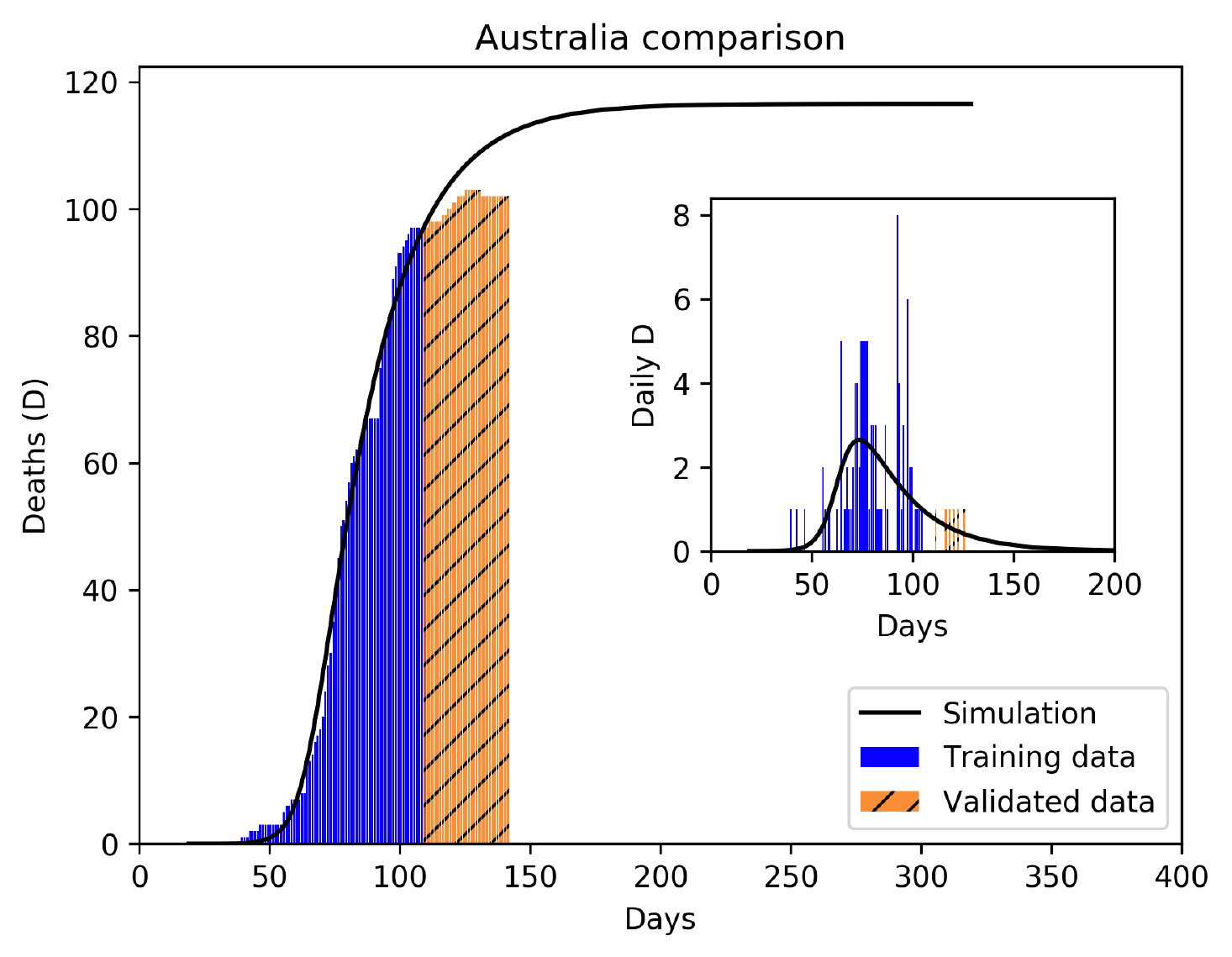}
      \caption{Australia mortality profiles.}
    \end{subfigure}
\caption{Infection (S8a) and mortality (S8b) epidemiology for Australia (Class B). The outsets all represent the cumulative statistics while the insets are for daily updates in the number of infected and death respectively.}
\label{fig_classB_australia}
  \end{figure}
  
       \begin{figure}[h!]
    \begin{subfigure}[t]{0.48\textwidth}
      \includegraphics[width=\textwidth,height=0.3\textheight]{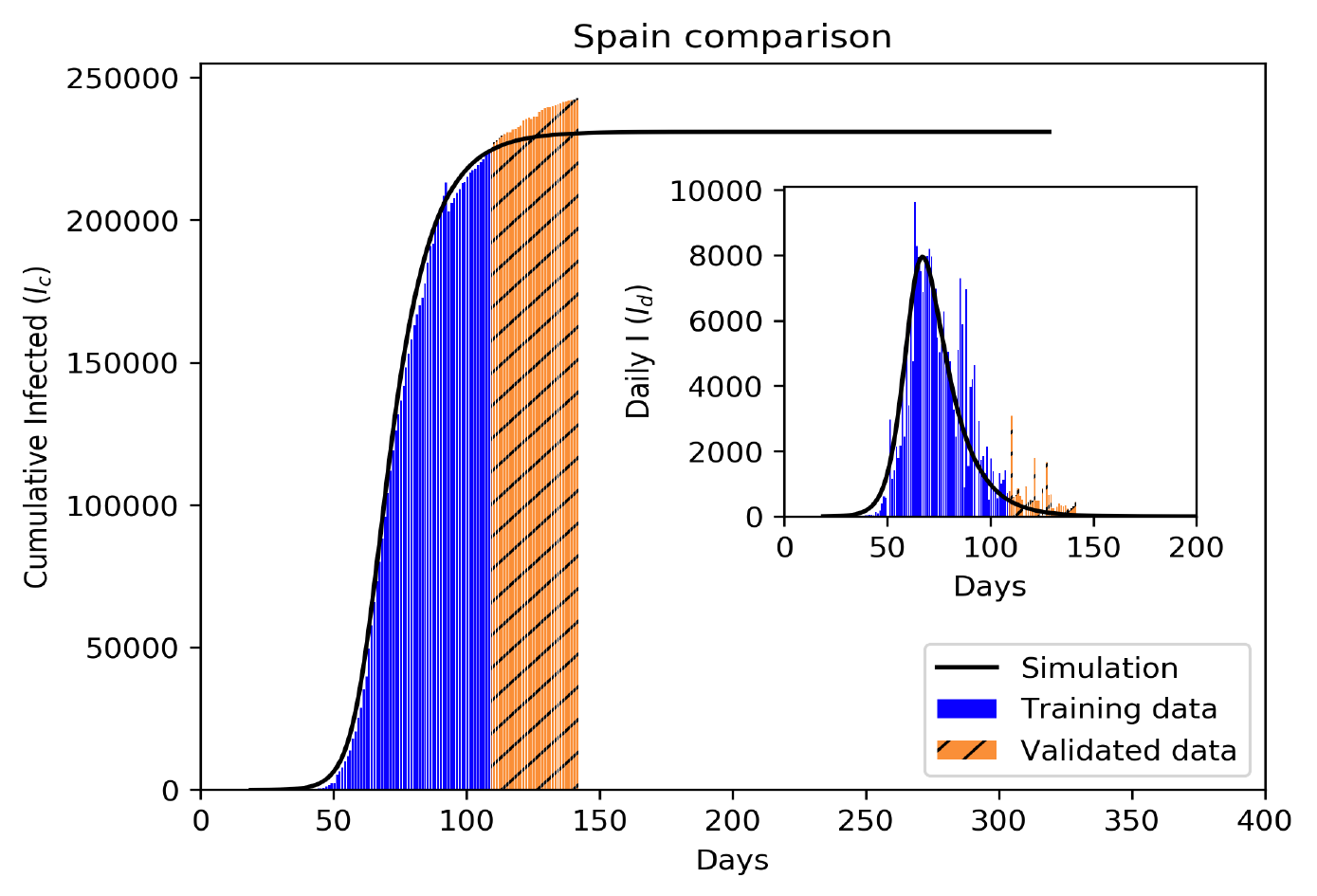}
      \caption{Spain infection profiles.}
    \end{subfigure}
    \hfill
    \begin{subfigure}[t]{0.48\textwidth}
      \includegraphics[width=\textwidth,height=0.3\textheight]{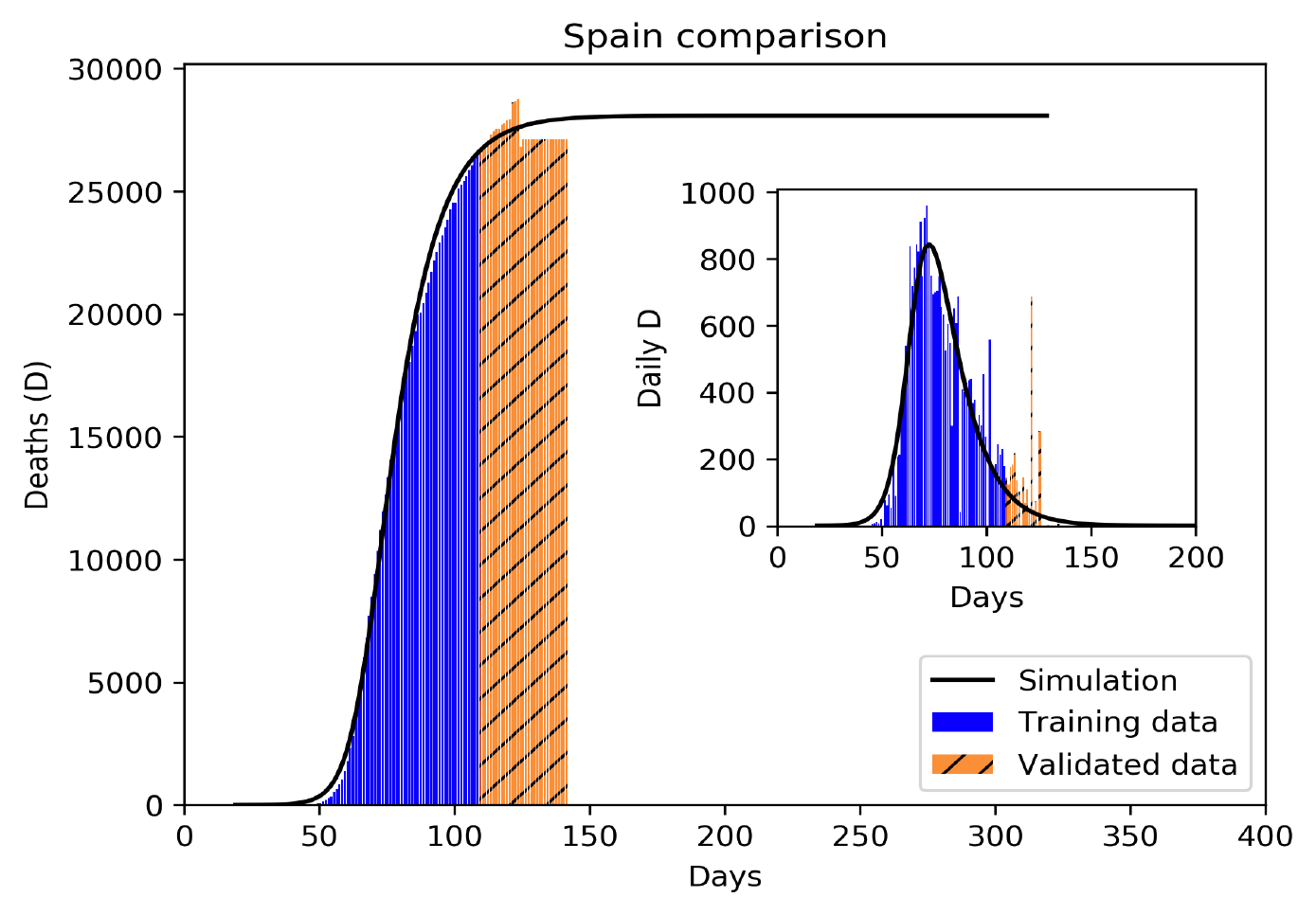}
      \caption{Spain mortality profiles.}
    \end{subfigure}
\caption{Infection (S9a) and mortality (S9b) epidemiology for Spain (Class C). The outsets all represent the cumulative statistics while the insets are for daily updates in the number of infected and death respectively.}
\label{fig_classC_spain}
  \end{figure}
  
         \begin{figure}[h!]
    \begin{subfigure}[t]{0.48\textwidth}
      \includegraphics[width=\textwidth,height=0.3\textheight]{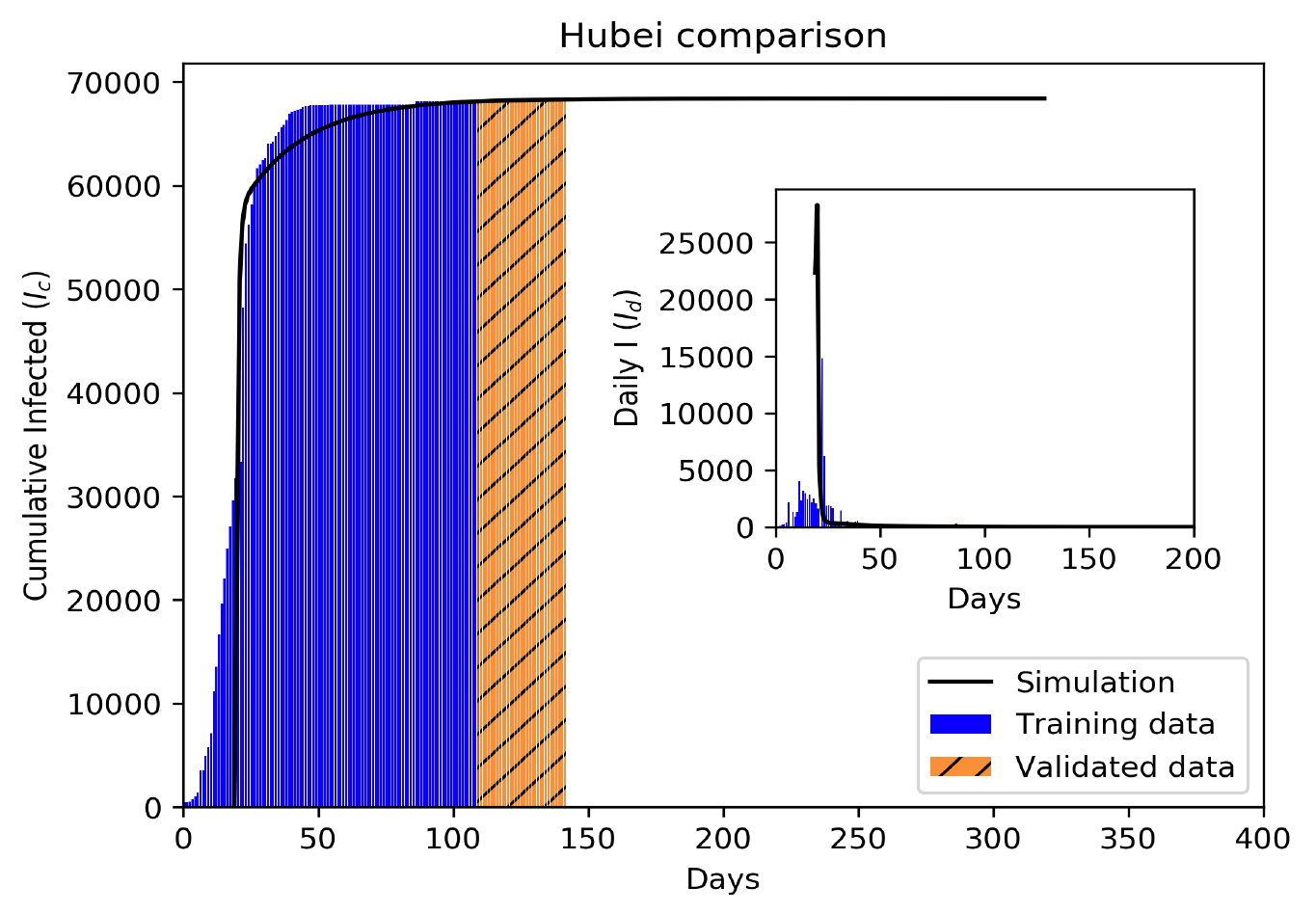}
      \caption{Hubei infection profiles.}
    \end{subfigure}
    \hfill
    \begin{subfigure}[t]{0.48\textwidth}
      \includegraphics[width=\textwidth,height=0.3\textheight]{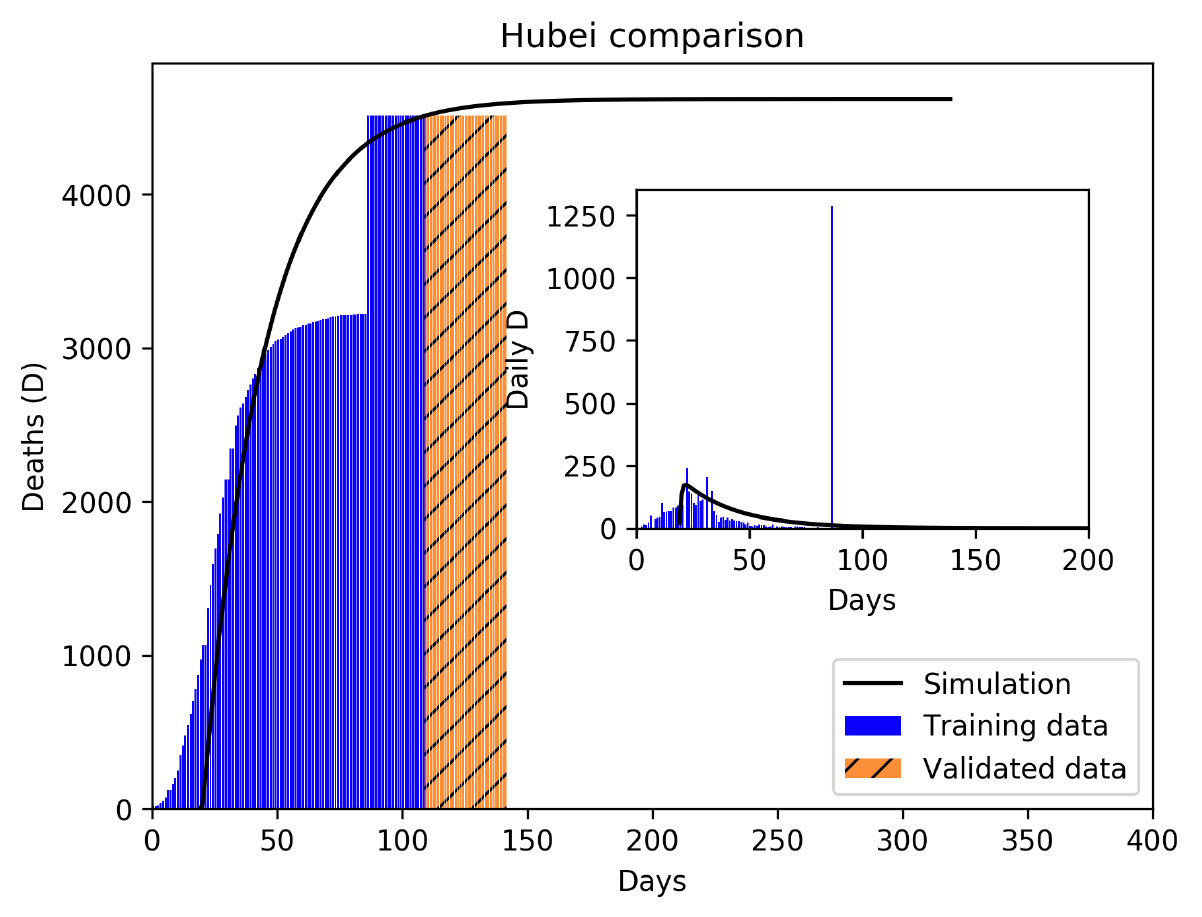}
      \caption{Hubei mortality profiles.}
    \end{subfigure}
\caption{Infection (S10a) and mortality (S10b) epidemiology for Hubei (Class C). The outsets all represent the cumulative statistics while the insets are for daily updates in the number of infected and death respectively.}
\label{fig_classC_hubei}
  \end{figure}
  
    \begin{figure}[h!]
     \begin{subfigure}[t]{0.48\textwidth}
      \includegraphics[width=\textwidth,height=0.3\textheight]{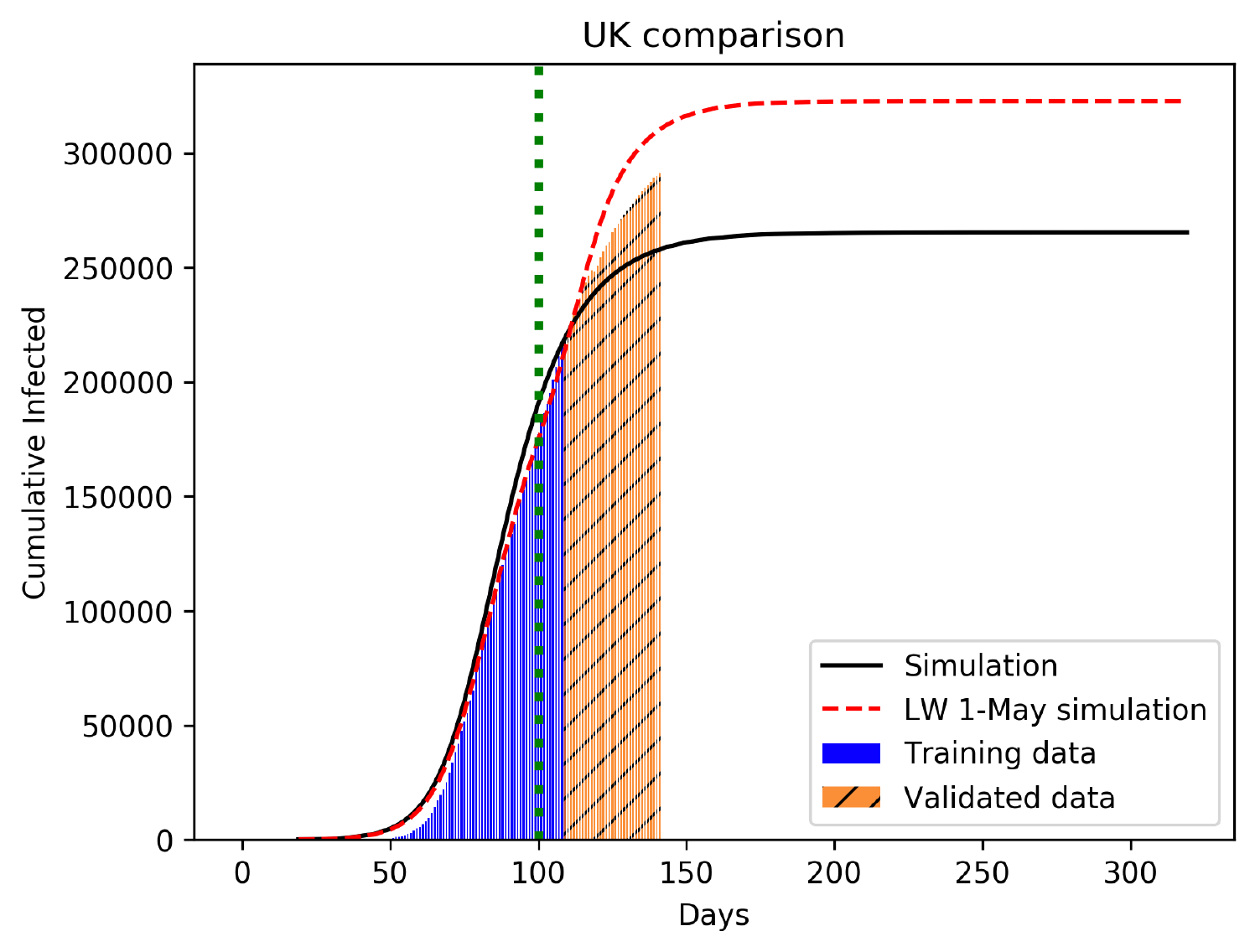}
      \caption{UK: 1 May 2020}
    \end{subfigure}
       \begin{subfigure}[t]{0.48\textwidth}
      \includegraphics[width=\textwidth,height=0.3\textheight]{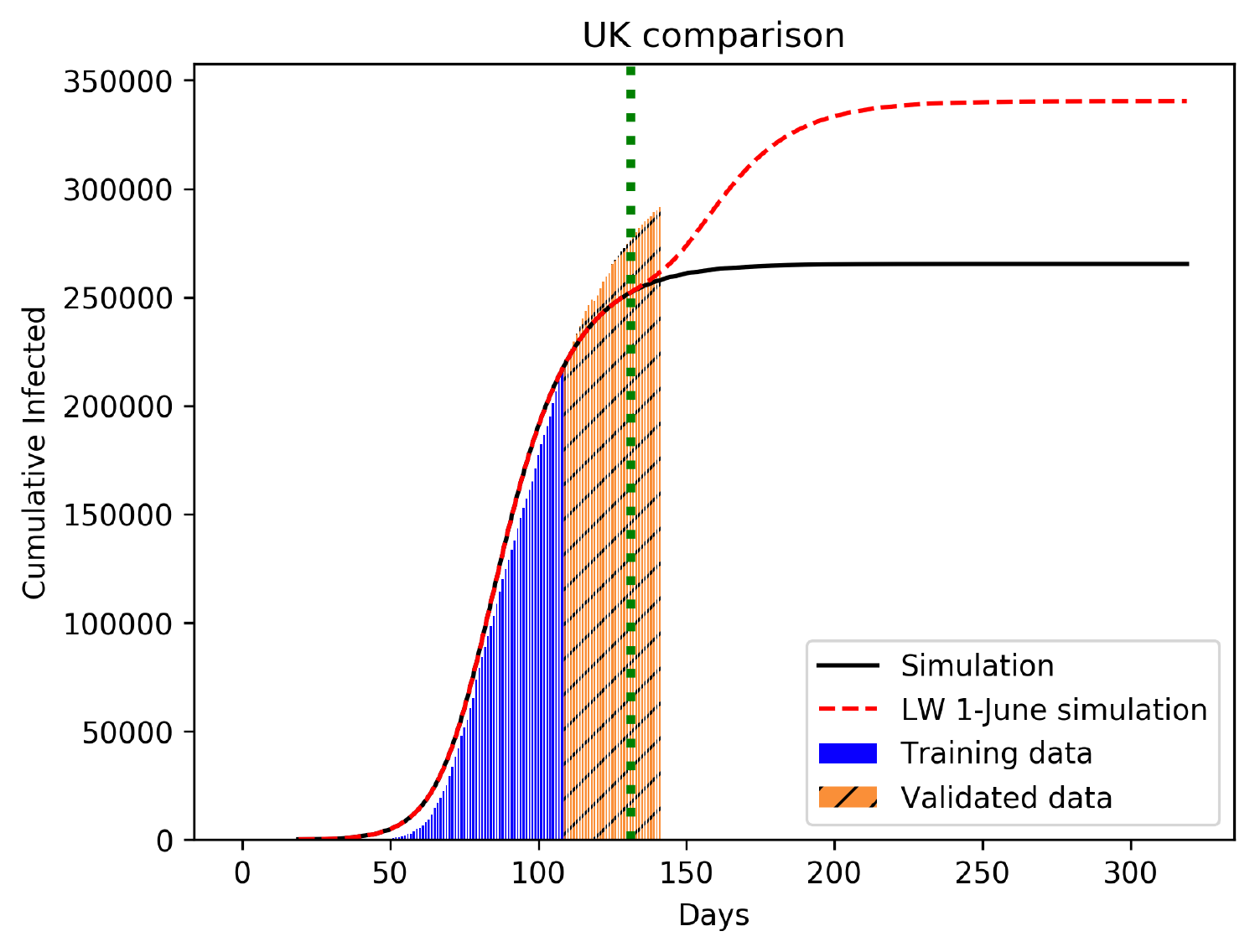}
      \caption{UK: 1 June 2020}
    \end{subfigure}
           \begin{subfigure}[t]{0.48\textwidth}
      \includegraphics[width=\textwidth,height=0.3\textheight]{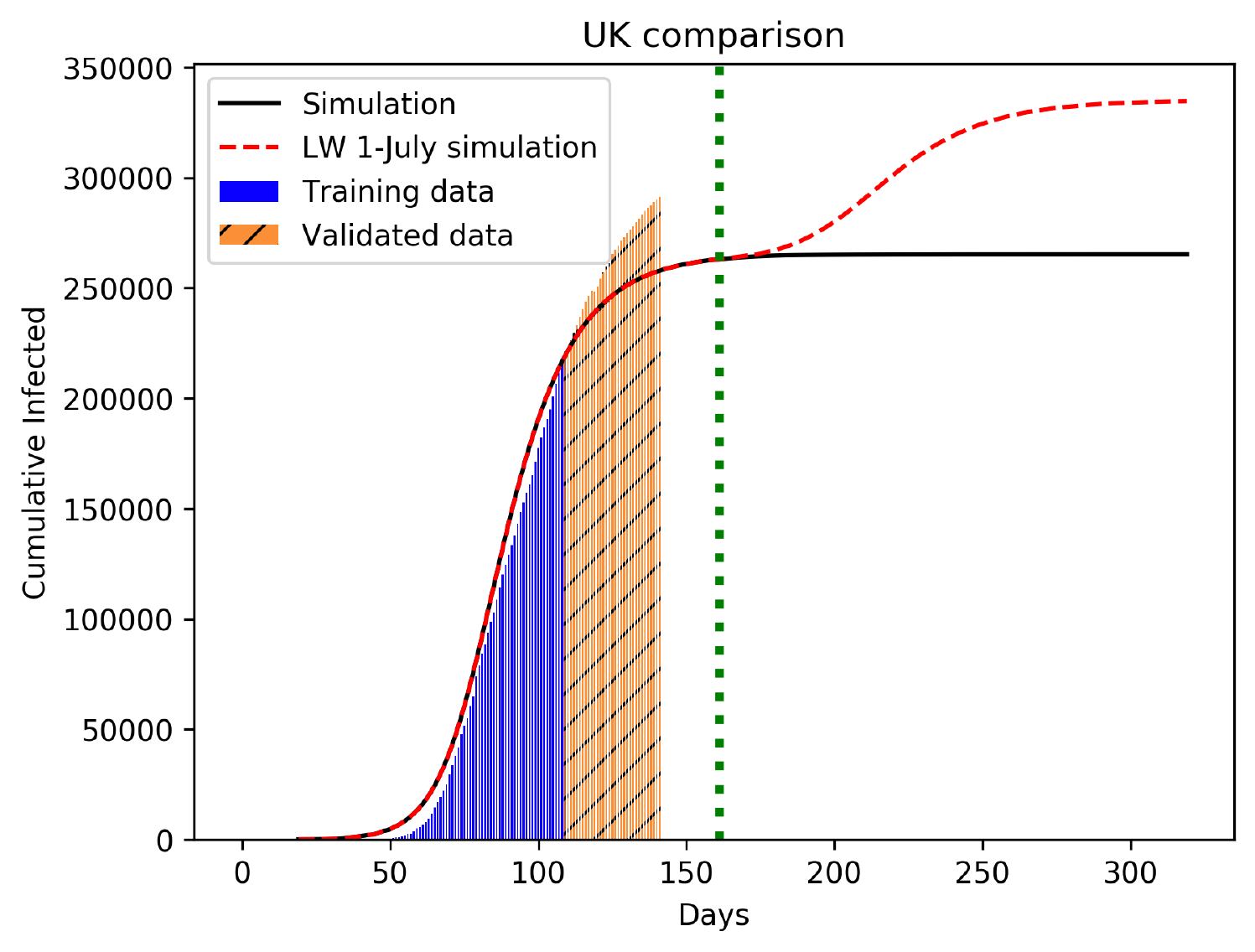}
      \caption{UK: 1 July 2020}
    \end{subfigure}
\caption{UK cumulative infected prediction for 3 different withdrawal dates - 1 May 2020, 1 June 2020, 1 July 2020.}
\label{fig_uk_lockdown}
  \end{figure}
    \begin{figure}[h!]
     \begin{subfigure}[t]{0.48\textwidth}
      \includegraphics[width=\textwidth,height=0.3\textheight]{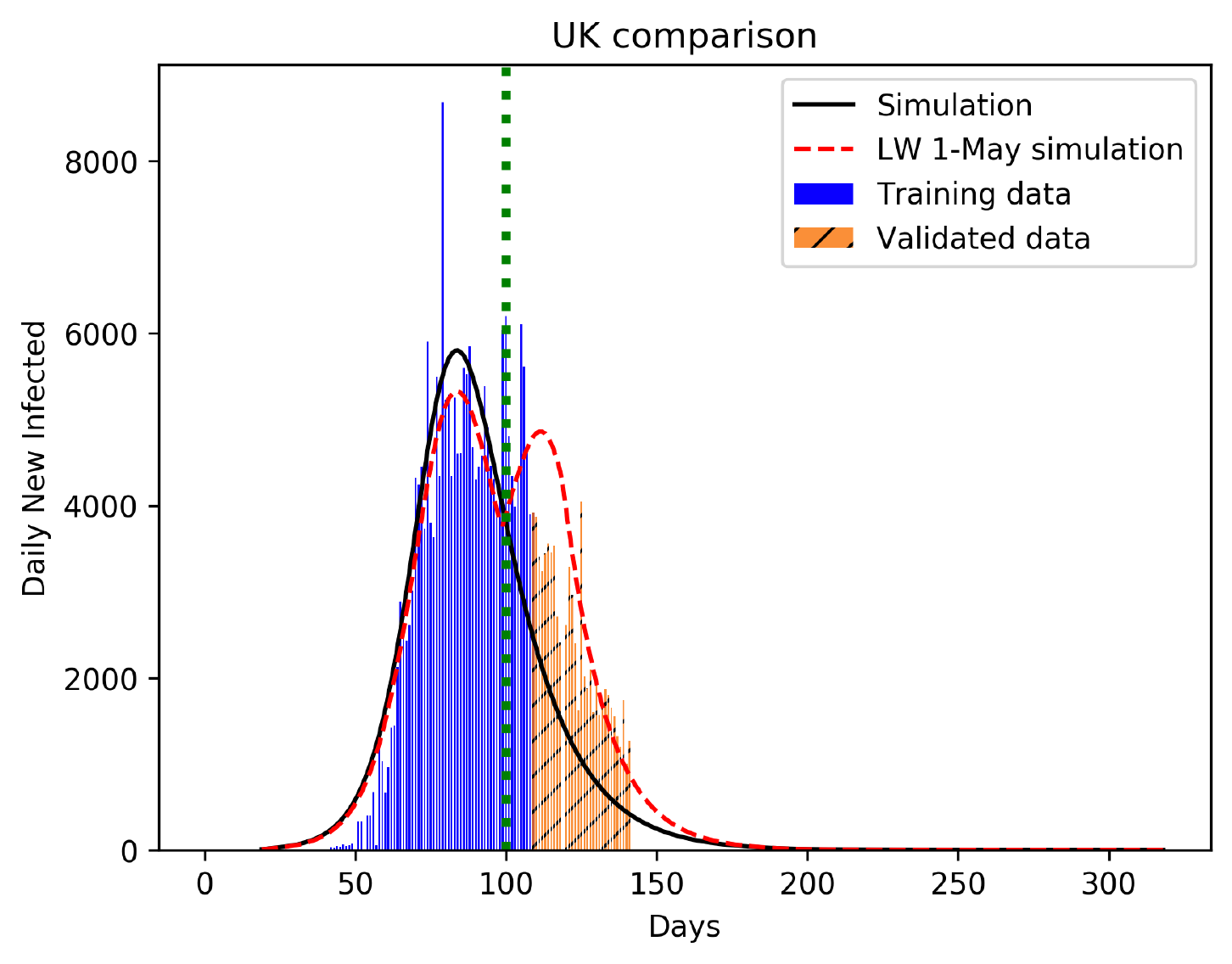}
      \caption{UK: 1 May 2020}
    \end{subfigure}
       \begin{subfigure}[t]{0.48\textwidth}
      \includegraphics[width=\textwidth,height=0.3\textheight]{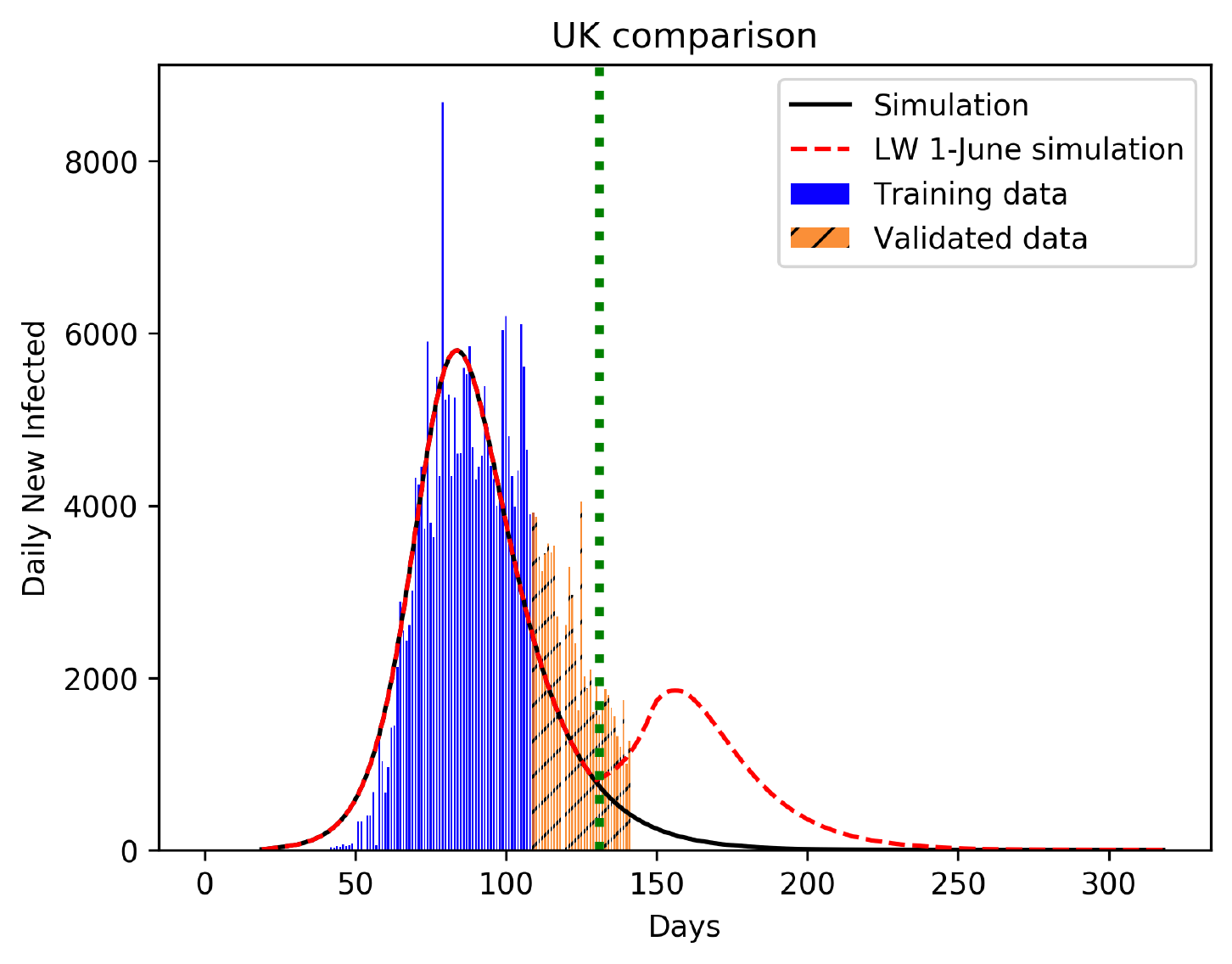}
      \caption{UK: 1 June 2020}
    \end{subfigure}
           \begin{subfigure}[t]{0.48\textwidth}
      \includegraphics[width=\textwidth,height=0.3\textheight]{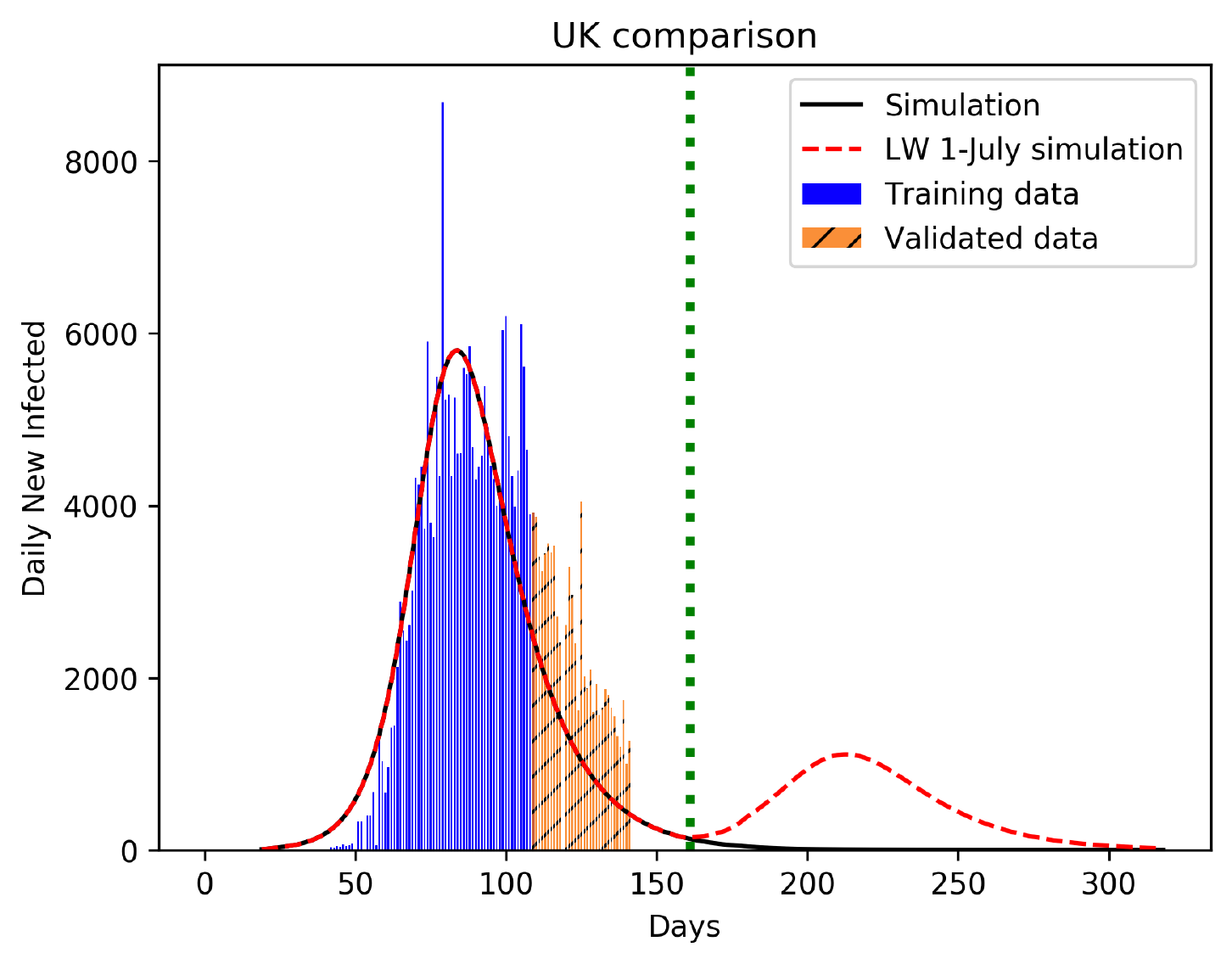}
      \caption{UK: 1 July 2020}
    \end{subfigure}
\caption{UK Lockdown prediction for 3 different withdrawal dates - 1 May 2020, 1 June 2020, 1 July 2020.}
\label{fig_uk_lockdown}
  \end{figure}

\begin{figure}[h!]
\includegraphics[width=6in]{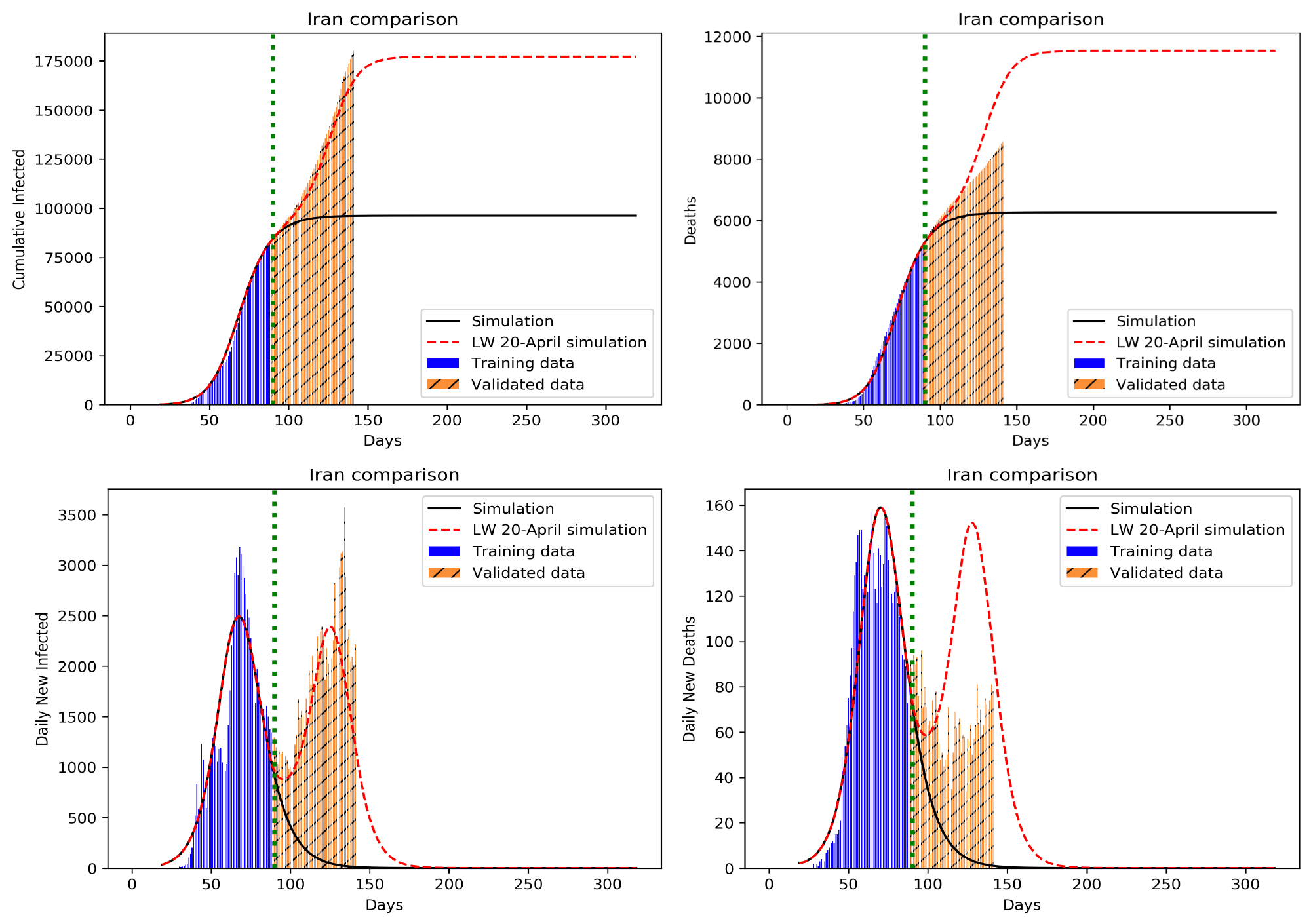}
 \caption{Iran cumulative infected and mortality prediction for lockdown withdrawal on 20 April 2020.}
\end{figure}

\begin{figure}[h!]
\includegraphics[width=6in]{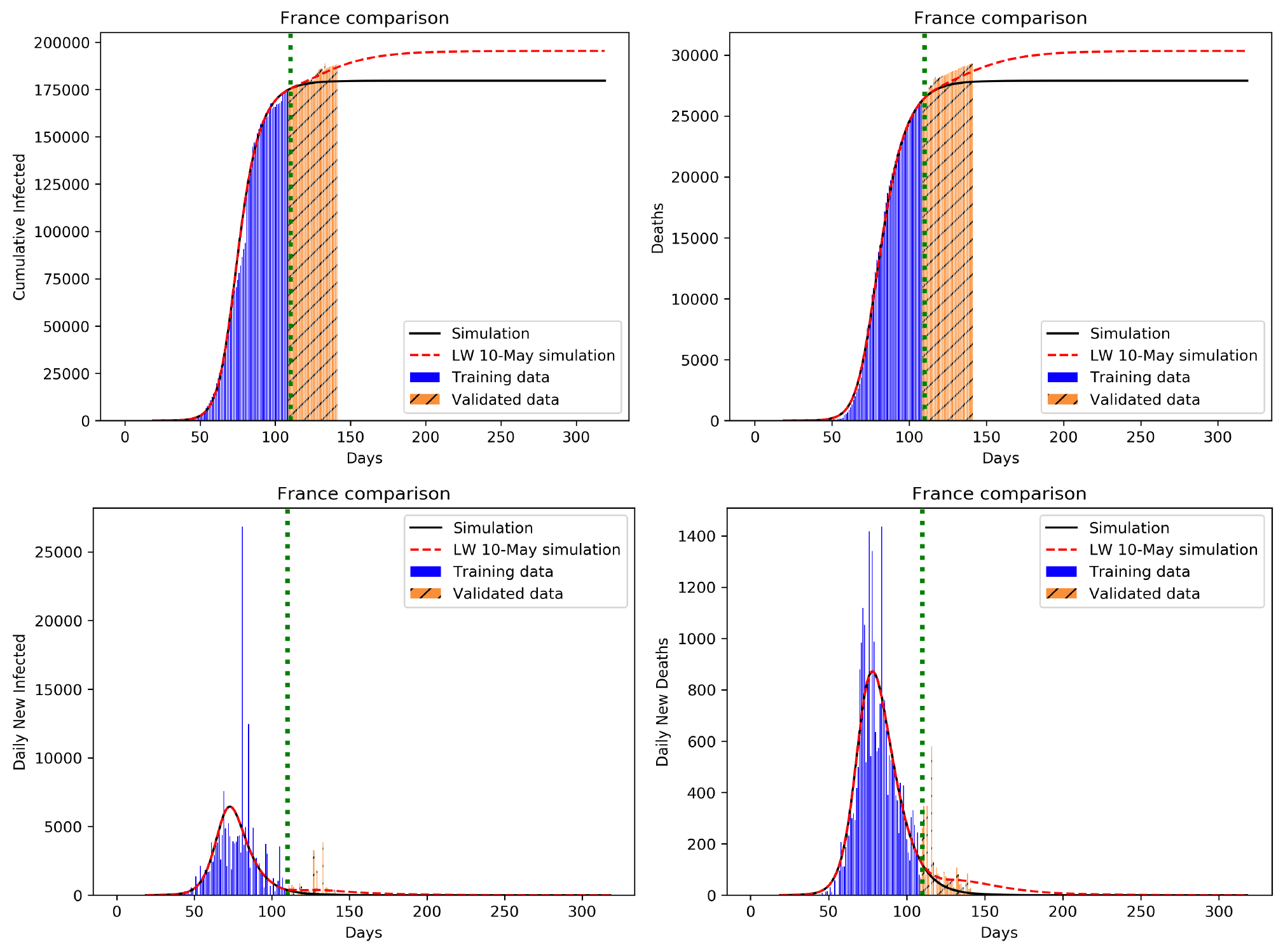}
 \caption{France cumulative infected and mortality prediction for lockdown withdrawal on 10 May 2020.}
\end{figure}

\begin{figure}[h!]
\includegraphics[width=6in]{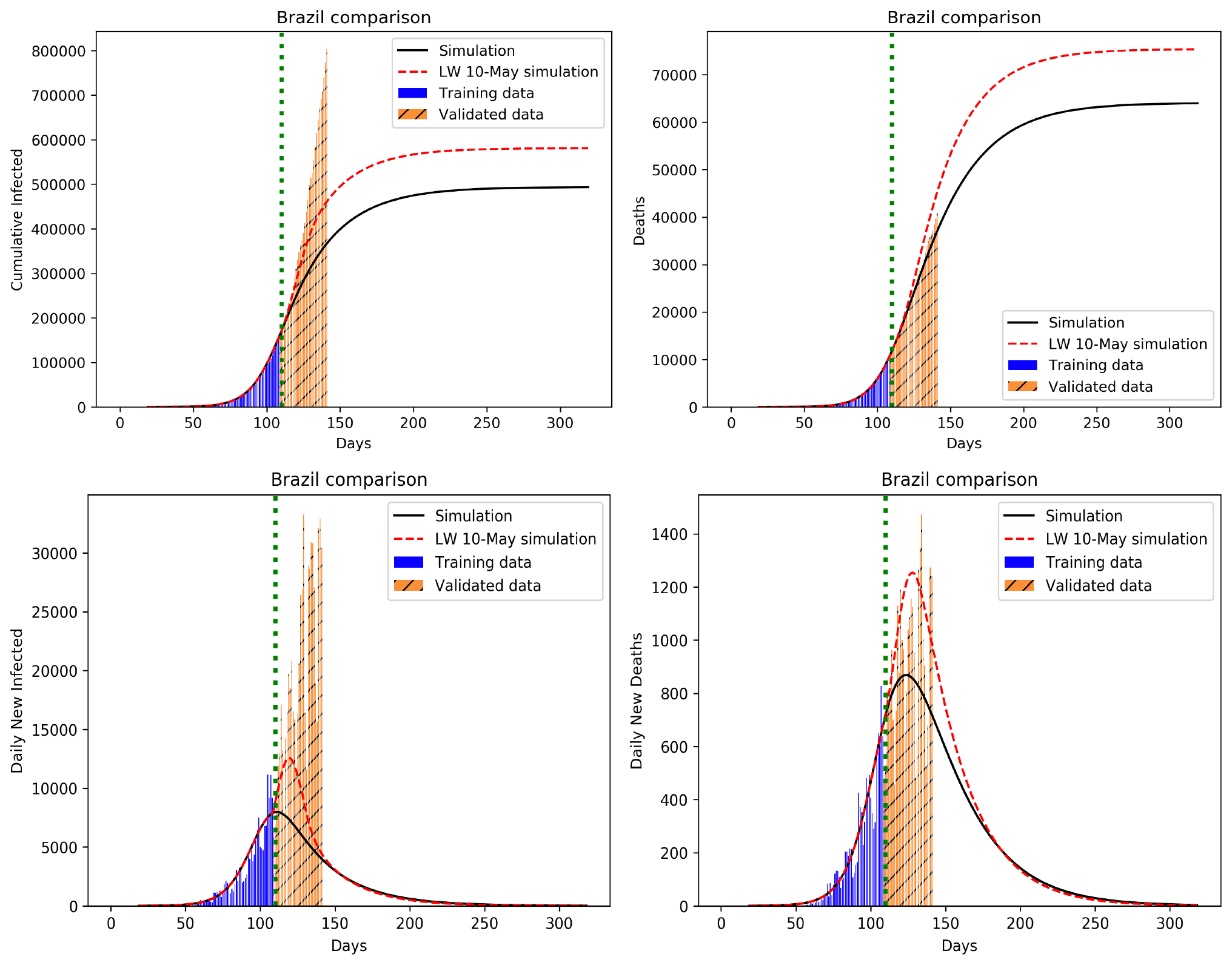}
 \caption{Brazil cumulative infected and mortality prediction for lockdown withdrawal on 10 May 2020.}
\end{figure}

\begin{figure}[h!]
\includegraphics[width=6in]{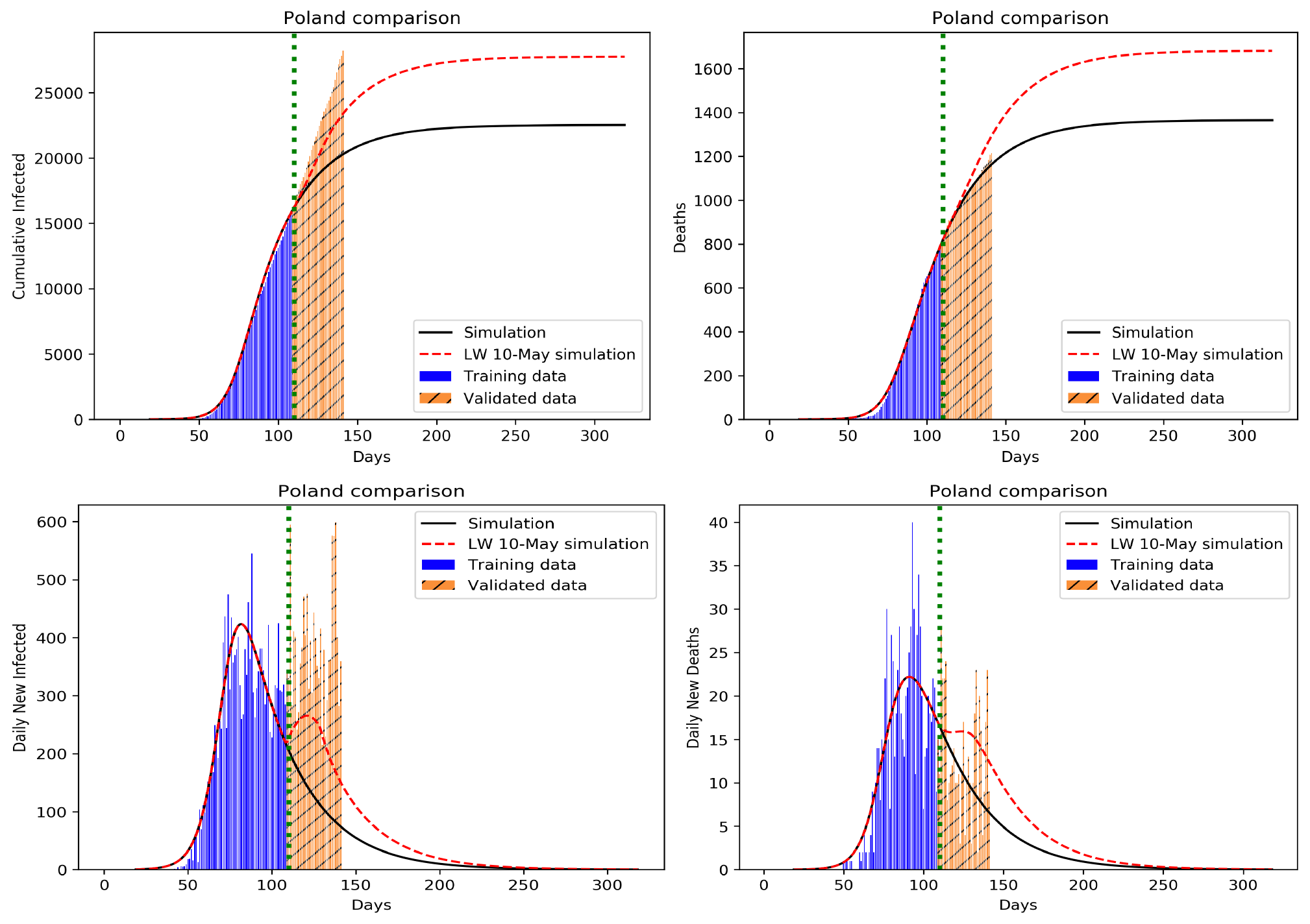}
 \caption{Poland cumulative infected and mortality prediction for lockdown withdrawal on 10 May 2020.}
\end{figure}

\begin{figure}[h!]
\includegraphics[width=6in]{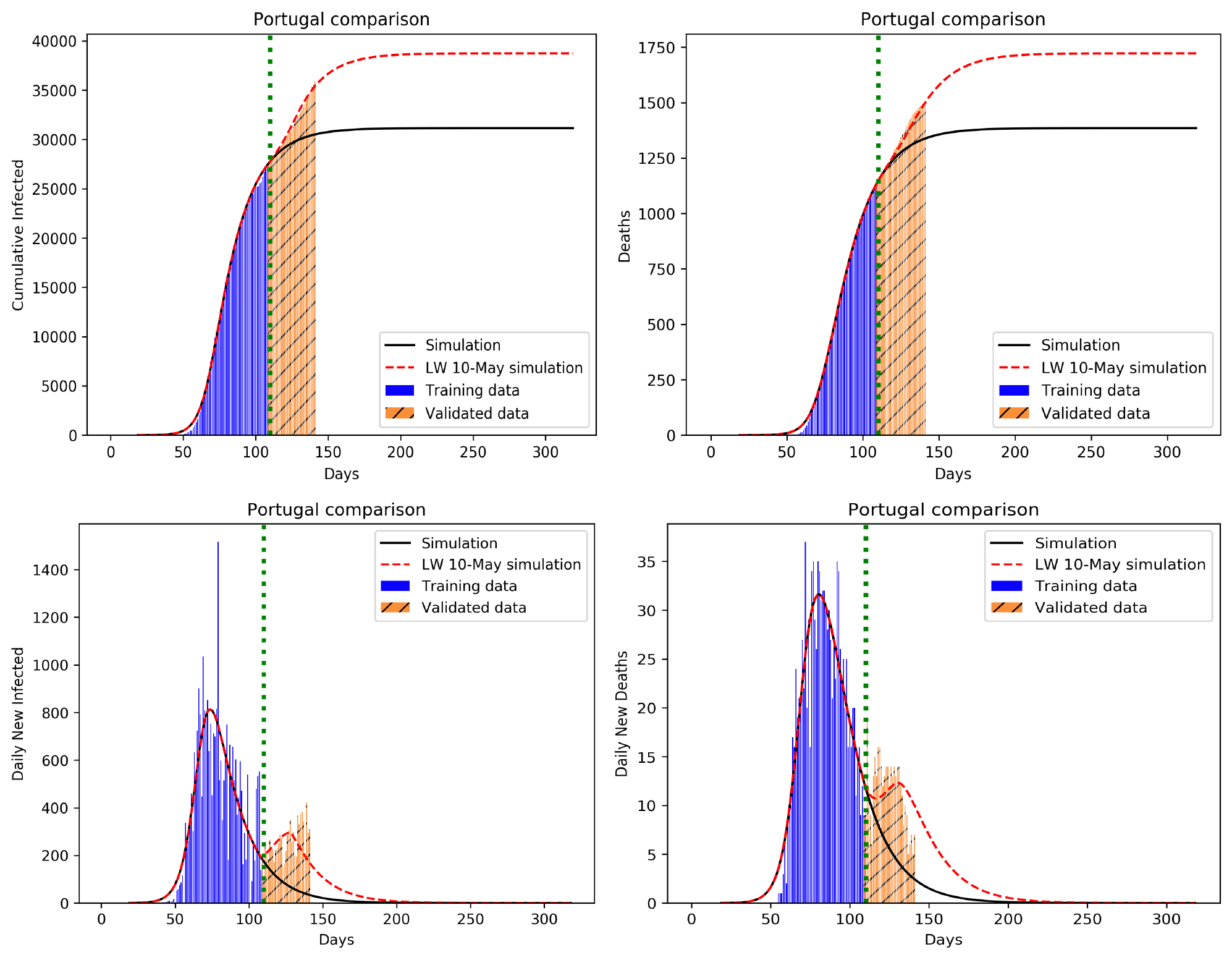}
 \caption{Portugal cumulative infected and mortality prediction for lockdown withdrawal on 10 May 2020.}
\end{figure}

      \begin{figure}[h!]
     \begin{subfigure}[t]{0.48\textwidth}
      \includegraphics[width=\textwidth,height=0.3\textheight]{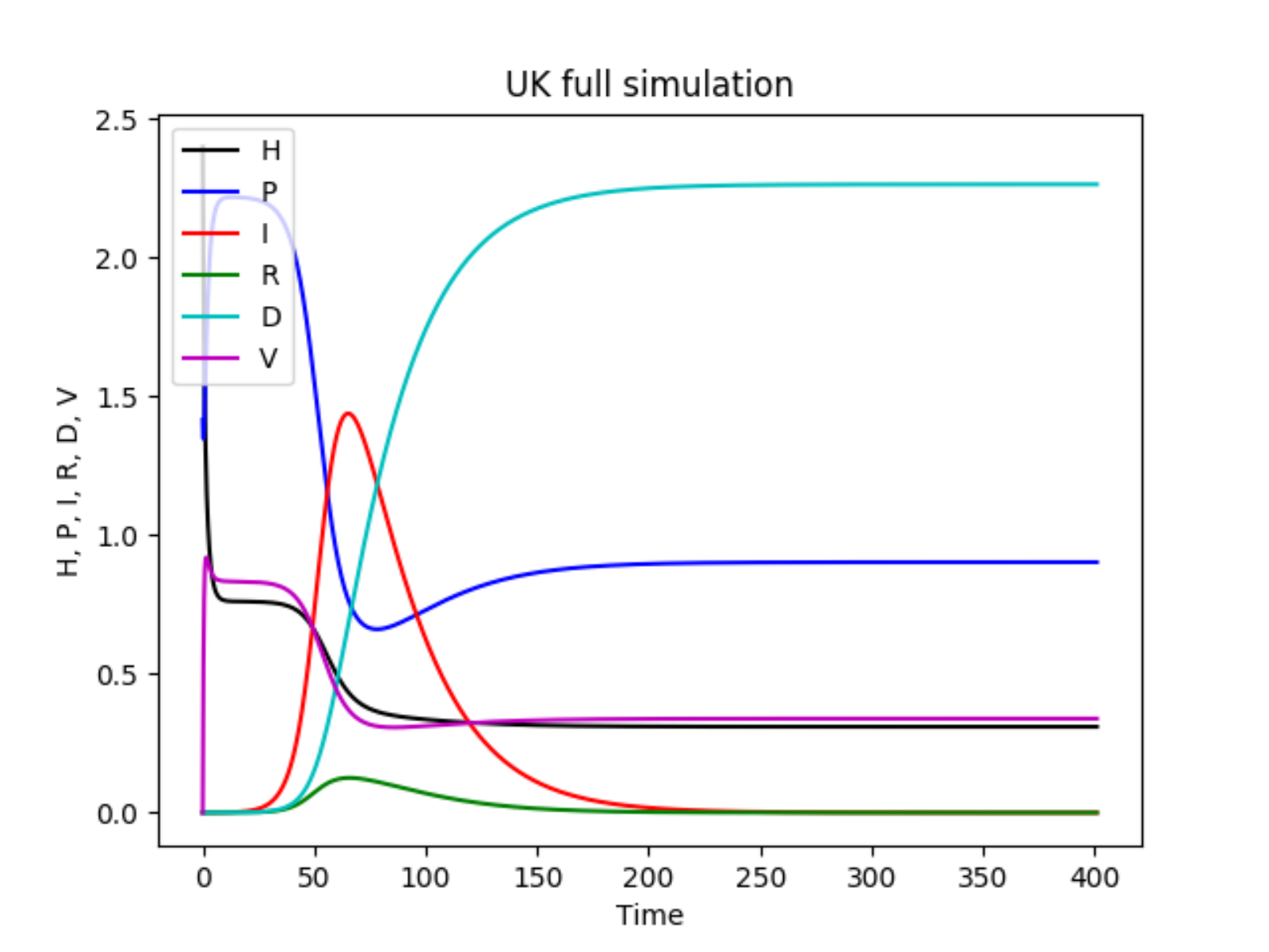}
      \caption{Full simulation plots including all 6 dimensions for the UK}
    \end{subfigure}
     \begin{subfigure}[t]{0.48\textwidth}
      \includegraphics[width=\textwidth,height=0.3\textheight]{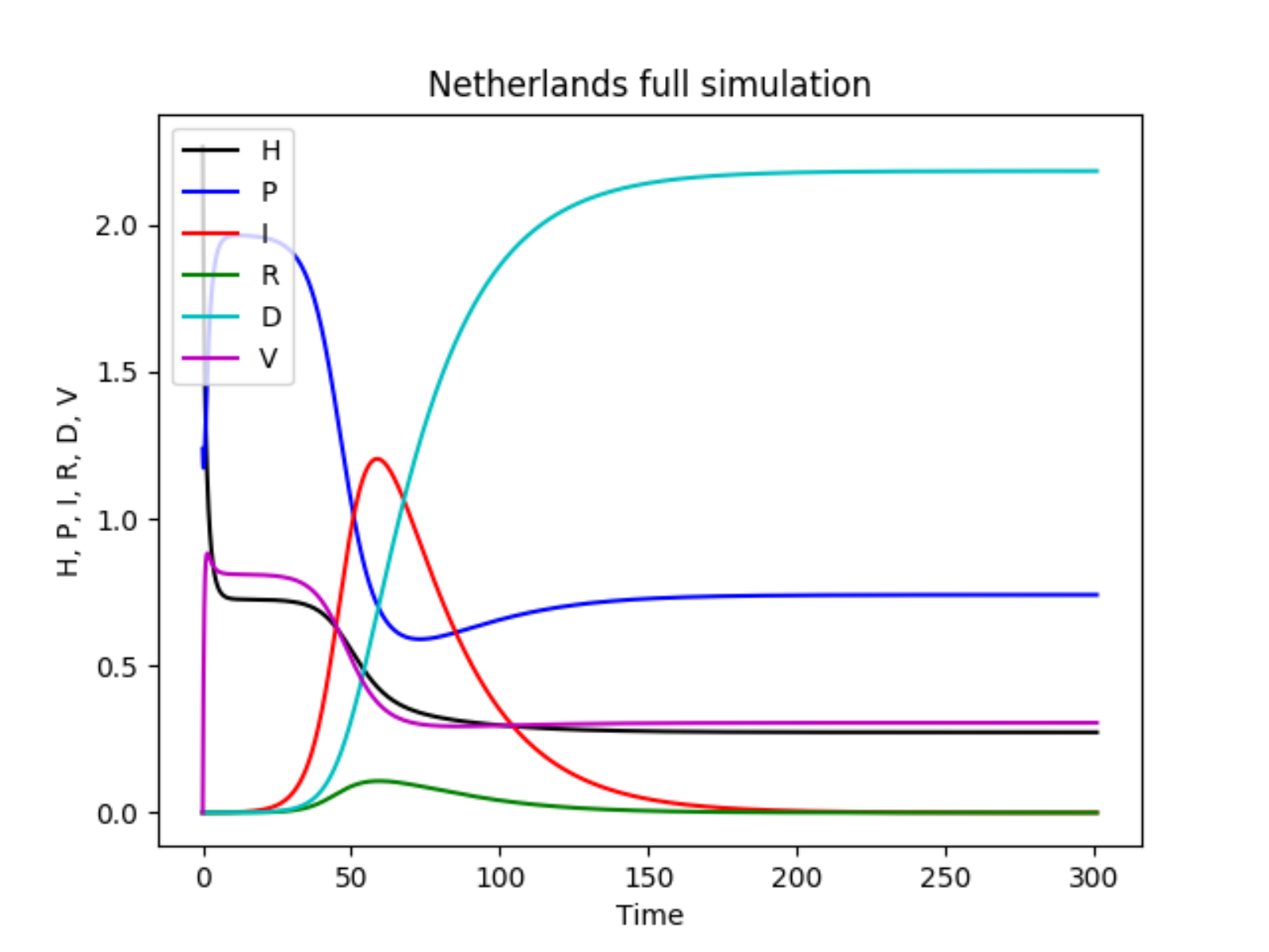}
      \caption{Full simulation plots including all 6 dimensions for the Netherlands}
    \end{subfigure}
     \begin{subfigure}[t]{0.48\textwidth}
      \includegraphics[width=\textwidth,height=0.3\textheight]{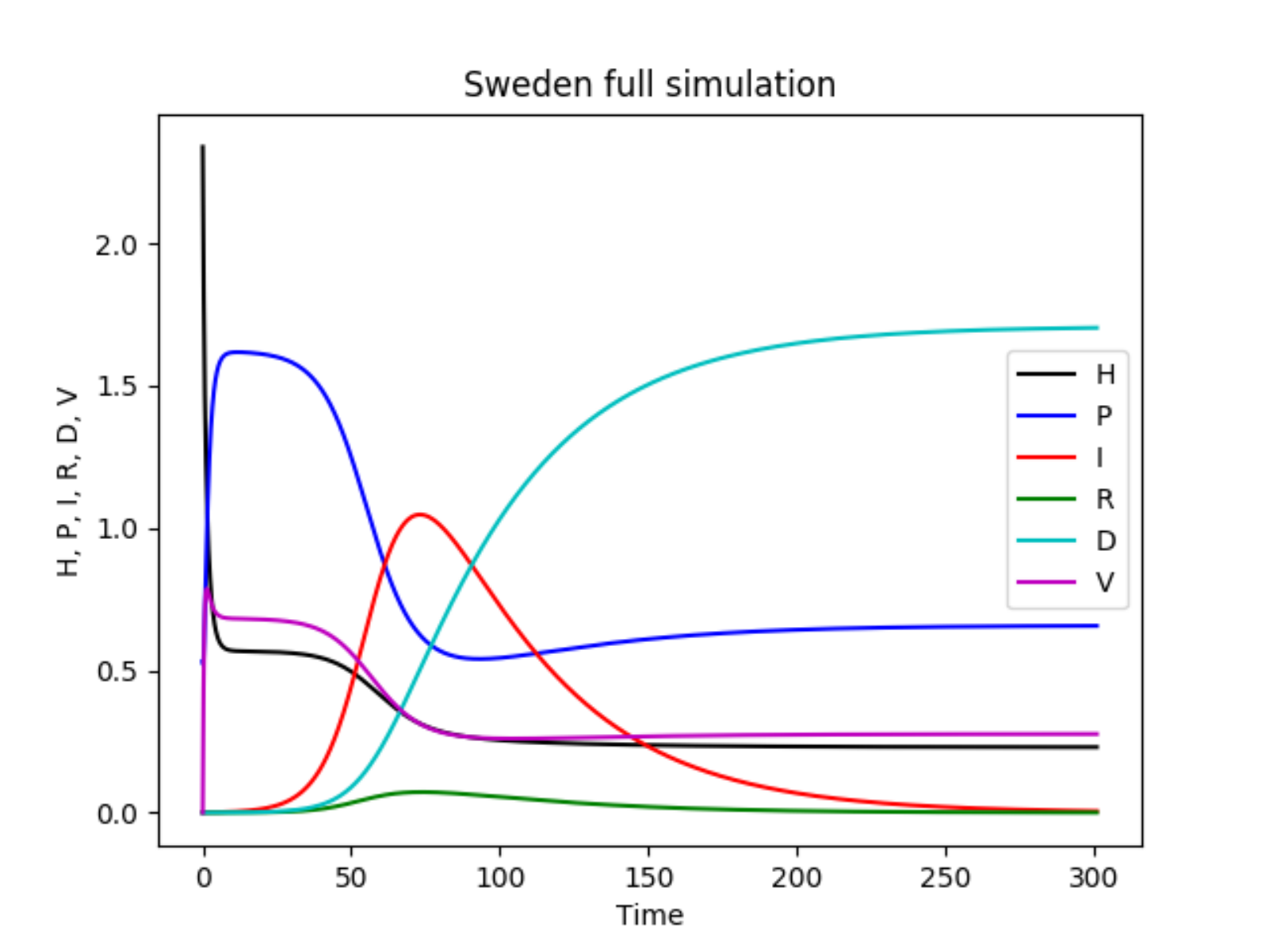}
      \caption{Full simulation plots including all 6 dimensions for Sweden}
    \end{subfigure}
         \begin{subfigure}[t]{0.48\textwidth}
      \includegraphics[width=\textwidth,height=0.3\textheight]{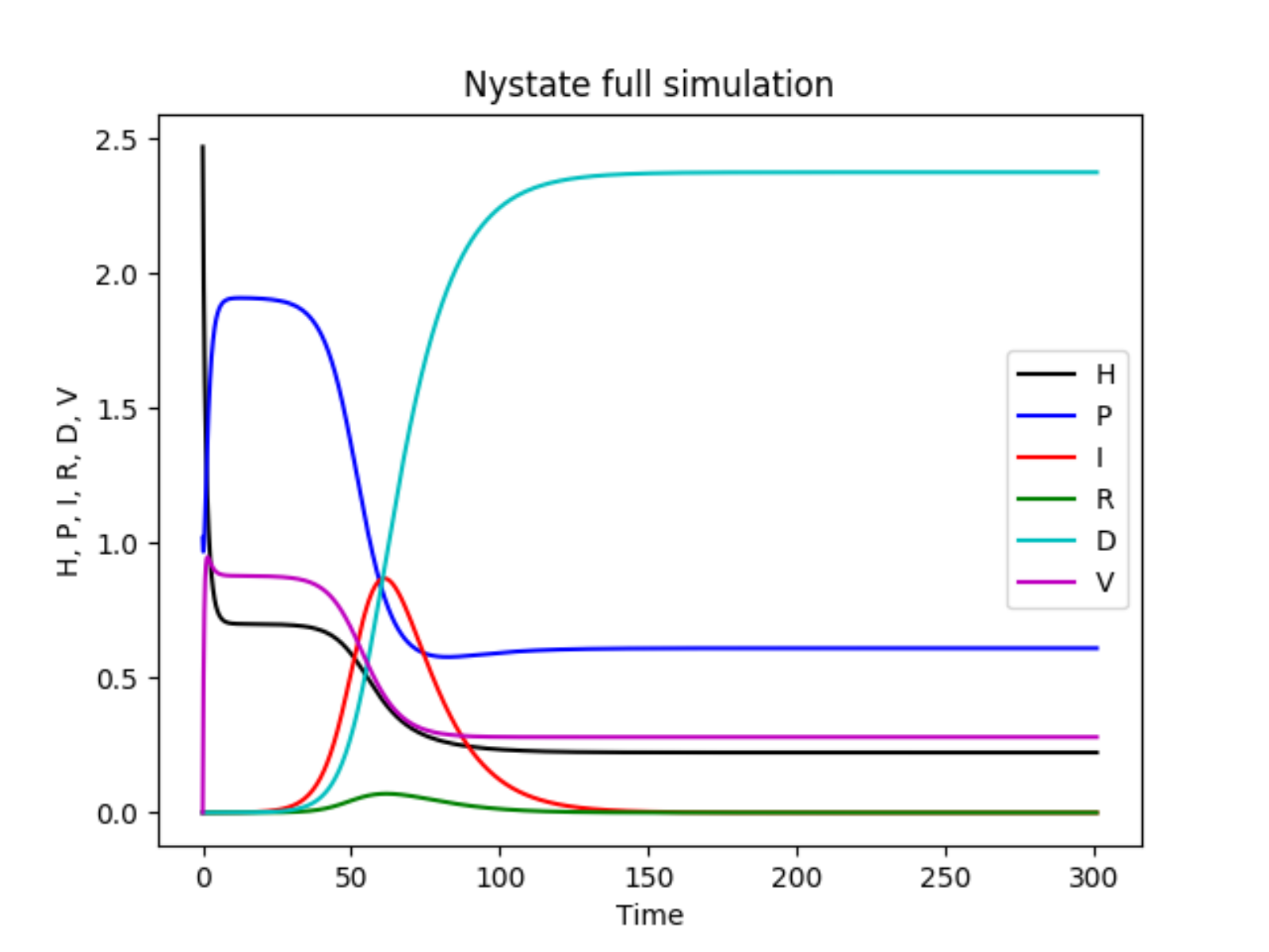}
      \caption{Full simulation plots including all 6 dimensions for New York State}
    \end{subfigure}
\caption{Full Simulation plots for Class A countries.}
\label{fig_fullsimu_classA}
  \end{figure}
  
        \begin{figure}[h!]
     \begin{subfigure}[t]{0.48\textwidth}
      \includegraphics[width=\textwidth,height=0.3\textheight]{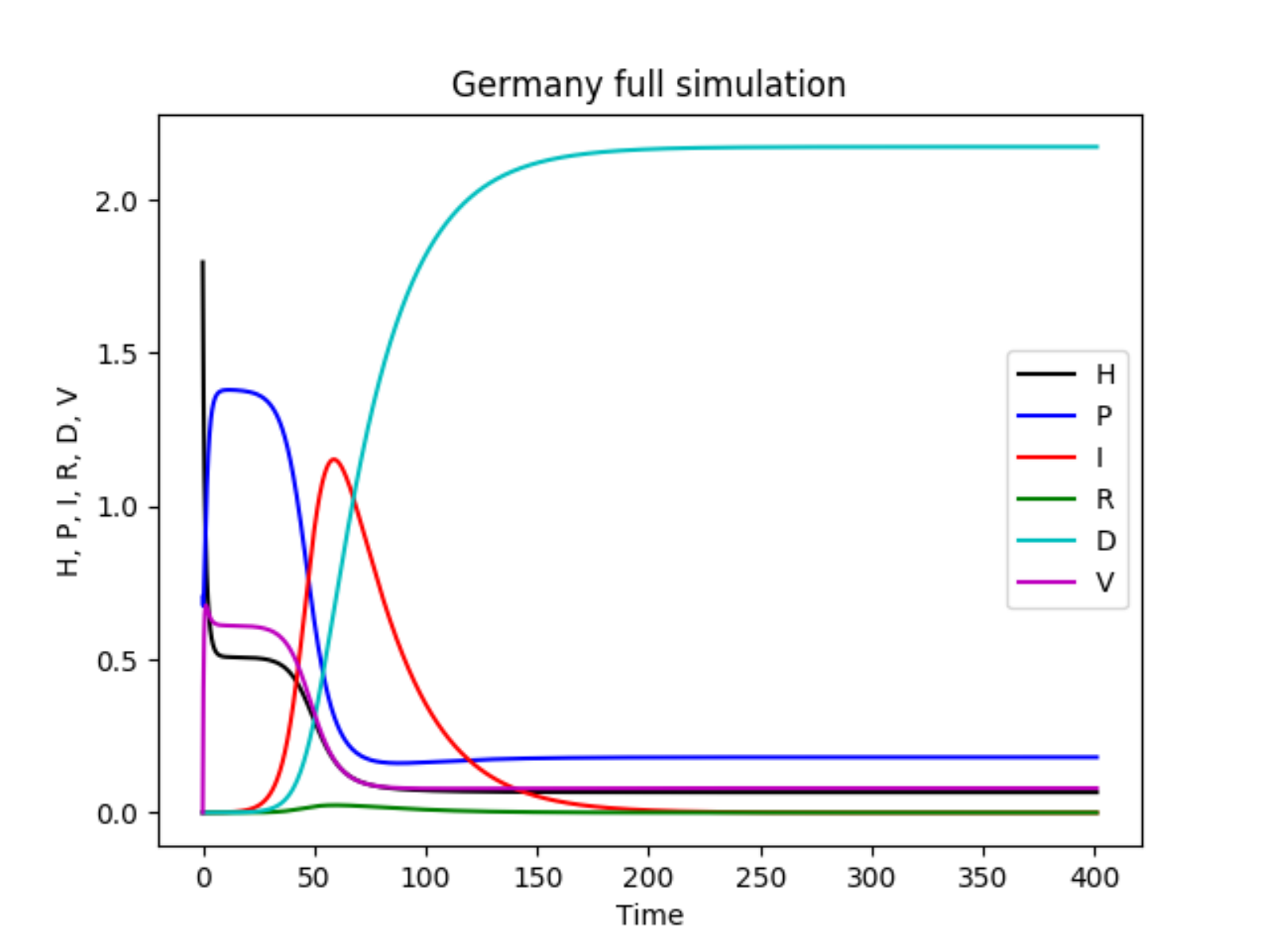}
      \caption{Full simulation plots including all 6 dimensions for Germany}
    \end{subfigure}
     \begin{subfigure}[t]{0.48\textwidth}
      \includegraphics[width=\textwidth,height=0.3\textheight]{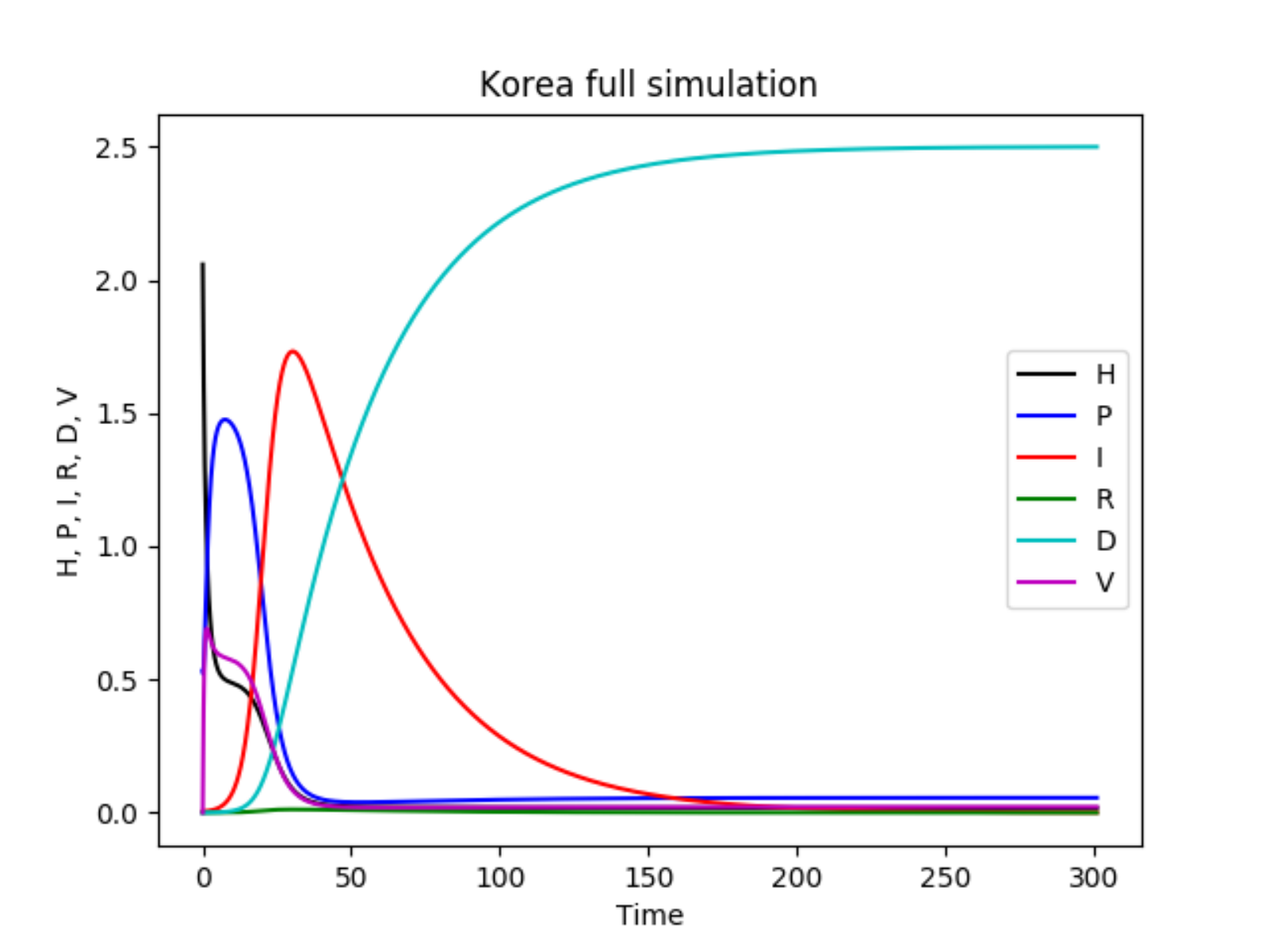}
      \caption{Full simulation plots including all 6 dimensions for Korea}
    \end{subfigure}
     \begin{subfigure}[t]{0.48\textwidth}
      \includegraphics[width=\textwidth,height=0.3\textheight]{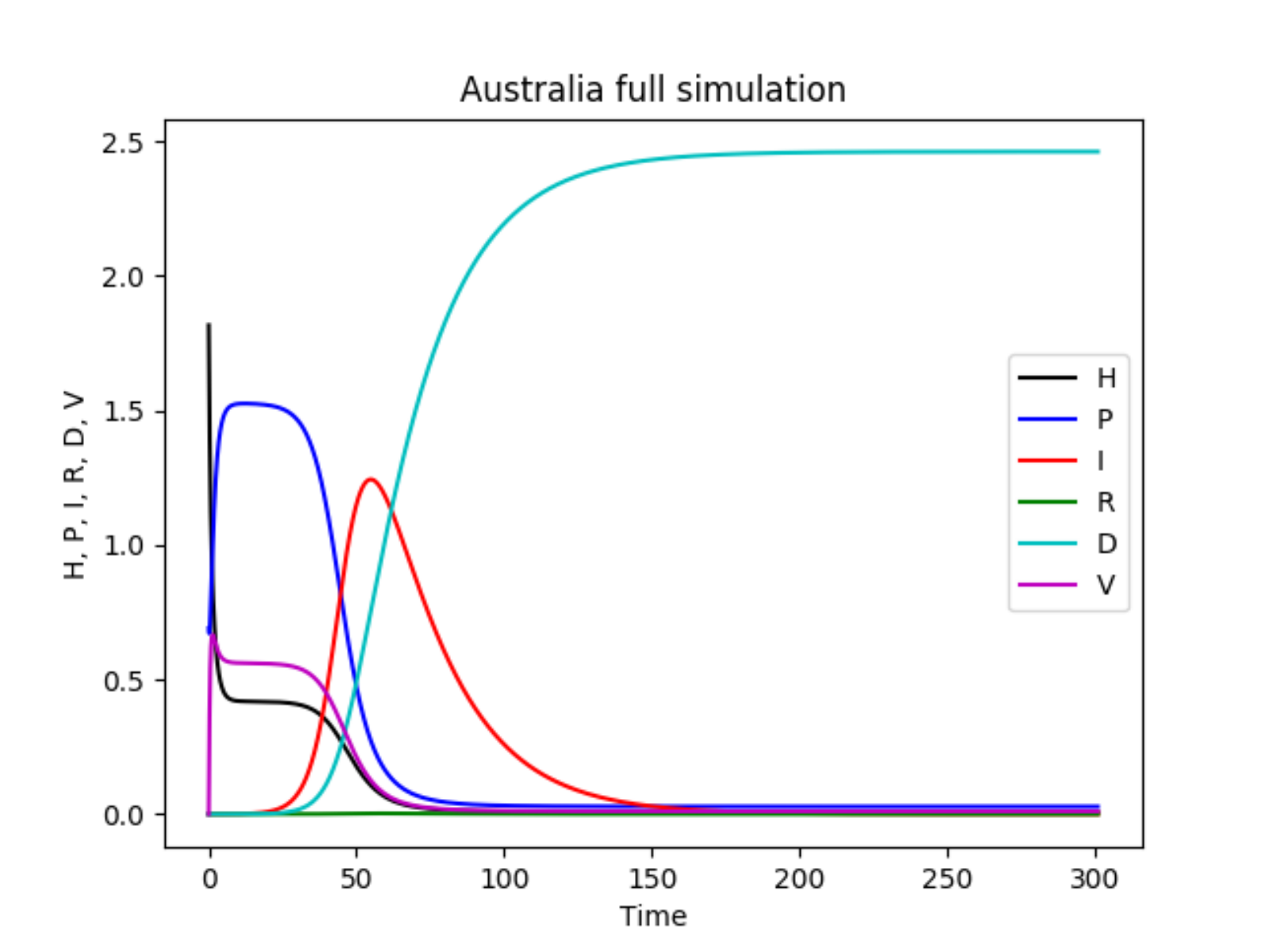}
      \caption{Full simulation plots including all 6 dimensions for Australia}
    \end{subfigure}
         \begin{subfigure}[t]{0.48\textwidth}
      \includegraphics[width=\textwidth,height=0.3\textheight]{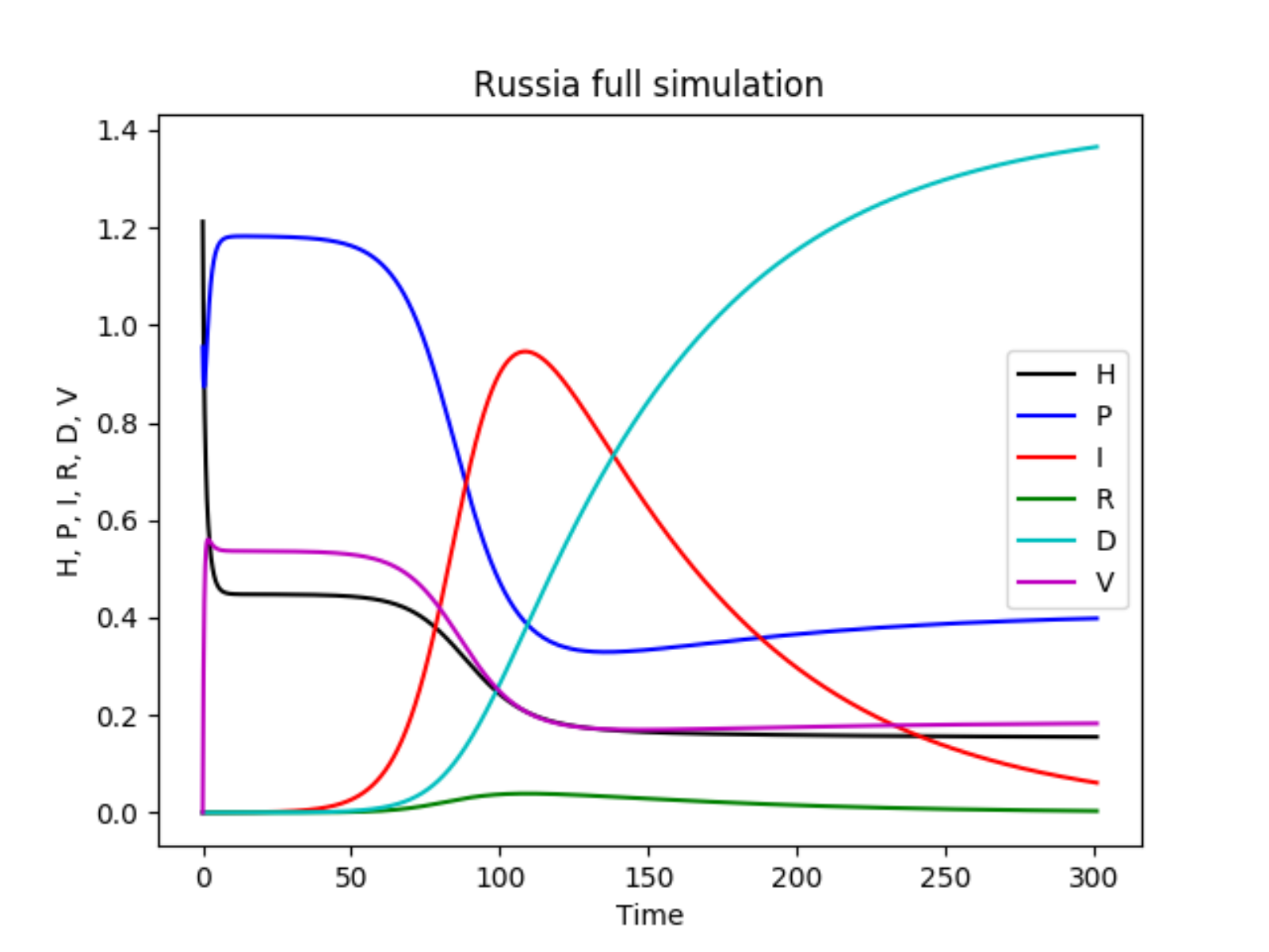}
      \caption{Full simulation plots including all 6 dimensions for Russia}
    \end{subfigure}
             \begin{subfigure}[t]{0.48\textwidth}
      \includegraphics[width=\textwidth,height=0.3\textheight]{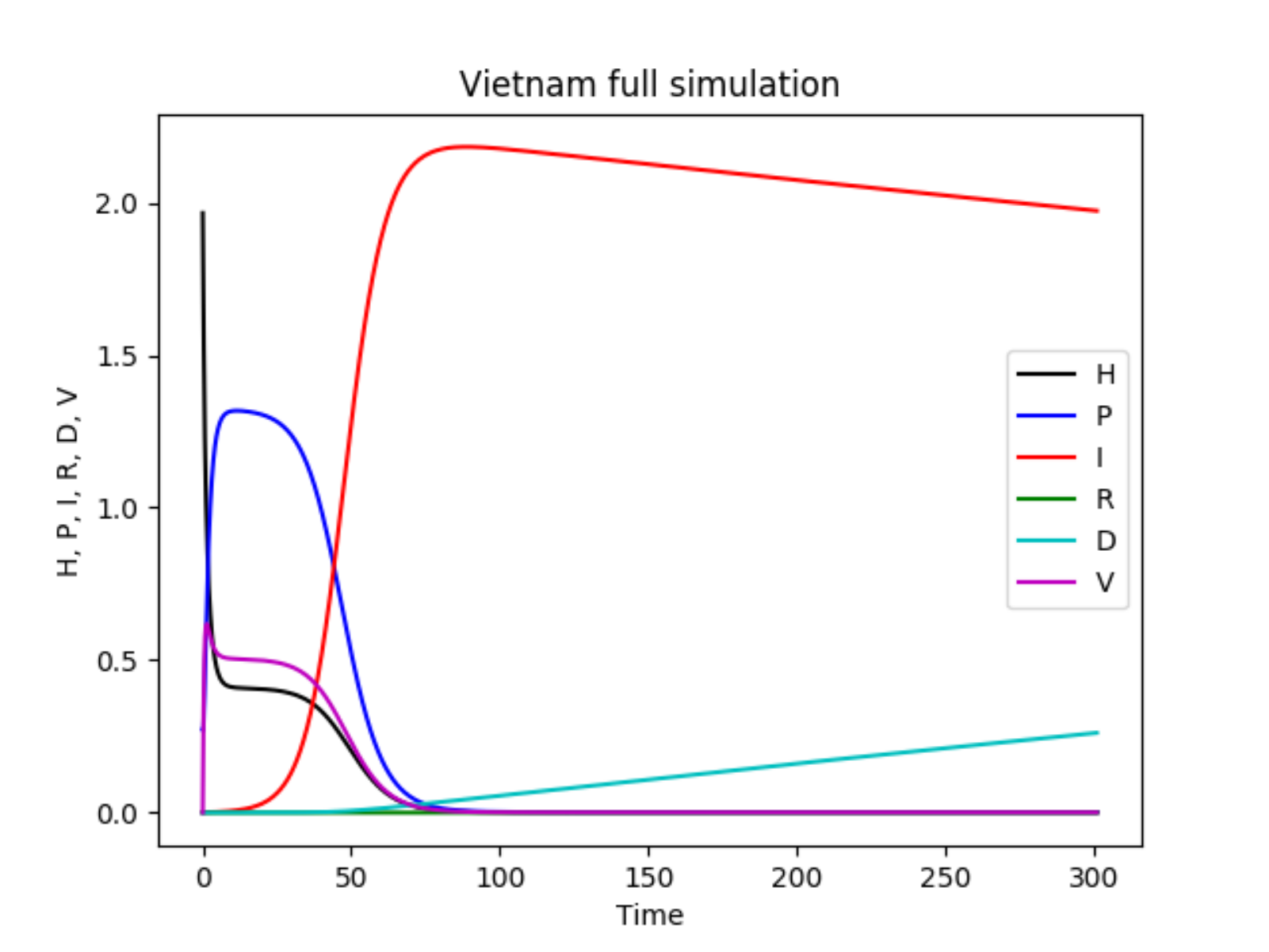}
      \caption{Full simulation plots including all 6 dimensions for Vietnam}
    \end{subfigure}
\caption{Full Simulation plots for Class B countries.}
\label{fig_fullsimu_classB}
  \end{figure}
  
          \begin{figure}[h!]
     \begin{subfigure}[t]{0.48\textwidth}
      \includegraphics[width=\textwidth,height=0.3\textheight]{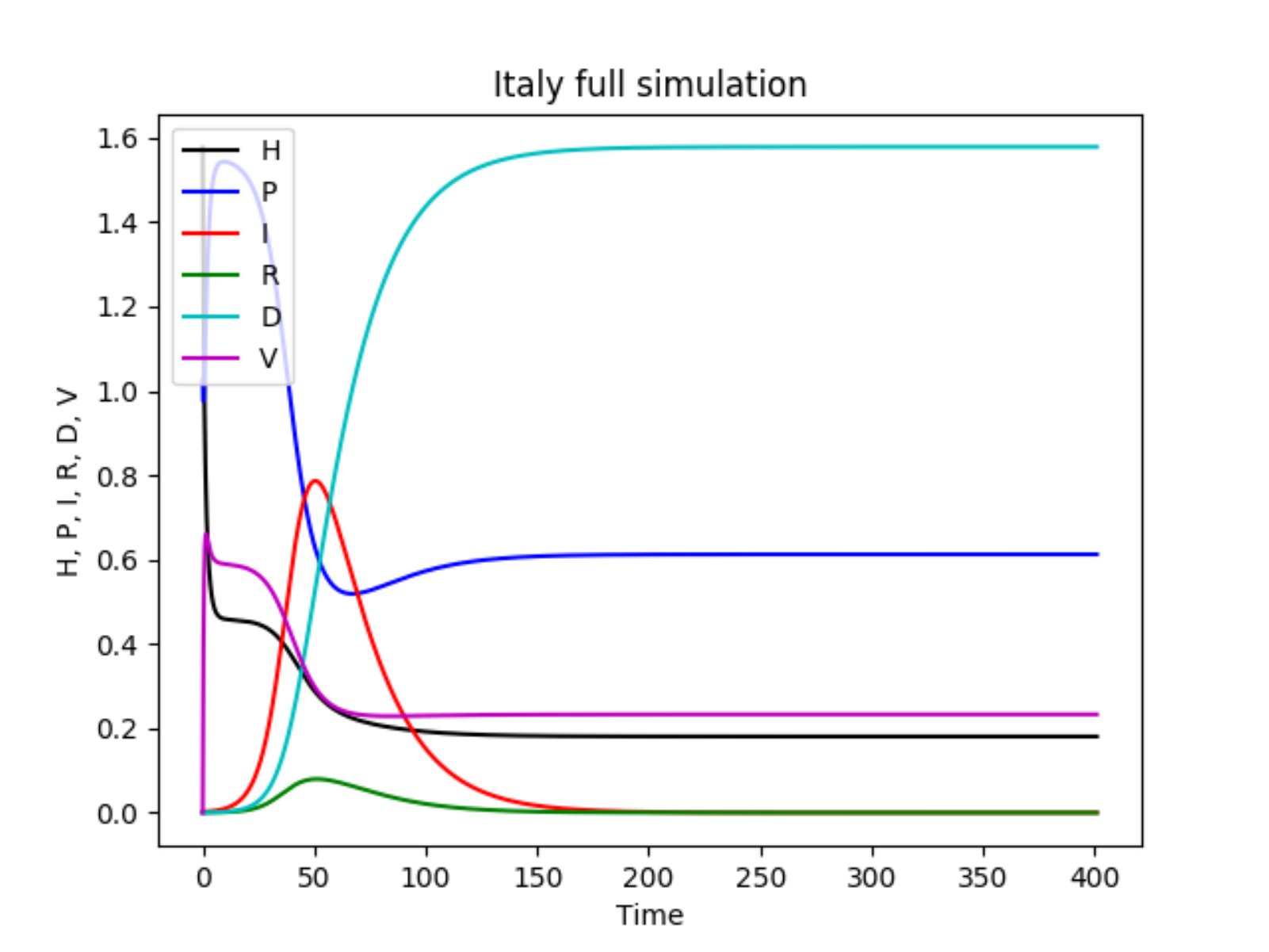}
      \caption{Full simulation plots including all 6 dimensions for Italy}
    \end{subfigure}
     \begin{subfigure}[t]{0.48\textwidth}
      \includegraphics[width=\textwidth,height=0.3\textheight]{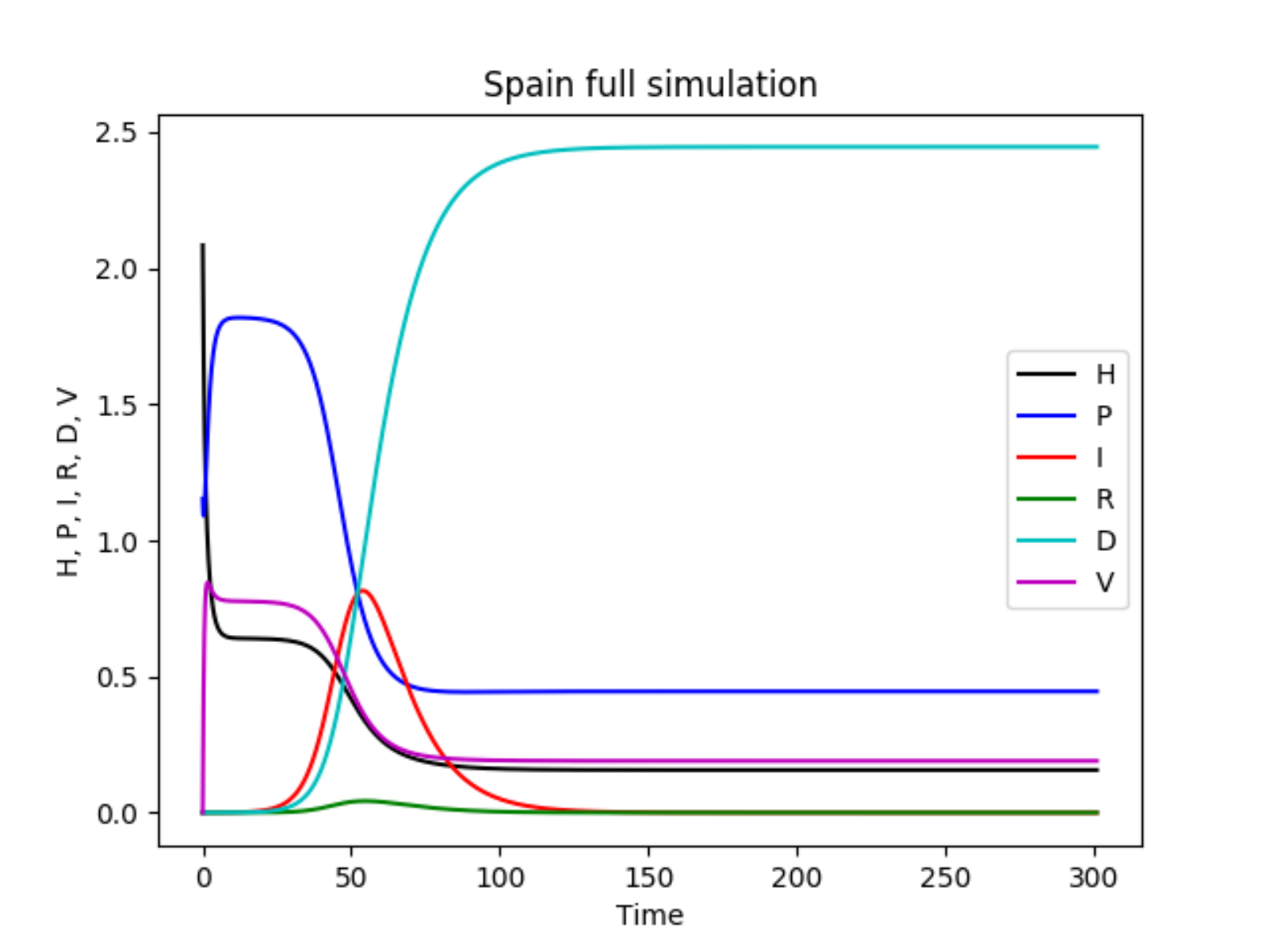}
      \caption{Full simulation plots including all 6 dimensions for Spain}
    \end{subfigure}
     \begin{subfigure}[t]{0.48\textwidth}
      \includegraphics[width=\textwidth,height=0.3\textheight]{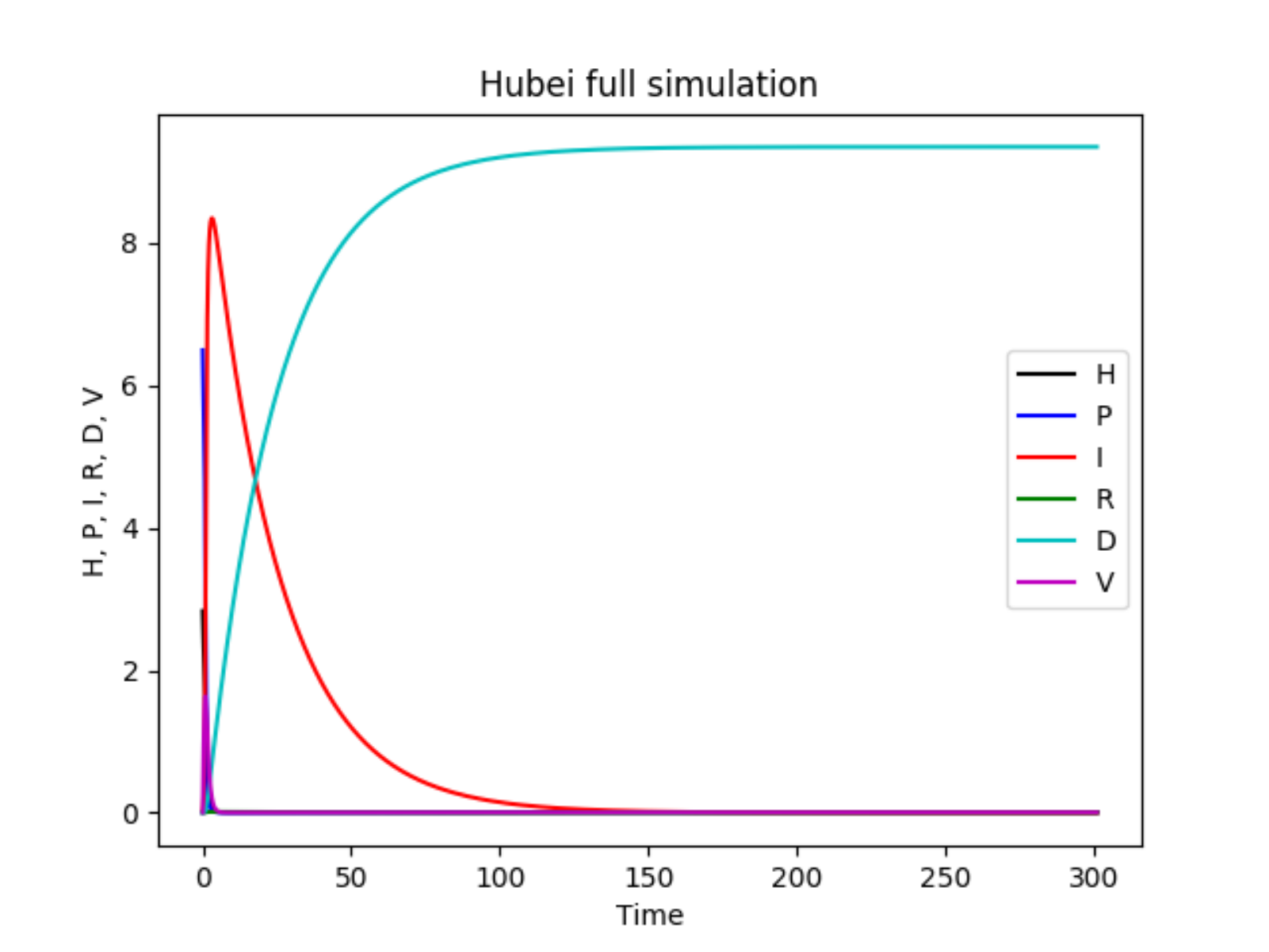}
      \caption{Full simulation plots including all 6 dimensions for Hubei}
    \end{subfigure}
\caption{Full Simulation plots for Class C countries.}
\label{fig_fullsimu_classC}
  \end{figure}

        \begin{figure}[h!]
     \begin{subfigure}[t]{0.48\textwidth}
      \includegraphics[width=\textwidth,height=0.3\textheight]{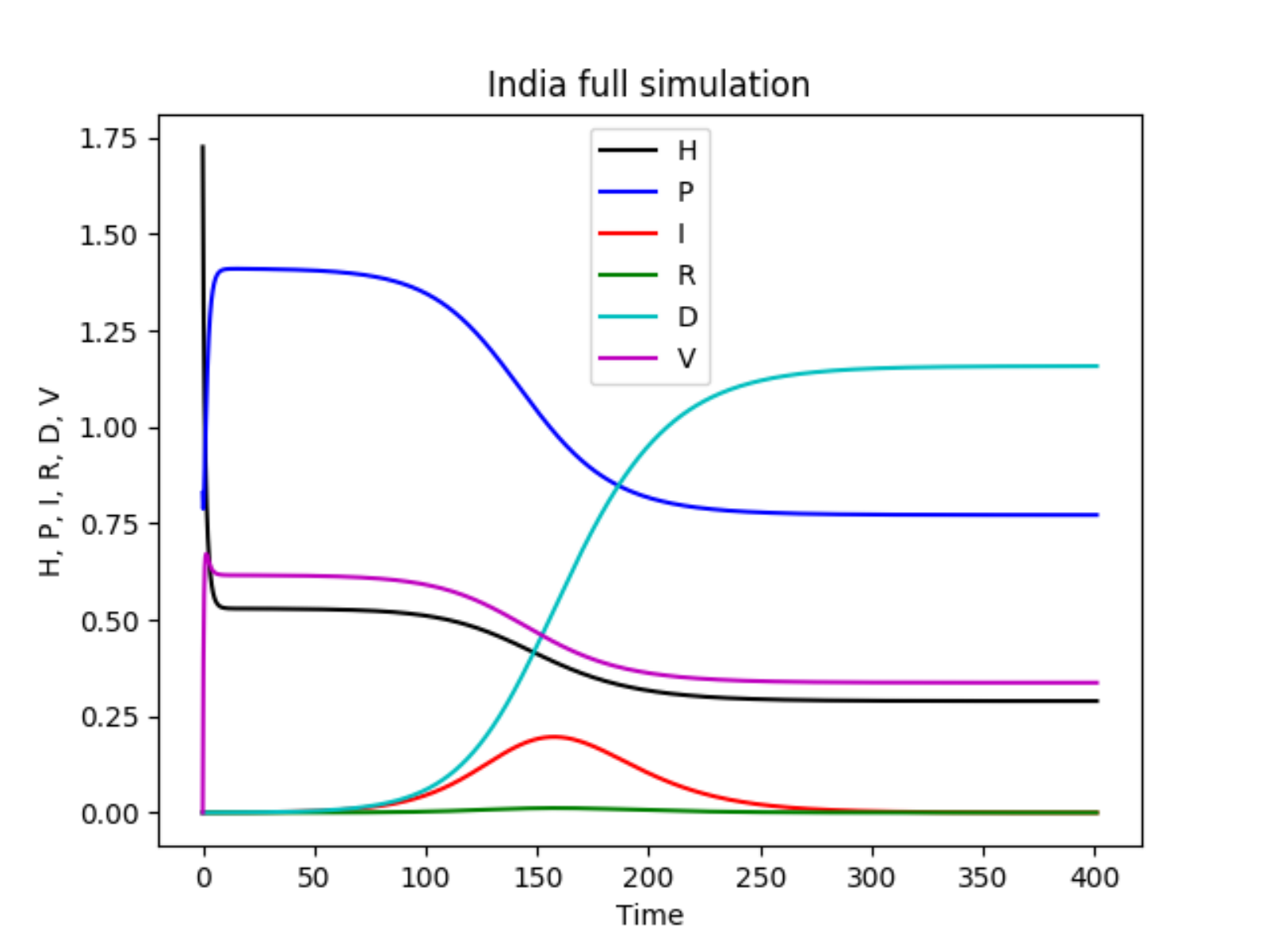}
      \caption{Full simulation plots including all 6 dimensions for India}
    \end{subfigure}
     \begin{subfigure}[t]{0.48\textwidth}
      \includegraphics[width=\textwidth,height=0.3\textheight]{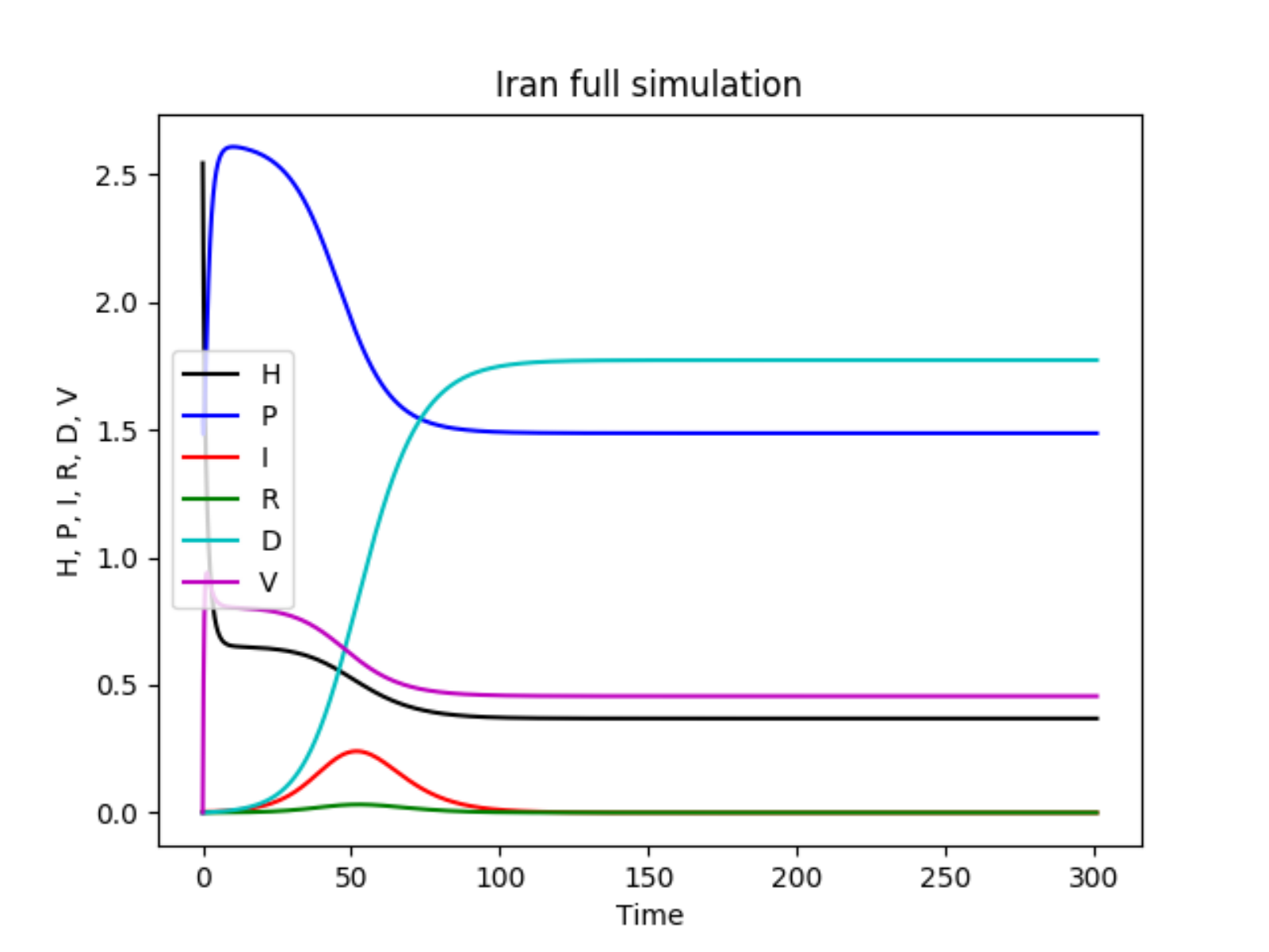}
      \caption{Full simulation plots including all 6 dimensions for Iran}
    \end{subfigure}
     \begin{subfigure}[t]{0.48\textwidth}
      \includegraphics[width=\textwidth,height=0.3\textheight]{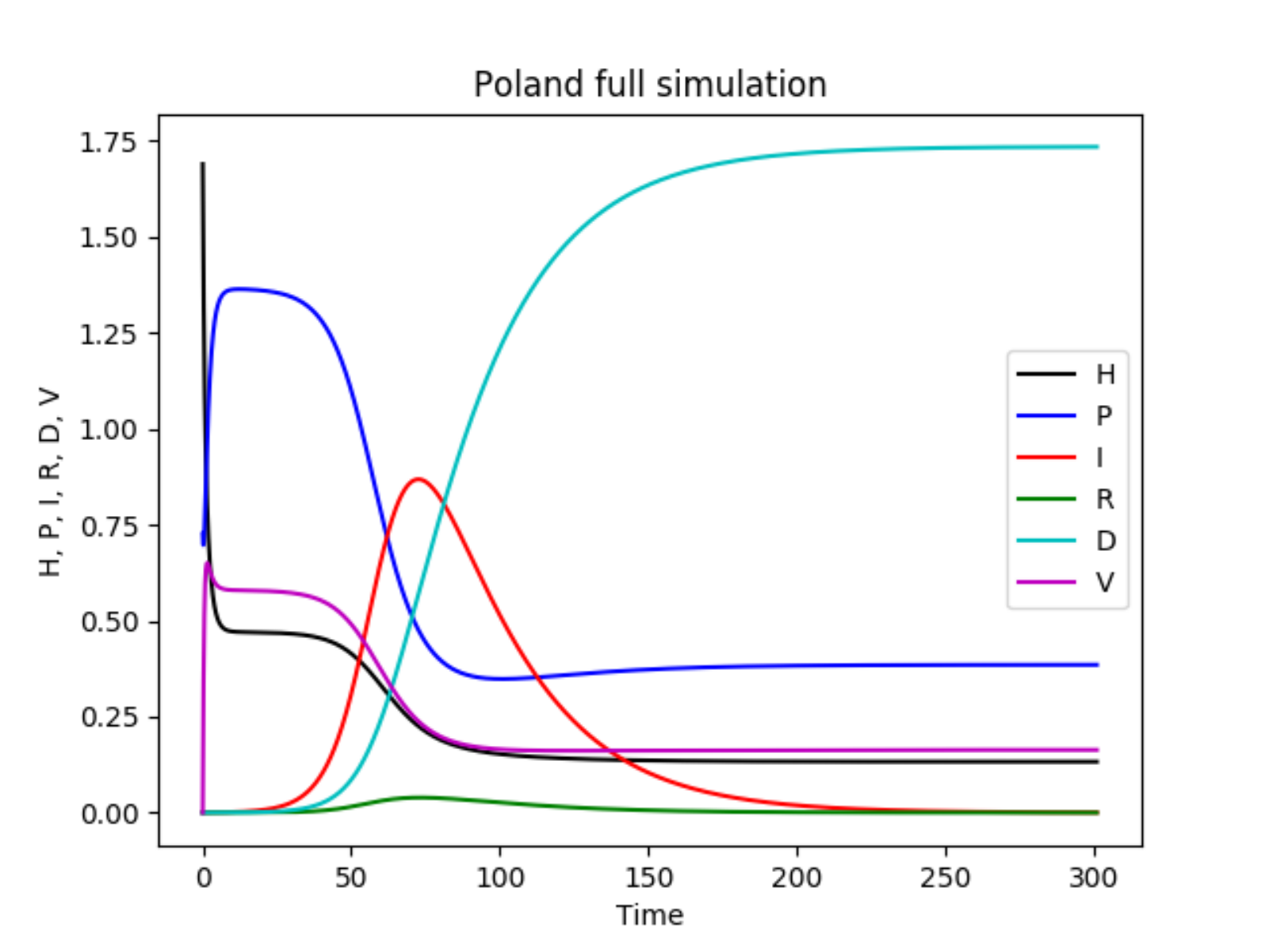}
      \caption{Full simulation plots including all 6 dimensions for Australia}
    \end{subfigure}
         \begin{subfigure}[t]{0.48\textwidth}
      \includegraphics[width=\textwidth,height=0.3\textheight]{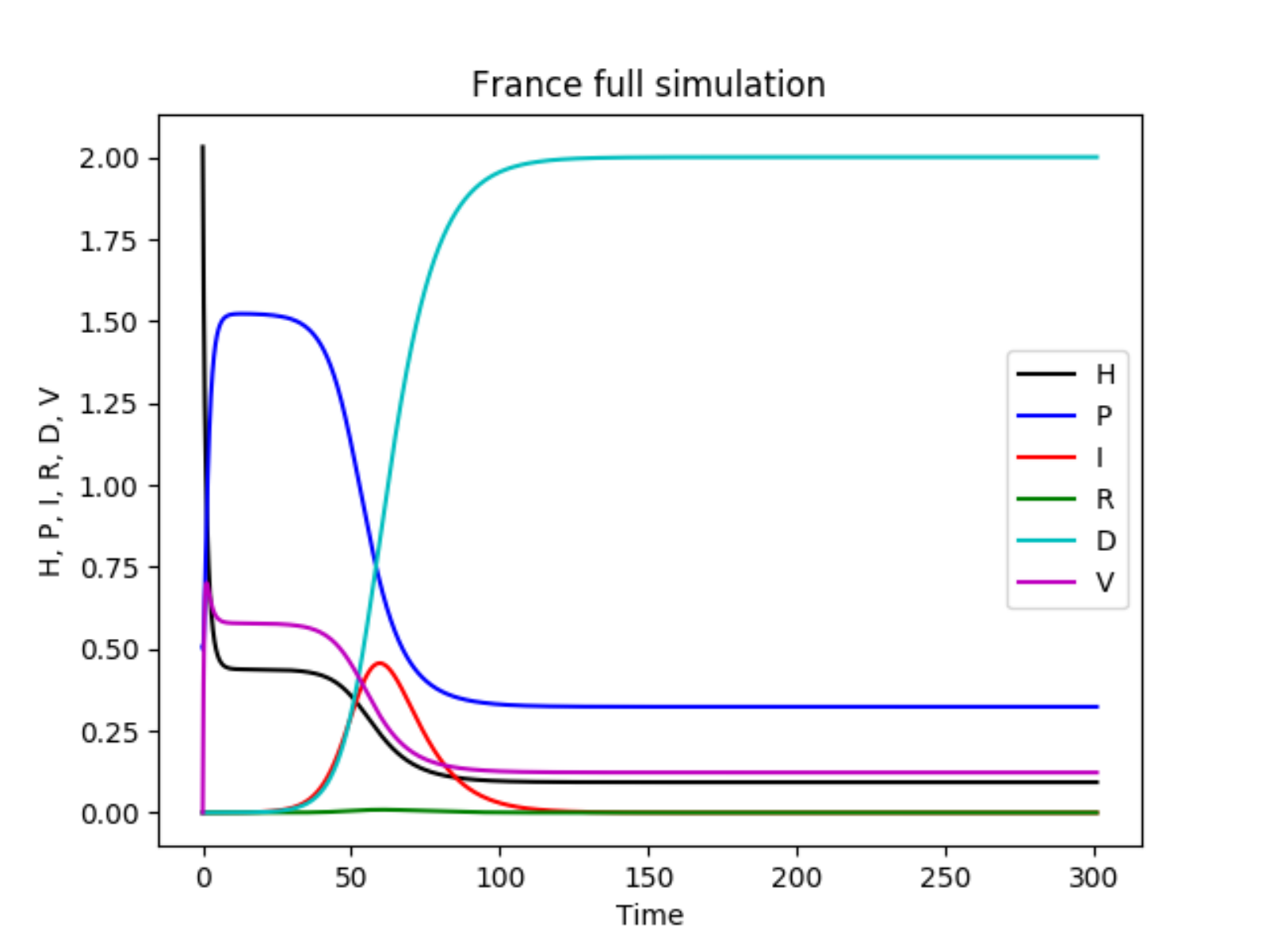}
      \caption{Full simulation plots including all 6 dimensions for France}
    \end{subfigure}
             \begin{subfigure}[t]{0.48\textwidth}
      \includegraphics[width=\textwidth,height=0.3\textheight]{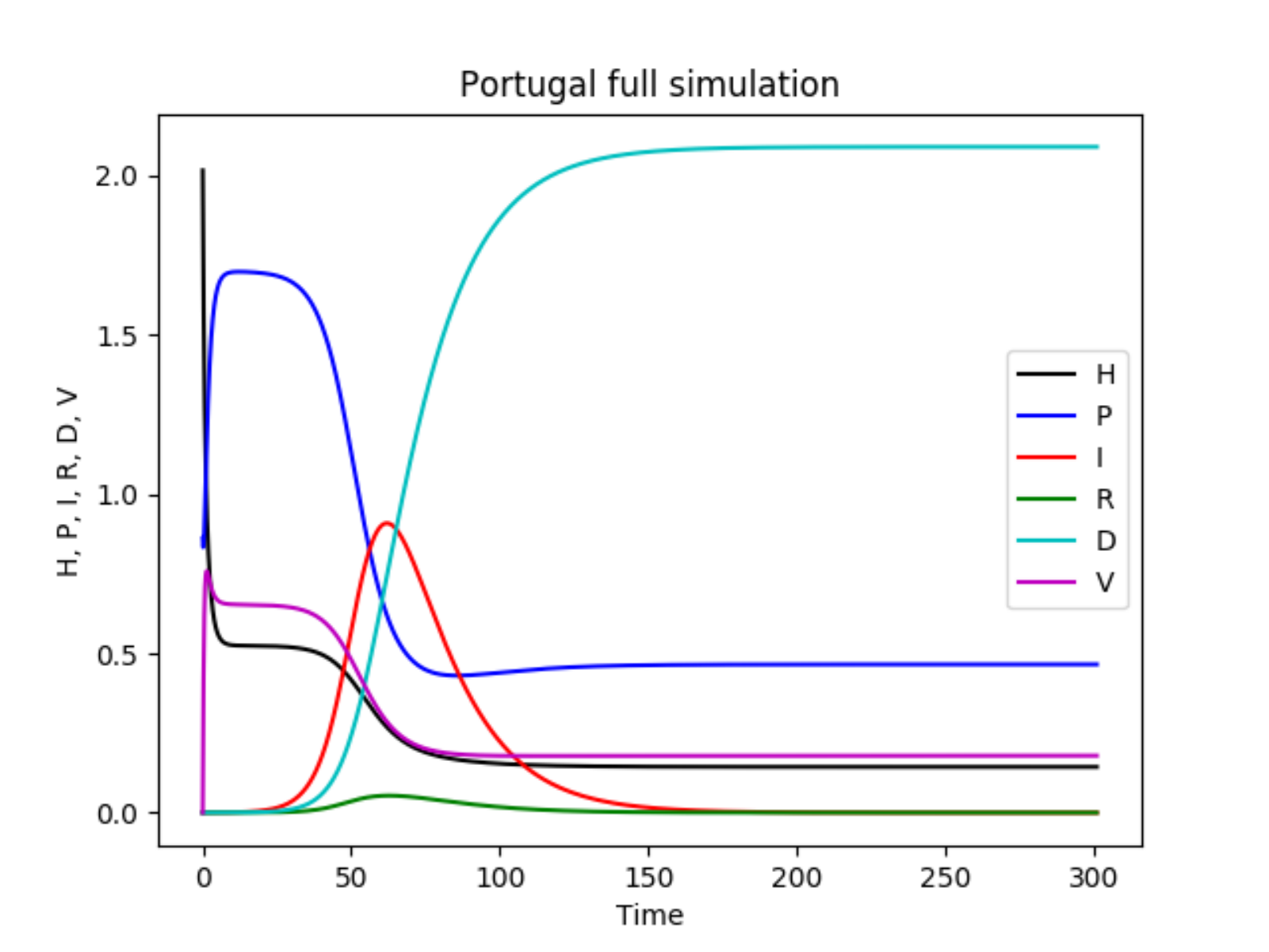}
      \caption{Full simulation plots including all 6 dimensions for Portugal}
    \end{subfigure}
                 \begin{subfigure}[t]{0.48\textwidth}
      \includegraphics[width=\textwidth,height=0.3\textheight]{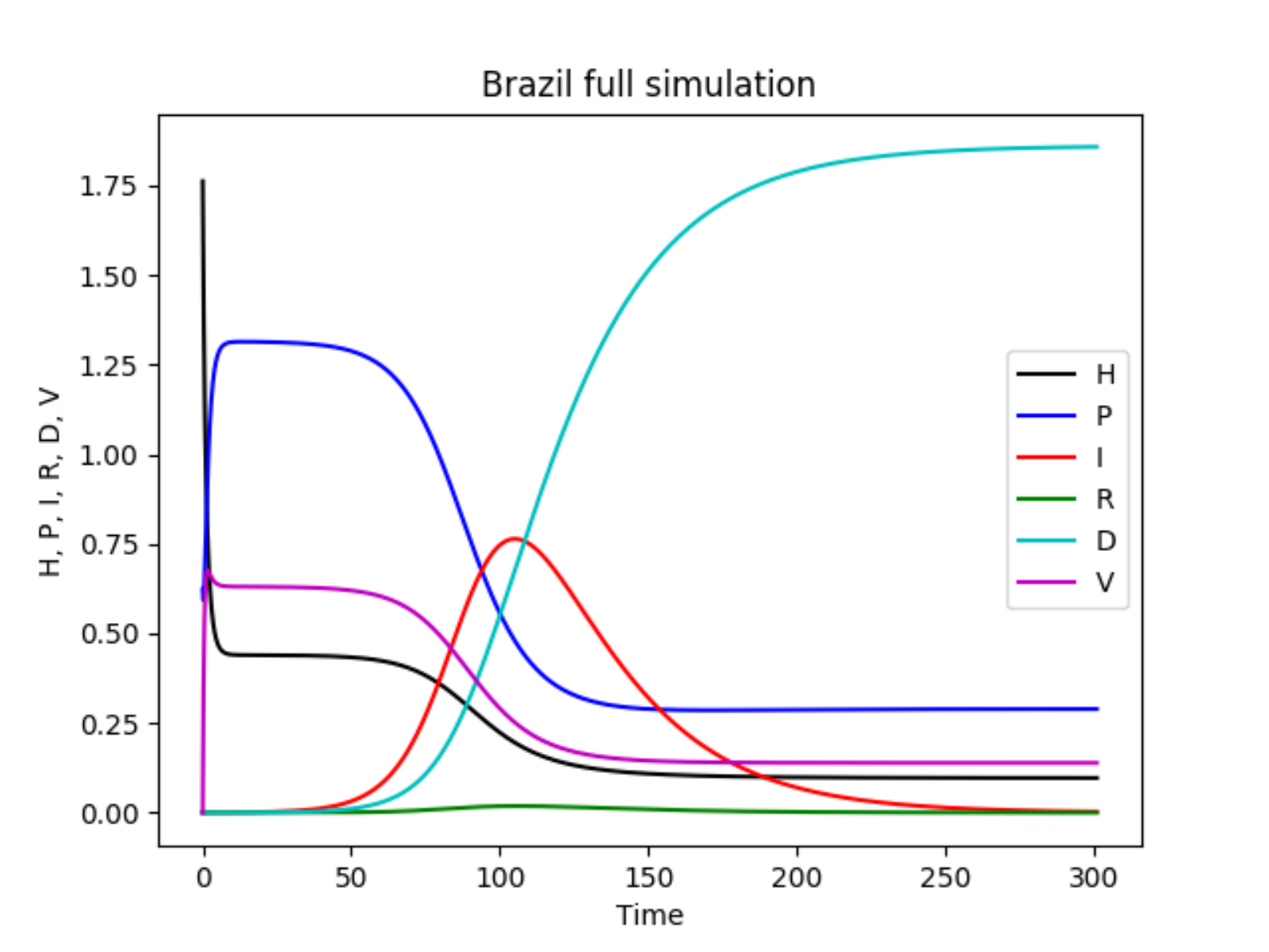}
      \caption{Full simulation plots including all 6 dimensions for Brazil}
    \end{subfigure}
\caption{Full Simulation plots for Class D countries.}
\label{fig_fullsimu_classD}
  \end{figure}

\clearpage

\subsection*{Appendix III: Comparative Estimation: Data versus Model Prediction Table}
\renewcommand\thetable{S2}
\begin{table}[h!]

\centering 
\resizebox{\linewidth}{!}{%
\begin{tabular}{|c|c|c|c|c|c|c|c|c|c|c|c|c|c|c|c|c|c|c|c|c|c|c|c|c|c|c|c|c|c|c|c|c}
\toprule \hline
 \backslashbox{\tabular{@{}l@{}}Days\endtabular}{ Country}  & \multicolumn{4}{c|}{Netherlands}& \multicolumn{4}{c|}{Sweden}& \multicolumn{4}{c|}{New York State, USA}& \multicolumn{4}{c|}{Belgium}\\
 \cline{2-17}
 & \multicolumn{2}{c|}{Infected} &\multicolumn{2}{c|}{Death}
 & \multicolumn{2}{c|}{Infected} &\multicolumn{2}{c|}{Death}
 & \multicolumn{2}{c|}{Infected} &\multicolumn{2}{c|}{Death}
 & \multicolumn{2}{c|}{Infected} &\multicolumn{2}{c|}{Death}
 \\ \cline{2-17}
 & Data & Simulation & Data & Simulation  & Data & Simulation & Data & Simulation & Data & Simulation & Data & Simulation& Data & Simulation& Data & Simulation\\
\midrule
\hline
31/05/20&102&151&5&26&775&266&65&45&1329&395&51&51&98&144&19&45\\
\hline
01/06/20&86&145&10&25&2214&260&74&44&1045&368&51&48&70&138&17&43\\
\hline
02/06/20&209&139&13&24&1080&254&20&43&1048&343&155&45&82&132&26&41\\
\hline
03/06/20&210&134&15&23&1056&248&77&42&1075&319&62&42&140&126&18&39\\
\hline
04/06/20&183&128&6&22&948&242&17&41&1108&297&44&39&165&120&14&37\\
\hline
05/06/20&239&123&2&21&843&237&3&41&781&277&94&36&154&115&15&36\\
\hline
06/06/20&165&118&3&20&403&231&35&40&702&258&43&34&122&110&11&34\\
\hline
07/06/20&164&113&15&19&791&226&23&39&683&240&41&31&89&105&13&33\\
\hline
08/06/20&184&109&11&19&890&221&78&38&674&224&84&29&132&100&10&31\\
\hline
09/06/20&164&105&2&18&1474&216&19&37&736&208&38&27&142&96&7&30\\
 \hline \bottomrule
\end{tabular}
}
\caption{Validation of Daily new Infected and Death: Netherlands, Sweden, New York State, Belgium (Class A)}
\end{table}


\renewcommand\thetable{S3}
\begin{table}[h!]

\centering 
\resizebox{\linewidth}{!}{%
\begin{tabular}{|c|c|c|c|c|c|c|c|c|c|c|c|c|c|c|c|c|c|c|c|c|c|c|c|c|c|c|c|c|c|c|c|c}
\toprule \hline
 \backslashbox{\tabular{@{}l@{}}Days\endtabular}{ Country}  & \multicolumn{4}{c|}{Korea}& \multicolumn{4}{c|}{Australia}& \multicolumn{4}{c|}{Russia}\\
 \cline{2-13}
 & \multicolumn{2}{c|}{Infected} &\multicolumn{2}{|c|}{Death}
 & \multicolumn{2}{c|}{Infected} &\multicolumn{2}{|c|}{Death}
 & \multicolumn{2}{c|}{Infected} &\multicolumn{2}{|c|}{Death}
 \\ \cline{2-13}
 & Data & Simulation & Data & Simulation  & Data & Simulation & Data & Simulation & Data & Simulation & Data & Simulation\\
\midrule
\hline
31/05/20&49&8&1&0&8&2&0&0&8858&9766&182&168\\
\hline
01/06/20&39&8&0&0&11&2&0&0&8529&9632&177&167\\
\hline
02/06/20&39&8&0&0&7&2&0&0&8823&9497&168&167\\
\hline
03/06/20&51&7&0&0&5&1&0&0&8718&9364&144&1660\\
\hline
04/06/20&57&7&0&0&7&1&0&0&8846&9231&197&165\\
\hline
05/06/20&38&7&0&0&6&1&0&0&8971&9100&134&164\\
\hline
06/06/20&38&7&1&0&2&1&0&0&8970&8970&112&163\\
\hline
07/06/20&50&6&2&0&7&1&0&0&8587&8841&171&162\\
\hline
08/06/20&45&6&0&0&11&1&0&0&8393&8714&216&161\\
\hline
09/06/20&56&6&1&0&4&1&0&0&8777&8588&172&159\\
 \hline \bottomrule
\end{tabular}
}
\caption{Validation of Daily new Infected and Death: Korea, Australia, Russia (Class B)}
\end{table}
\renewcommand\thetable{S4}
\begin{table}[h!]
\centering 
\resizebox{\linewidth}{!}{%
\begin{tabular}{|c|c|c|c|c|c|c|c|c|c|c|c|c|c|c|c|c|c|c|c|c|c}
\toprule \hline
 \backslashbox{\tabular{@{}l@{}}Days\endtabular}{ Country}  & \multicolumn{4}{c|}{Spain}& \multicolumn{4}{c|}{Hubei Province, China}\\ \cline{2-9}
 & \multicolumn{2}{c|}{Infected} &\multicolumn{2}{c|}{Death}
 & \multicolumn{2}{c|}{Infected} &\multicolumn{2}{c|}{Death}
 \\ \cline{2-9}
 & Data & Simulation & Data & Simulation  & Data & Simulation & Data & Simulation  \\
\midrule
\hline
31/05/20&294&95&0&21&0&4&0&1\\
\hline
01/06/20&394&88&1&20&0&4&0&1\\
\hline
02/06/20&334&82&5&18&0&4&0&1\\
\hline
03/06/20&318&75&1&17&0&3&0&1\\
\hline
04/06/20&332&70&1&15&0&3&0&1\\
\hline
05/06/20&240&64&1&14&0&3&0&1\\
\hline
06/06/20&167&60&0&13&0&3&0&1\\
\hline
07/06/20&249&55&0&12&0&3&0&1\\
\hline
08/06/20&314&51&0&11&0&3&0&1\\
\hline
09/06/20&427&47&0&10&0&3&0&1\\
 \hline \bottomrule
\end{tabular}
}
\caption{Validation of Daily new Infected and Death: Spain, Hubei (Class C)}
\end{table}


\renewcommand\thetable{S5}
 \begin{table}[h!]

\centering 
\resizebox{\linewidth}{!}{%
\begin{tabular}{|c|c|c|c|c|c|c|c|c|c|c|c|c|c|c|c|c|c|c|c|c|c|c|c|c|c|c|c|c|c|c|c|c}
\toprule \hline
 \backslashbox{\tabular{@{}l@{}}Days\endtabular}{ Country}  & \multicolumn{4}{c|}{Poland}& \multicolumn{4}{c|}{Iran}& \multicolumn{4}{c|}{Francs}& \multicolumn{4}{c|}{Portugal}& \multicolumn{4}{c|}{Brazil}\\
 \cline{2-21}
 & \multicolumn{2}{c|}{Infected} &\multicolumn{2}{|c|}{Death}
 & \multicolumn{2}{c|}{Infected} &\multicolumn{2}{|c|}{Death}
 & \multicolumn{2}{c|}{Infected} &\multicolumn{2}{|c|}{Death}
 & \multicolumn{2}{c|}{Infected} &\multicolumn{2}{|c|}{Death}
 & \multicolumn{2}{c|}{Infected} &\multicolumn{2}{|c|}{Death}\\ \cline{2-21}
 & Data & Simulation & Data & Simulation  & Data & Simulation & Data & Simulation & Data & Simulation & Data & Simulation& Data & Simulation & Data & Simulation& Data & Simulation & Data & Simulation \\
\midrule
\hline
31/05/20&230&231&18&15&3117&2222&64&150&0&375&107&59&195&279&12&12&28936&8346&1262&1242\\
\hline
01/06/20&292&222&23&15&3134&2152&70&147&3856&370&81&59&366&269&11&12&28633&7777&1349&1229\\
\hline
02/06/20&361&213&2&15&3574&2073&59&144&0&365&43&59&331&258&8&12&30925&7269&1473&1212\\
\hline
03/06/20&362&204&20&14&2886&1987&63&140&552&359&46&58&377&247&10&12&30830&6815&1005&1192\\
\hline
04/06/20&576&196&16&14&2269&1896&75&136&529&354&31&58&382&237&9&11&27075&6409&904&1170\\
\hline
05/06/20&575&189&4&14&2364&1803&72&131&293&347&13&57&342&226&5&11&18912&6043&525&1146\\
\hline
06/06/20&599&181&9&14&2043&1707&70&126&98&341&53&56&192&216&6&11&15654&5712&679&1120\\
\hline
07/06/20&400&175&17&13&2095&1612&74&120&141&334&84&56&421&206&7&11&32091&5413&1272&1094\\
\hline
08/06/20&282&168&23&13&2011&1517&81&115&397&327&23&55&294&196&5&10&32913&5141&1274&1066\\
\hline
09/06/20&359&162&9&13&2218&1415&78&109&358&320&27&54&310&187&7&10&30412&4892&1239&1039\\
 \hline \bottomrule
\end{tabular}
}
\caption{Validation of Daily new Infected and Death: Poland, Iran, France, Portugal, Brazil (Class D)}
 \end{table}

\end{document}